\documentclass{amsart}
\usepackage{graphicx}
\theoremstyle{definition}

\theoremstyle{remark}

\numberwithin{equation}{section}
\newcommand{\norm}[1]{\left\Vert#1\right\Vert}
\newcommand{\abs}[1]{\left\vert#1\right\vert}
\newcommand{\set}[1]{\left\{#1\right\}}
\newcommand{\Real}{\mathbb R}
\newcommand{\eps}{\varepsilon}
\newcommand{\To}{\longrightarrow}
\newcommand{\BX}{\mathbf{B}(X)}
\newcommand{\A}{\mathcal{A}}
\begin{document}

\title{Aggregation of Composite Solutions:\\
   strategies, models,
%   applied
   examples}
\author{Mark Sh. Levin}%

%%%%%%%%%%%%%%%%%%%%%%%%%%%%%%%%%%%%%%%%%%%%%%%%%%%%%%%%%%%%%%
% \address{Sumskoy Proezd 5-1-103, Moscow 117208, Russia;
% Mail address: POB 102, Moscow 117208, Russia}%
%%%%%%%%%%%%%%%%%%%%%%%%%%%%%%%%%%%%%%%%%%%%%%%%%%%%%%%%%%%%%%
%%%%%%%%%%%%%%%%%%%%%%%%%%%%%%%%%%%%%%%%%%%%%%%%%%%%%%%%%%%%%%
 \address{Home address: Mark Sh. Levin, Sumskoy Proezd 5-1-103, Moscow 117208, Russia;
 Mail address: POB 102, Moscow 117208, Russia;
 Http: //www.mslevin.itp.ru/}%
%%%%%%%%%%%%%%%%%%%%%%%%%%%%%%%%%%%%%%%%%%%%%%%%%%%%%%%%%%%%%%
%\address{Inst. for Information Transmission Problems, Russian Academy of Sciences}%
\email{mslevin@acm.org}%

%\thanks{}%
%\subjclass{}%
 %

%\date{}%
%\dedicatory{}%
%\commby{}%
% ----------------------------------------------------------------
\begin{abstract}
 The paper addresses  aggregation issues for
 composite (modular) solutions.
 A systemic view point is suggested for various aggregation
 problems.
 Several solution structures are considered:
 sets,
%  multisets,
 set morphologies, trees, etc.
 Mainly, the aggregation approach is targeted
 to set morphologies.
 The aggregation problems are based on basic structures as
 substructure, superstructure, median/consensus, and
 extended median/consensus.
 In the last case, preliminary structure is built
 (e.g., substructure, median/consensus)
 and addition of solution elements  is considered
 while taking into account profit of the additional elements and total resource constraint.
%
% The examined aggregation problems have not been previously discussed in the literature.
%
 Four aggregation strategies are
 examined:
 (i) extension strategy
 (designing a substructure of initial solutions
   as ``system kernel''
  and extension of the substructure by additional elements);
 (ii) compression strategy
  (designing a superstructure of initial solutions
  and deletion of some its elements);
 (iii) combined strategy; and
  (iv) new design strategy to build a new solution over
 an extended domain of solution elements.
 Numerical real-world examples
 (e.g., telemetry system,
 communication protocol,
 student plan,
% hierarchical
  security system,
 Web-based information system,
 investment,
 educational courses)
% multi-source information retrieval)
% (hierarchical telemetry system)
 illustrate the suggested aggregation approach.
\end{abstract}

\maketitle

%%%%%%%%%%%%%%%%%%%%%%%%%%%%%%%%%%%%%%%%%%%%%%%%%%%%
\tableofcontents

\newcounter{cms}
\setlength{\unitlength}{1mm}

% ----------------------------------------------------------------
\section{Introduction}

 Traditional approaches to generation of new design solutions
 are based on modifications/improvements
 of existing products/systems
 (e.g., \cite{cross00}).
 Often one existing product/system is used as a basic solution for
 modification/improvement
 (e.g., \cite{lev98}, \cite{lev06}, \cite{levprob07},
 \cite{lev10a},
 \cite{lev11arch},
 \cite{levsaf10a},
 \cite{levsaf11},
 \cite{levand11}) (Fig. 1.1).

\begin{center}
%\begin{picture}(80,57)
\begin{picture}(80,25)

\put(00,00){\makebox(0,0)[bl]{Fig. 1.1. Modification/improvement
of
% an
 existing system
 }}

%===============================================

\put(10,14){\oval(20,16)}

%\put(40,10){\oval(33,7)}

\put(03,17){\makebox(0,0)[bl]{Existing}}
\put(03,12.5){\makebox(0,0)[bl]{product/}}
\put(03,09){\makebox(0,0)[bl]{system}}

%\put(85,06){\line(1,0){20}} \put(85,41){\line(1,0){20}}
%\put(85,06){\line(0,1){23}} \put(105,06){\line(0,1){23}}

\put(20,14){\vector(1,0){06}}

%--------------------------- Aggregation process

\put(26,06){\line(1,0){25}} \put(26,22){\line(1,0){25}}
\put(26,06){\line(0,1){16}} \put(51,06){\line(0,1){16}}

\put(26.5,6.5){\line(1,0){24}} \put(26.5,21.5){\line(1,0){24}}
\put(26.5,6.5){\line(0,1){15}} \put(50.5,6.5){\line(0,1){15}}

\put(28,16.5){\makebox(0,0)[bl]{Modification/}}
\put(28,12.5){\makebox(0,0)[bl]{improvement}}
\put(28,8.5){\makebox(0,0)[bl]{process}}

\put(51,10){\vector(1,0){06.5}} \put(51,14){\vector(1,0){06}}
\put(51,18){\vector(1,0){06.5}}

%===============================================

\put(67,14){\oval(20,16)} \put(67,14){\oval(19,15)}

\put(61,17){\makebox(0,0)[bl]{New}}
\put(61,12.5){\makebox(0,0)[bl]{design}}
\put(61,09){\makebox(0,0)[bl]{solution}}

%\put(36,9.2){\makebox(0,0)[bl]{\(S^{agg}\)}}

\end{picture}
\end{center}

 In this article, the generation of a new modular design solution
 is examined as aggregation of a set of initial existing
 modular (composite) design solutions.
%
% The paper addresses  aggregation issues for
% composite (modular) solutions.
%  which are presented as set morphologies.
%
% Generally,
 The considered aggregation process is depicted in Fig. 1.2:

~~

 {\it Find an aggregated composite solution} \(S^{agg}\)
 {\it from a set of initial composite solutions}
 \(\{S_{1},...,S_{\tau},...,S_{m}\}\),
 i.e.,~~~
 \(\{ S_{1},...,S_{\tau},...,S_{m} \}  \Longrightarrow
 S^{agg}\).

\begin{center}
%\begin{picture}(80,57)

\begin{picture}(105,52)

\put(16,00){\makebox(0,0)[bl]{Fig. 1.2. Illustration for
aggregation process}}

%---------------------Initial space

\put(40,38.5){\oval(80,19)}

%\put(50,58){\makebox(0,0)[bl]{Extended domain}}
%\put(50,54){\makebox(0,0)[bl]{of}}
%\put(50,50){\makebox(0,0)[bl]{design elements}}

%-------------------Initial solutions

\put(18,33.5){\oval(30,5)}

\put(04.5,32){\makebox(0,0)[bl]{Initial solution \(S_{1}\)}}

%--

\put(61,34){\oval(34,6)}

\put(47,32.5){\makebox(0,0)[bl]{Initial solution \(S_{m}\)}}

%--

\put(40,42.5){\oval(30,7)}

\put(26.5,41){\makebox(0,0)[bl]{Initial solution \(S_{\tau}\)}}

%------------

\put(16,38){\makebox(0,0)[bl]{{\bf . ~. ~.}}}
\put(59,38.5){\makebox(0,0)[bl]{{\bf . ~. ~.}}}

%--

\put(18,31){\vector(1,-1){6}}

\put(40,39){\vector(0,-1){14}}

\put(61,31){\vector(-1,-1){6}}

%===============================================

%\put(86,40){\makebox(0,0)[bl]{Base of}}
\put(86,36){\makebox(0,0)[bl]{Additional}}
\put(86,32){\makebox(0,0)[bl]{system}}
\put(86,28){\makebox(0,0)[bl]{components}}
\put(86,24){\makebox(0,0)[bl]{(design}}
\put(86,20){\makebox(0,0)[bl]{elements)}}

\put(85,18){\line(1,0){20}} \put(85,41){\line(1,0){20}}
\put(85,18){\line(0,1){23}} \put(105,18){\line(0,1){23}}

\put(85,22){\vector(-1,0){25}}

%--------------------------- Aggregation process

\put(20,18){\line(1,0){40}} \put(20,25){\line(1,0){40}}
\put(20,18){\line(0,1){07}} \put(60,18){\line(0,1){07}}

\put(20.5,18.5){\line(1,0){39}} \put(20.5,24.5){\line(1,0){39}}
\put(20.5,18.5){\line(0,1){06}} \put(59.5,18.5){\line(0,1){06}}

\put(24.5,20){\makebox(0,0)[bl]{Aggregation process}}

\put(40,18){\vector(0,-1){04}}

%--

\put(40,10){\oval(44,8)} \put(40,10){\oval(43,7)}

\put(20.5,8.6){\makebox(0,0)[bl]{Aggregated solution \(S^{agg}\)}}

\end{picture}
\end{center}

%\section{Design Strategies}

 Note, the considered aggregation process has to be based on
 an analysis of engineering and/or management application(s).
  An expert-based scheme (framework) for problem solving
 (Fig. 1.3)
 is used as a basic framework for
 aggregation of composite (modular) solutions.

 Here it is necessary to point out main properties of the
 expert-based approach:

 {\it 1.} Each operation/step of the solving process has to be
 executed while taking into account an applied expert-based
 analysis of the problem situation or subsituation.

 {\it 2.} At each step of the solving process an expert opinion is
 the most important and the domain expert can correct the solving
 scheme and intermediate and/or resultant solutions.

\begin{center}

\end{center}

%
% It is reasonable to consider two aggregation situations.
%
% Thus, the first phase of the process is targeted to the
% analysis of applied situation.
% problem.
%

 Generally,  two aggregation cases have to be considered:

 {\it Case 1}. Usage of elements from \(m\) initial solutions,

 {\it Case 2.} Usage
 of elements from \(m\) initial solutions and
  additional design elements from an extended design domain
 as well.

 In {\it case 1} three solving strategies can be examined.
 First, the evident strategy consists in an analysis
 of the set of \(m\) initial solutions
 and building a system substructure (``system kernel'' \(K\),
 e.g., a ``basic system part'' as substructure, median/consensus of the initial solutions)
 that can be extended or modified.
%
% A basic aggregation situation can be examined as an analysis of
% \(m\) initial composite (modular) solutions
% and building an aggregation solution from the elements of the
% composite solutions.
%
% In this case, the analysis of \(m\) initial solutions leads to a
% ``system kernel'' \(K\)
% (e.g., substructure, median/consensus of the initial solutions)
% that can be extended.
%
 Several methods can be used to build the ``system kernel'' \(K\):
 (a) substructure of the initial solutions,
 (b) median structure (or consensus/agreement) for the initial solutions,
 (c) extended substructure,
 (d) extended median structure (or extended consensus).
 (e) selection of the ``best'' system element for each specified
 system part.

 Here expert judgment of expert(s) domain can be used
 at each stage of the solving process.

 Second, a superstructure for
 set of \(m\) initial solutions has to be designed \(\Omega\)
 (e.g., combining the initial solutions, ``covering'' the initial solutions).
 Then, the superstructure is compressed via deletion of the
 less important elements.

 In {\it case 2},
  it is possible to consider the following:

  to build a generalized domain
 of design alternatives (i.e., solution elements)
 and to design a new solution
 (including usage of new design elements).

 Here the extended design domain may be obtained via the following
  two ways:
  (a) covering/approximaiton of \(m\) initial solutions by a new
  design alternative domain,
  (b) examination of a new design alternative domain.
%

 % As a result,
 In the article,
% aggregation problems are examined for composite solutions which have
  the following structures will be examined as basic ones:
 (i) sets
 (e.g., \cite{knuth98}),
%
% (ii) multisets
% (e.g., \cite{dovier98}, \cite{knuth98}, \cite{syr01}),
%
 (ii) rankings
 (e.g., \cite{char11}, \cite{cook92}, \cite{cook96}, \cite{lev98}),
 (iii) set morphologies
  (e.g., \cite{ayr69}, \cite{jon81},
  \cite{lev98}, \cite{lev06},
    \cite{levprob07}, \cite{lev09},  \cite{zwi69}),
 (iv) trees
 (e.g., \cite{gar79}, \cite{harel87}, \cite{knuth98}).

 General structure (i.e., system model)
% \(\Lambda\)
 ({\bf \(\Lambda\)}) is:
 ~~~~ {\bf \(\Lambda\)}\(= \langle \)
% \(\overline{\rho}\) (\(\Lambda^{\alpha},\Lambda^{\beta})=\)
%
 {\bf T},{\bf P},{\bf D},{\bf R},{\bf I}
 \( \rangle\),
 where the following parts are considered
 (\cite{lev98}, \cite{lev06}, \cite{levprob07}):
 (i) tree-like system model {\bf T},
 (ii) set of leaf nodes as basic system parts/components {\bf P},
 (iii) sets of DAs for each leaf node {\bf D},
 (iv) DAs rankings (i.e., ordinal priorities) {\bf R}, and
 (v) compatibility estimates between DAs {\bf I}.
 Fig. 1.4 illustrates the generalized architecture (structure) of
 examined
 modular systems/solutions ({\bf \(\Lambda\)}).
% (\cite{lev98}, \cite{lev06}, \cite{levprob07}):
%
 Note the following significant structures are considered as well:
%%%%%%%%%%%%%%%%%%%%%%%%%%%%%%%%%%%%%%%%
%
 (a) morphology:
 ~~~~ {\bf \(\Phi \)} \(= \langle \) {\bf P},{\bf D} \( \rangle \);
 (b) morphological set:
 ~~~~ {\bf \(\overline{\Phi} \)} \(= \langle \)
  {\bf P},{\bf D},{\bf R},{\bf I} \( \rangle \);
%
%%%%%%%%%%%%%%%%%%%%%%%%%%%%%%%%%%%%%%%%%%%%%%%%%%
 (c) morphological structure (tree):
 ~~~~ {\bf \(\widetilde{\Phi} \)}\(= \langle \)
 {\bf T},{\bf P},{\bf D},{\bf R}
 \( \rangle \);
%%%%%%%%%%%%%%%%%%%%%%%%%%%%%%%%%%%%%%%%%%%%
%
 (d) morphological structure with compatibility
 (i.e., with compatibility of DAs):
 ~~~~ {\bf \(\widehat{\Phi} \)}\(= \langle \)
 {\bf T},{\bf P},{\bf D},{\bf R},{\bf I}
 \(\rangle \).

\begin{center}

\end{center}

 In Table 1.1, building problems for basic
 substructures/superstructures are briefly pointed out.

% Table 1 describes the basic target substructure/superstructure
% and corresponding bibliography references.

%
 Finally, two basis system problems are faced:
 (a) revelation/design of ``system kernel'',
% (e.g., substructure, median/ consensus),
%
 (b) building an extended design alternative domain.
%
%
%%%%%%%%%%%%%%%%%%%%%%%%%%%%%%%%%%%%%%%%%%%%%%%%%%%%%%%%%%%%%%
% It is necessary to note that the considered aggregation problem is based and
% targeted to applied (engineering/management) decision making
% problems.
%
% Evidently, corresponding strategies have to be used.
% and can be based on some corresponding strategies.
% Simple strategies consist in searching for subsets/super sets or
% their approximation.
%%%%%%%%%%%%%%%%%%%%%%%%%%%%%%%%%%%%%%%%%%%%%%%%%%%%%%%%%%%%%%%
%
%%%%%%%%%%%%%%%%%%%%%%%%%%%%%%%%%%%%%%%%%%%%%%% Apr. 13, 2011
% A composite strategy may involve two phases:
%
% {\it Phase 1:}~ Revelation of a basic set (subset,
% median/consensus or their approximation).
%
% {\it Phase 2:}~ Addition of elements to the basic set
% (via knapsack problem, multiple choice problem, HMMD).
%%%%%%%%%%%%%%%%%%%%%%%%%%%%%%%%%%%%%%%%%%%%%%%%%%%%%%%%%%%%%
%
 Four solving aggregated strategies are considered:

 (1) extension strategy:
 designing ``system kernel'' based on set of initial solutions
  (e.g., substructure, median/consensus)
  and extension of
  ``system kernel'' by additional elements;

 (2) compression strategy:
  designing a superstructure of initial solutions
  and deletion of some its elements; and

 (3) combined strategy
 (extension, deletion, and
 replacement operations for system elements over a preliminary aggregated solution), and

  (4) design strategy:
 building an extended design domain of solution elements
 and designing a new solution.

  The above-mentioned strategies are based on combinatorial
  models (as underlaying problems):
  multicriteria ranking/selection, knapsack problem,
  multiple choice knapsack problem, combinatorial morphological
  synthesis.

  The suggested approaches are illustrated through
  numerical examples including applied examples
  (e.g.,
   telemetry system,
   communication protocol,
  hierarchical security system,
  plan of students art activity,
  combinatorial investment,
 Web-based information system,
 notebook,
 educational courses).

\begin{center}

\end{center}

%%%%%%%%%%%%%%%%%%%%%%%%%%%%%%%%%%%%%%%%%%%%%%%%%%%%%%%%%%%%%%%%%%%%%%%

\section{Auxiliary Problems and Aggregation Strategies}

\subsection{Basic Auxiliary Problems}

%
% Here
% The aggregation approach is based on basic structures as
% substructure, superstructure, median/consensus, and
% extended median/consensus.
% In last case,
%%% a preliminary structure is built
%%% (e.g., substructure, median/consensus)
%%% and addition of solution elements  is used
% considered
%%% while taking into account profit of the additional element and total resource constraint.
%
% Additions- NB!
%
 In the case of two initial solutions
 as
 element sets \(A_{1}\) and \(A_{2}\),
 Fig. 2.1 illustrates
 substructure
 \(\widetilde{S}_{A_{1} A_{2}} \subseteq (A_{1} \& A_{2}) \)
 and superstructure
 \(\overline{S}_{A_{1} A_{2}} \supseteq (A_{1} \bigcup A_{2}) \)
 (via Venn-diagram).

\begin{center}
\begin{picture}(80,35)
\put(00,00){\makebox(0,0)[bl]{Fig. 2.1. Illustration for
substructure, superstructure}}

%-------------------- SuperSet

\put(00,29){\makebox(0,0)[bl]{Superstructure \(\overline{S}_{A_{1}
A_{2}} \supseteq (A_{1} \bigcup A_{2/}) \) }}

\put(09,29){\vector(0,-1){10}}

\put(40,15){\oval(80,18)} \put(40,15){\oval(79.4,17.4)}

%-------------------- Set A1

\put(31,15){\oval(38,15)}

\put(17,14){\makebox(0,0)[bl]{Set \(A_{1}\)}}

%-------------------- Set A2

\put(49,15){\oval(38,12)} \put(49,15){\oval(38.4,12.4)}

\put(55,14){\makebox(0,0)[bl]{Set \(A_{2}\)}}

%-------------------- Subset

\put(40,15){\oval(10,9.6)}

\put(35,25){\makebox(0,0)[bl]{Substructure \(\widetilde{S}_{A_{1}
A_{2}} \subseteq (A_{1} \& A_{2}) \) }}

\put(49,25){\vector(-1,-1){05.2}}

\put(35,15){\line(1,0){10}}

%%%%%%%%%%%%%%%%%%%%%%%%%%%%%%%%%%%%

\put(35.5,16){\line(1,0){9}}

\put(36,17){\line(1,0){8}}

\put(36.5,18){\line(1,0){7}}

\put(37,19){\line(1,0){6}}

\put(35.5,14){\line(1,0){9}}

\put(36,13){\line(1,0){8}}

\put(36.5,12){\line(1,0){7}}

\put(37,11){\line(1,0){6}}

%----------------------------

%\put(60,45){\circle*{2}}
%\put(60,45){\circle{4}}

%---------------

%\put(31,07){\line(0,1){11}} \put(89,07){\line(0,1){11}}

\end{picture}
\end{center}

 Further, let \(S= \{ S_{1},...,S_{i},...,S_{n} \} \) be a set of initial
 solutions (structures).
%
%
%-----------------------------------------------------------------
%
%  Let us define a metric/proximity of solutions.
 Let
% \(S= \{ S_{1},...,S_{i},...,S_{n} \} \) be a set of initial
% structures and
  function
 \( \rho (S_{i_{1}},S_{i_{2}} ) \),
  \( i_{1},i_{2} \in \{1,2,...,n\} \)
 be a proximity or a metric for the solutions.
%
% ----------------------------------------------------------------
 The following main basic auxiliary problems can be examined:

 ~~

 {\it Problem 1.} Find  a maximum substructure:

 \[\widetilde{S} = \arg ~ \max_{ \{ S' \} } ~ ( | S' | ), ~~
  \forall S' \in \bigcap_{i=1}^{n} ~ S_{i}.\]
 {\it Problem 2.} Find a minimum  superstructure
 \[ \overline{S} = \arg ~ \min_{\{ S'' \} } ~ ( | S'' | ),~~
  \forall S'' \in \bigcup_{i=1}^{n} ~ S_{i}.\]
%
%
%%%%%%%%%%%%%%%%%%%%%%%%%%%%%%%%%%%%%%%%%%%%%%%%%%%%%%%%%%%%%%%
 Now let us consider definitions of  medians
%  (generalized median) \(M^{*}\)
 for the above-mentioned set of initial sets \(S\)
 (e.g.,  \cite{hig00}, \cite{jiang04}, \cite{nicolas03}, \cite{solnon07}):

 (a) median (``generalized median'')
 \(M^{g}\)
  is:
 \[ M^{g} = \arg ~ min_{M \in D}  ~
 ( \sum_{i=1}^{n} ~ \rho (M, D_{i})  ),\]
  where \(D\) (\(D\supseteq S  \)) is a set of structures
  of a specified kind
  (searching for the median is usually NP complete problem);

 (b) simplified case of median
 (an approximation)
  as ``set median'' \(M^{s}\) over set \(S\):
 \[ M^{s} = \arg ~ min_{M \in S}  ~
 ( \sum_{i=1}^{n} ~ \rho (M, S_{i})  ).\]
  Here a representative from \(S =
  \{S_{1},...,S_{i},...,S_{n}\}\)\
  is searched for.
%
% Note
 Computation of proximity \(\rho (M, S_{i}\))
 is usually NP-complete problem as well.
 Note a similar
 ``closest string problem''
  is widely
 applied in bioinformatics
 (e.g.,  \cite{gomes08}, \cite{gramm08},
 \cite{gramm03},
 \cite{kelsey10},
 \cite{li02}, \cite{ma08},
 \cite{wang08}).
%
%%%%%%%%%%%%%%%%%%%%%%%%%%%%%%%%%%%%%%%%%%%%%%%%%%%%%%
%
 Finally, the following  problem 3 and problem 4 can be
 considered:

 ~~

 {\it Problem 3.} Find ``set median'' \(M^{s}\).

 ~~

 {\it Problem 4.} Find ``median''
 (``generalized median'')
  \(M^{g}\).

 ~~

 {\it Problem 5.} Find an extended median/consensus structure
 via addition to
 (or correstion/editing of)
  the basic median/structure
 some elements while some resource constraint(s).
 Here some elements are added to the median set (problem 3 or 4)
 while taking into account profit and required resource for the
 addition.

 ~~

 % ----------------------------------------------------------------

 The problems are considered for the following kinds of structures:
 (i) sets,
%
% (ii) multisets,
%
 (ii) {\it set morphologies},
 (iii) trees,
 and
 (iv) trees with {\it set morphologies}.
 In the case of vector-like metric/proximity
  \( \rho (S_{i_{1}},S_{i_{2}} ) \),
 Pareto-efficient solutions
 are searched for in problem 3, problem 4, and problem 5.

\subsection{Building of ``System Kernel''}

 The basic auxiliary problem consists in designing the
 ``system kernel''.
%%% This process is based on formal models and/or man-machine procedures
%%% (including expert judgment):
% building the following structures
% (on the basis of initial solutions):
%%% (a) substructure of initial solutions,
%%% (b) median/consensus of initial solutions,
%%% (c) extended version of the above-mentioned structures.
%
%%%%%%%%%%%%%%%%%%%%%%%%%%%%%%%%%%%%%%%%%%%%%%%%%%%%%%%%%%%%%
%
 Several methods can be used to build the ``system kernel'' \(K\):

 (i) substructure of the initial solutions,

 (ii) median structure (or consensus/agreement) for the initial solutions,

 (iii) extended substructure  (or extended median structure, extended
 consensus),
 and

 (iv) a two-stage framework:
 (a) specifying a set of basic system parts
 (as a subset of the system parts/components),
 (b) selection of the ``best'' system elements
 for each specified basic system part above.

 Sometimes, a subsolution
 (i.e., ``system kernel'' as substructure)
 is a very small subset.
 In this case, it is reasonable to use
 a special (more ``soft'') method to select elements for ``system kernel''.
 Let \(S= \{ S_{1},...,S_{i},...,S_{n} \} \) be a set of initial
 solutions (structures).
 Then, element \(e\) will be included into (added to) ``system kernel''
 if \( \eta_{i} \geq \alpha \) (e.g., \(\alpha \geq 0.5\))
 where \(\eta_{e}\) is the number of initial solutions which involve
 element \(e\).
 The usage of this rule will lead to an extension of the
 basic method for building the system substructure.

\subsection{Aggregation Strategies}

 Four basic aggregation strategies are examined:
 (i) extension strategy (extension of ``system kernel''),
 (ii) deletion strategy (compression of a superstructure for initial solutions),
 (iii) combined strategy (i.e., extension, deletion, and
 replacement operations for system elements over a preliminary aggregated solution), and
 (iv) design strategy (new system design).

% The extension aggregation strategy can be based on the following
% methods:
%/
% (1) addition of system component(s),
% i.e., increasing of system functionality through additional
% system function(s);
%
% (2) usage of design alternatives (i.e., several system elements
% for the same system function) to increase system redundancy,
% adaptability.

%
 The extension strategy is the following:

~~

 {\bf Type I.} Extension strategy:

 {\it Phase 1.1.} Analysis of applied problem, initial solutions,
 resources, solution elements.

  {\it Phase 1.2.} Revealing  from \(m\) initial solutions
  a basis as a subsolution or ``system kernel'' \(K\)
  (e.g., subset of elements, substructure, median/consensus).

  {\it Phase 1.3.} Forming a set of additional solution elements
 which were not included  into the basis above.

  {\it Phase 1.4.} Selection of the most important
  elements from the set of additional elements
 while taking into account the following:
 (i) profit of the selected elements,
 (ii) total resource constraint(s),
 (iii) compatibility among the selected elements and elements of
 the basis (i.e., ``system kernel'').

  {\it Phase 1.5.} Analysis of the obtained aggregated
  solution(s).

~~

 Fig. 2.2 illustrates the extension strategy.

\begin{center}
\begin{picture}(80,54)
\put(011,00){\makebox(0,0)[bl]{Fig 2.2. Scheme of extension
strategy}}

%-----

\put(13,45.5){\makebox(0,0)[bl]{Initial solutions \(\{
S_{1},..,S_{\tau},...,S_{m} \}\)}}

\put(00,44){\line(1,0){80}} \put(00,50){\line(1,0){80}}
\put(00,44){\line(0,1){06}} \put(80,44){\line(0,1){06}}

%-----

\put(18,44){\vector(0,-1){4}}
\put(03,34){\makebox(0,0)[bl]{``System kernel'' \(K\)}}
\put(18,35){\oval(36,10)}

%-----
\put(62,44){\vector(0,-1){4}}

\put(47,36){\makebox(0,0)[bl]{Element-candidates}}
\put(47,32){\makebox(0,0)[bl]{for addition W }}

\put(44,30){\line(1,0){36}} \put(44,40){\line(1,0){36}}
\put(44,30){\line(0,1){10}} \put(80,30){\line(0,1){10}}

%-----

\put(62,30){\vector(0,-1){4}}

\put(18,30){\vector(0,-1){4}}

%--

\put(19,20){\makebox(0,0)[bl]{Extension strategy (type I)}}

\put(04,18){\line(1,0){72}} \put(04,26){\line(1,0){72}}
\put(00,22){\line(1,1){4}} \put(00,22){\line(1,-1){4}}
\put(80,22){\line(-1,1){4}} \put(80,22){\line(-1,-1){4}}

%-----

\put(40,18){\vector(0,-1){4}}
\put(20,08.5){\makebox(0,0)[bl]{Aggregated solution \(S^{agg}\)}}
\put(40,10){\oval(70,08)}

%\put(40,20){\vector(0,-1){4}}
%\put(16,10){\makebox(0,0)[bl]{Aggregated solution \(S^{agg}\)}}
%\put(40,11){\oval(70,10)}

%\put(05,06){\line(1,0){70}} \put(05,16){\line(1,0){70}}
%\put(00,11){\line(1,1){5}} \put(00,11){\line(1,-1){5}}
%\put(80,11){\line(-1,1){5}} \put(80,11){\line(-1,-1){5}}

%%%%%%%%%%%%%%%%%%%%%%%%%%%%%%%%%%%%%%%%%%%%%%%%%%%%%%%%%%%%%%%%%%%%%%%%
%\put(63,20){\oval(8,5.6)}
%----------
%\put(6,31){\circle*{2}} \put(34,31){\circle*{2}}
%\put(63,31){\circle*{2}}
%%%%%%%%%%%%%%%%%%%%%%%%%%%%%%%%%%%%%%%%%%%%%%%%%%%%%%%%%%%%%%%%%%%%%%%%%%%%
%----------------

\end{picture}
\end{center}

 The second (compression) strategy is based on preliminary union of all elements
  of \(m\) initial solutions and deletion of the non-important
  elements to satisfy
%  a
  some resource constraint(s) (Fig. 2.3):

~~

 {\bf Type II.} Compression (deletion) strategy:

 {\it Phase 2.1.} Analysis of applied problem, initial solutions,
 resources, solution elements.

 {\it Phase 2.2.} Union of all elements from \(m\) initial solutions
  to form a basic supersolution (e.g., superset of elements, superstructure).

 {\it Phase 2.3.} Forming a set of element candidates to delete.

 {\it Phase 2.4.} Selection of the most non-important
  elements from the set of  element candidates
 while taking into account the following:
 (i) integrated profit of the compressed solution,
 (ii) total resource constraint(s),
 (iii) compatibility among the selected elements
 of the compressed solution.

 {\it Phase 2.5.} Analysis of the obtained aggregated
  solution(s).

~~

\begin{center}

\end{center}

 The third strategy is consists in possible combination of
 addition operations, deletion operation,  and
 correction (replacement) operations.
 The third strategy is (Fig. 2.4):

\begin{center}
%
\end{center}

 {\bf Type III.} Combined strategy:

 {\it Phase 3.1.} Analysis of applied problem, initial solutions,
 resources, solution elements.

 {\it Phase 3.2.} Revealing  from \(m\) initial solutions
  a basis as a subsolution or ``system kernel'' \(K\)
  (e.g., subset of elements, substructure, median/consensus).

  {\it Phase 3.3.} Forming the following:

  (a) a set of additional solution elements
  as candidates for addition
  \(W \)
 which were not included  into the basis above
 (\( | W \bigcap K | = 0\)),

 (b) a subset  \( B \subseteq K\) as  candidates for deletion, and

 (c) a set of element pair
 \(C =   \{(a_{u},b_{v})| (a_{u} \in K) \& ( b_{v} \overline{\in} K)  \}\),
 thus a set of correction operations is considered
 as replacement of \(a_{u}\) by \(b_{v}\).

  {\it Phase 3.4.}  Selection of the most important
  operations (including element addition, element deletion, and
  element replacement)
 while taking into account the following:
 (i) profit of the operations,
 (ii) total resource constraint(s),
 (iii) compatibility among the selected elements in the resultant solution.

 {\it Phase 3.5.} Analysis of the obtained aggregated solution(s).

~~~

 The fourth strategy is targeted to usage of additional elements
 from an extended design domain (``design space'').
% which were not included into \(m\) initial solutions.
 Here the resultant aggregated solution may involve elements which
 were not belonging to \(m\) initial solutions.
 The fourth strategy is:

~~

 {\bf Type IV.} Strategy of extended design domain:

 {\it Phase 4.1.} Analysis of applied problem, initial solutions,
 resources, solution elements.

 {\it Phase 4.2.} Extension of the union of all elements from \(m\) initial solutions
  to form an extended design element domain (``design space'').

 {\it Phase 4.3.} New design of the composite solution over
  the obtained design element domain
   while taking into account the following:
 (i)  profit of the designed solution,
 (ii) total resource constraint(s),
 (iii) compatibility among the selected
 (i.e., resultant)
  elements
 of the compressed solution.

 {\it Phase 4.4.} Analysis of the obtained aggregated solution(s).

~~~

 Fig. 2.5 illustrates the fourth strategy. Given three initial
 solutions \(S_{1}\), \(S_{2}\), and \(S_{3}\).
 The resultant aggregated solution \(S^{agg}\)
 can involve elements from solution \(S_{1}\), \(S_{2}\)
 and
 elements from the extended design domain.

% In this article, the first strategy is mainly examined.
%

\begin{center}
\begin{picture}(80,36)
\put(0,00){\makebox(0,0)[bl]{Fig. 2.5. Illustration for
aggregation design strategy}}

%  \put(00,39){\line(1,0){110}}

%---------------------Design space

\put(40,19){\oval(80,26)}

\put(50,28){\makebox(0,0)[bl]{Extended domain}}
\put(50,24){\makebox(0,0)[bl]{of}}
\put(50,20){\makebox(0,0)[bl]{design elements}}

%-------------------Initial solutions

\put(18,11){\oval(30,6)}
\put(16,09.5){\makebox(0,0)[bl]{\(S_{1}\)}}

\put(60,12){\oval(36,8)} \put(58,10){\makebox(0,0)[bl]{\(S_{3}\)}}

\put(35,26){\oval(26,8)}
\put(33,24.5){\makebox(0,0)[bl]{\(S_{2}\)}}

%--

\put(27.5,19){\oval(8.6,20)} \put(27.5,19){\oval(7.6,19)}
\put(24,17.5){\makebox(0,0)[bl]{\(S^{agg}\)}}

\end{picture}
\end{center}

%====================================================================

%~~

%======================================================================================================

% ----------------------------------------------------------------
\section{Examined Structures, Substructure, Superstructure}

  Generally, the following basic kinds of sets are examined:
 sets, multisets, lists, complete lists, and trees
 (e.g.,  \cite{dovier98}, \cite{knuth98}, \cite{syr01}).
 In this material,
 our main examination is targeted to special kinds of composite
 structures:
 ~ {\it morphological structures}.

% ----------------------------------------------------------------
\subsection{Sets}

 Evidently, sets are
% the
  basic structures.
 Fig. 3.1 depicts a numerical example of two initial sets \(A\) and
 \(B\).

\begin{center}

\end{center}

 Fig. 3.2 contains examples of subset \(\widetilde{S}_{AB}\)
 and superset \(\overline{S}_{AB}\).

\begin{center}
%
\end{center}

%%%%%%%%%%%+++++++++++++++++++++++++++++++++++++++++++++++++++

%\subsection{Multi-set Case}

 Now let us consider the case of \(m\) sets.
%
% The extended median/consensus for \(m\) sets
%
 Let ~\( \{A_{1},...,A_{i},...,A_{m}\}\)~
 be  initial  sets.
 Then, a subset is:
 ~\(\widetilde{S}_{\{A_{i}\}} ~ \subseteq ~ \bigcap_{i=1}^{m}
 A_{i}\).
 A superset is:
  ~\(\overline{S}_{\{A_{i}\}} ~ \supseteq  ~\bigcup_{i=1}^{m}
 A_{i}\).

% ----------------------------------------------------------------
%\subsection{Aggregation of multisets}

\subsection{Rankings (Layered Sets)}

%%%%%%%%%%%%%%%%%%%%%%%%%%%%%%%%%%%%%%%%%%%%%%%%%%%%%%%%%%%%%%%%%%%%

  Here ranking is examined as a layered set.
 Let ~\(A = \{ 1,...,i,...,n \}\)~ be a set of elements/items.
 Ranking (a partial order/partition of set \(A\))
 is considered as linear ordered subsets of ~\(A\)~ (Fig. 3.3):~~
 \( A = \bigcup_{k=1}^{m} A(k)\),~ \(|A(k_{1}) \cap
 A(k_{2})| = 0\)~~
 if~~ \(k_{1}\neq k_{2}\),
 ~~\( i_{2}\preceq i_{1}\)~~ \(\forall i_{1} \in A(k_{1})\),
 ~\(\forall i_{2} \in H(k_{2})\), ~ \(k_{1} \leq k_{2}\).
 Set ~\(A(k)\) is called layer ~\(k\), and each item
 ~\(i \in A\)~
 gets priority \(r_{i}\)
 that equals the number of the corresponding layer.
 The described partition of \(A\) is be called as
%%%%%%%%%%%%%% NB
  partial ranking,
%   \cite{dinu06},
  stratification,
%  \cite{levmih88},
  layered set
%    \cite{bellev90}.
%
 (e.g., \cite{bellev90}, \cite{dinu06}, \cite{lev88}, \cite{lev98a},
 \cite{lev98}, \cite{lev06}, \cite{levmih88},
  \cite{roy96}, \cite{zap02}).
 Evidently, a linear order of elements from \(A\)
 is ranking as well.
 Many years ranking problems (or sorting) have been intensively
 used and studied in various domains
 (e.g., \cite{bellev90}, \cite{brin99}, \cite{dinu06},
 \cite{char11}, \cite{kemeny59}, \cite{kemeny60},
  \cite{lev06}, \cite{levmih88}, \cite{roy96}, \cite{zap02},
   \cite{zop09}).

\begin{center}
\begin{picture}(69,47)
\put(10,0){\makebox(0,0)[bl]{Fig. 3.3. Scheme of ranking}}

\put(15,25){\oval(29,35)}

%\put(5,32){\makebox(0,0)[bl]{Set of items}}

\put(3,27){\makebox(0,0)[bl]{Initial elements}}

%\put(5,32){\makebox(0,0)[bl]{Set of items}}

%\put(5,32){\makebox(0,0)[bl]{\(V=\)}}
\put(1,21){\makebox(0,0)[bl]{\(A=\{1,...,i,...,n\}\)}}
%--

\put(34,35){\makebox(0,0)[bl]{\(\Longrightarrow \)}}

%\put(34,35){\makebox(0,0)[bl]{\(\Longrightarrow \)}}
%\put(34,30){\makebox(0,0)[bl]{\(\Longrightarrow \)}}

\put(34,24){\makebox(0,0)[bl]{\(\Longrightarrow \)}}

%\put(34,20){\makebox(0,0)[bl]{\(\Longrightarrow \)}}

\put(34,13){\makebox(0,0)[bl]{\(\Longrightarrow \)}}
%---

\put(55,42){\oval(20,6)}
\put(52.4,40.5){\makebox(0,0)[bl]{\(A_{1}\)}}

\put(49.5,39){\vector(0,-1){4}} \put(53,39){\vector(0,-1){4}}
\put(56.5,39){\vector(0,-1){4}} \put(60,39){\vector(0,-1){4}}

%--

\put(52.1,33){\makebox(0,0)[bl]{{\bf . . .}}}

%%%%%%

%\put(49.5,42){\vector(0,-1){4}} \put(53,42){\vector(0,-1){4}}
%\put(56.5,42){\vector(0,-1){4}} \put(60,42){\vector(0,-1){4}}

%\put(55,35){\oval(20,6)}

%\put(52.4,33.5){\makebox(0,0)[bl]{\(A_{k}\)}}

%%%%%%%%%%%%%%%%%%%%%%%%%%%%%%%%%%%%%

\put(49.5,32){\vector(0,-1){4}} \put(53,32){\vector(0,-1){4}}
\put(56.5,32){\vector(0,-1){4}} \put(60,32){\vector(0,-1){4}}

\put(55,25){\oval(20,6)}

\put(52.4,23.5){\makebox(0,0)[bl]{\(A_{k}\)}}

\put(49.5,22){\vector(0,-1){4}} \put(53,22){\vector(0,-1){4}}
\put(56.5,22){\vector(0,-1){4}} \put(60,22){\vector(0,-1){4}}

%--

\put(52.1,16){\makebox(0,0)[bl]{{\bf . . .}}}

\put(49.5,15){\vector(0,-1){4}} \put(53,15){\vector(0,-1){4}}
\put(56.5,15){\vector(0,-1){4}} \put(60,15){\vector(0,-1){4}}

\put(55,8){\oval(20,6)}

\put(52.4,6.5){\makebox(0,0)[bl]{\(A_{m}\)}}
%--

\end{picture}
\end{center}

\subsection{Multisets and Morphological Sets}

 Now let us consider two kinds of basic structures:
 multiset and morphological set.

 Let \(U\) be a universe (collection of all relevant
 items) and \(N\) be a set of non-negative integers
 (e.g., ~\(N = \{1,...,\tau,...,m\}\) ).
 Formally, multiset is a mapping:
 \[\Gamma:~~ U \longrightarrow N.\]
 Clearly, three basic operations can be considered for sets and
 multisets:

 {\it 1.} unions:~ \(A \bigcup B\)~ (for sets \(A\) and \(B\), analogically for
 multisets),

 {\it 2.} intersections:~ \(A \bigcap B\)~
  (for sets \(A\) and \(B\), analogically for multisets),
  and

  {\it 3.} complements:~
  \(A  \backslash B\)~
  (for sets \(A\) and \(B\), analogically for multisets).

 A macroset is a (finite or infinite) set of multisets over a
 finite alphabet.

%\section{Set Morphology Structure}

 Generally, {\it morphological set}
% {\it set morphology}
  is defined as follows:

~~

 {\bf Definition 3.1.} {\it Morphological set }
 is a structure consisting of:

 {\it 1.} a finite set of integers ~\(N=\{ 1,...,\tau,...,m\}\)~
 (each integer \(\tau\) corresponds to a system part);

 {\it 2.} set of elements (alternatives) for each system part
 \(\tau\):~~

 \(A_{\tau} = \{ A_{\tau 1},...,A_{\tau \xi},...,A_{\tau q_{\tau} }  \}\),
 where \( A_{\tau \xi} \) is a design alternative;
%  for system part \(\xi\);

 {\it 3.} preference relation over elements of  ~\(A_{\tau}\)~
 (or estimates upon a set of specified criteria or
 resultant ordinal priorities for each alternative
 ~\(p(A_{\tau \xi})\)); and

 {\it 4.} weighted (by ordinal scale) binary relation of {\it compatibility}
 for each pair of system parts
 ~\((\alpha, \beta ) \in N\)~ over elements of alternative sets
 ~\( A_{\alpha}, A_{\beta}\):~
 \(R_{A_{\alpha},A_{\beta}}\).

%====================================================Apr6-2011

 A composite system consisting of \(m\) parts is examined
 (Fig. 3.4) (\cite{lev98},\cite{lev06},\cite{lev09}).

\begin{center}

\end{center}

%====================================================================

 %
 Thus, {\it system morphology} \(S_{A}\)
 (or morphology  {\(\Phi \)} )
 is defined as  follows
 ~\(A = \{ A_{1},...,A_{\tau},...,A_{m}\}\)
 (Fig. 3.5).
%
% This structure is close to a multiset
% (e.g., \cite{dovier98},
% \cite{knuth98},
% \cite{syr01}).

\begin{center}
%
\end{center}

 Further, numerical examples for two morphologies are
 presented:
 morphology \(A^{1}\) (Fig. 3.6),
  morphology \(A^{2}\) (Fig. 3.7),
 sub-morphology \(\widetilde{S}_{A^{1},A^{2}}\) (Fig. 3.8), and
 super-morphology \(\overline{S}_{A^{1},A^{2}}\) (Fig. 3.9).

\begin{center}
%
\end{center}

 In addition,
 tables of ordinal compatibility estimates have to be considered.
 The aggregation of estimates (by element) can be based on
 a special operation (e.g., minimal value of element).
 Table 3.1 presents an illustrative numerical
 example for compatibility estimates and their aggregation.

\begin{center}
%
\end{center}

% ----------------------------------------------------------------
\subsection{Trees and Morphological Structures}

 Now let us consider trees
 and morphological structures (as morphological trees).
 Fig. 3.10 illustrates a numerical example for two trees
 \(T_{1}\) and \(T_{2}\).

\begin{center}
%
\end{center}

 Fig. 3.11 depicts corresponding examples of supertree
 \(\overline{T}\)
 and  subtree
 \(\widetilde{T}\).

\begin{center}
%
\end{center}

 Now let us consider trees with morphologies.
 Fig. 3.12
 illustrates
 two trees with morphologies
 \(\Theta_{1}\) and \(\Theta_{2}\).

\begin{center}
%
\end{center}

 Fig. 3.13 illustrates examples of supertree and  subtree
 with morphologies
 \(\overline{\Theta}\) and
 \(\widetilde{\Theta}\).

\begin{center}
%
\end{center}

% NB!: Properties of subtrees NB!

 Evidently, tables of compatibility estimates and
 their aggregation can be considered here as well.

% ----------------------------------------------------------------
\section{Preliminary Illustrative Example for Notebook}

 The preliminary illustrative applied example is targeted to
 representation and aggregation of
 three initial personal computers (notebooks).
 The considered general morphological structure of the
 notebook is presented in Fig. 4.1:

 0. Notebook \(S\).

 1. Hardware \(H\):

    1.1. Basic computation \(C\):

       1.1.1. Mother board \(B\):
      \(B_{1}\) (P67A - C43(B3) ATX Intel),
      \(B_{2}\) (MSI 870A-G54 ATX AMD),
      \(B_{3}\) (ASRoot P67 EXTREME 4(B3) ATX Intel);
 %    \(B_{4}\) ();

      1.1.2. CPU \(U\):
      \(U_{1}\) (Intel Pentium dual-core processor T 2330),
      \(U_{2}\) (Celeron dual-core processor 2330),
      \(U_{3}\) (Intel core 2 T 7200);
%     \(U_{4}\) ();

      1.1.3. RAM \(R\):
      \(R_{1}\) (1 GB DDR A-DATA),
      \(R_{2}\) (2 GB DDR2 KINGSTON),
      \(R_{3}\) (2 GB DDR3 A-DATA),
      \(R_{4}\) (2 GB DDR3 HYPER X KINGSTON);

     1.2. Hard drive \(V\):
      \(V_{1}\) (100 GB HDD),
      \(V_{2}\) (120 GB HDD),
      \(V_{3}\) (160 GB HDD),
      \(V_{4}\) (200 GB HDD);

     1.3. Video/graphic cards \(J\):
      \(J_{1}\) (NVIDIA GeForce CTS 300M),
      \(J_{2}\) (GT 400M Series),
      \(J_{3}\) (ATI Radion HD 5000 M Series);
%     \(J_{4}\) ();

       1.4. Communication equipment (modems) \(E\):
      \(E_{1}\) (Internal Modem \& Antenna),
      \(E_{2}\) (None).
%      \(E_{3}\) (),
%      \(E_{4}\) ().

 2. Software \(W\):

    2.1. Operation system and safety \(Y\):

         2.1.1. OS \(O\):
          \(O_{1}\) (Windows XP),
          \(O_{2}\) (Windows Vista);
         \(O_{3}\) (Linux).
%
% WIndows 8, Vista

%
         2.1.2. Safety software \(F\):
    \(F_{1}\) (Norton AntiVirus),
    \(F_{2}\) (AntiVirus Kaspersky).
%
%  Norton, Kaspersky

    2.2. Information processing and Internet \(I\):

         2.2.1. Data support and processing \(D\):
      \(D_{1}\) (Microsoft Office),
      \(D_{2}\) (None).
%
%  MicroSoft Office (Word, PowerPoint, Excel
%

         2.2.2. Internet access (browser) \(A\):
   \(A_{1}\) (Microsoft Internet Explorer),
   \(A_{2}\) (Mozilla);
   \(A_{3}\) (Microsoft Internet Explorer \& Mozilla).
%   \(A_{3}\) (Google Chrome);
%   \(A_{4}\) (Safari);
%%%%%%%%%%%%%%%%%%%%%%%%%%%%%%%
%  Firefox 3+
%  Opera 9.5+
%  Chrome 2.0+
%  Safari 7+
%  Internet Explorer 7+
%%%%%%%%%%%%%%%%%%%%%%%%%%%%%%%%
% Internet Explore, Mozilla, Skype  ...

    2.3. Professional software \(Z\):

         2.3.1. Information processing (e.g., engineering
         software) \(G\):
    \(G_{1}\) (Matlab),
   \(G_{2}\) (LabView);
    \(G_{3}\) (MatCad),
   \(G_{4}\) (None);
%  \(G_{4}\) (Mathematica);
%
%  MatLab, LabView, MatCad, mathematica

%
         2.3.2. Special software development environment \(P\):
   \(P_{1}\) (C++),
   \(P_{2}\) (JAVA);
   \(P_{3}\) (Delphi),
   \(P_{4}\) (None);
%   \(P_{3}\) (Fortran),
%
% Fortran, Java, C, Delphi

%
         2.3.3. Special editors \(L\):
   \(L_{1}\) (LaTex),
   \(L_{2}\) (None);
%   \(L_{3}\) (),
%   \(L_{4}\) ();
%
% LateX,

%
         2.3.4. Games \(Q\):
   \(Q_{1}\) (Tetris),
   \(Q_{2}\) (Solitaire),
   \(Q_{3}\) (Chess).
%   \(Q_{4}\) (..).
%
% LateX,

\begin{center}

\end{center}

 Now
 a simplified illustrative example for four initial solutions
 \(S_{1}\), \(S_{2}\),
 \(S_{3}\), and \(S_{4}\)
 is examined
 (Fig. 4.2, Fig. 4.3, Fig. 4.4, Fig. 4.5).
 Here the tree-like structure is not changed and only leaf nodes
 are considered.
 Compatibility estimates between design alternatives
 are not considered.
 A framework of aggregation process for four notebooks
 is depicted in Fig. 4.6
 (including two alternative methods to build the ``system kernel'').

\begin{center}
%
\end{center}

 Substructure for the considered solutions  \(\widetilde{S}\)
 is depicted in Fig. 4.7,
 superstructure \(\overline{S}\) is depicted in Fig. 4.8.
 Further,
% First,
 let us consider
  ``system kernel'' \(K\)
 as an extension of substructure  \(\widetilde{S}\)
 (Fig. 4.9).

\begin{center}
%
\end{center}

 Further, the aggregation strategy
 as modification of ``system kernel'' \(K\)
  can be applied.
 A set of candidate modification operations are the  following:

 {\it 1.} addition operations:
 {\it 1.1.} addition for \(U\):
 \(U_{1}\) or   \(U_{2}\) or  \(U_{3}\),
  {\it 1.2.} addition for \(F\):
 \(F_{1}\) or  \(F_{2}\),
 {\it 1.3.} addition for \(P\):
 \(P_{2}\) or   \(P_{3}\) or  \(P_{4}\);

 {\it 2.} correction  operations:
 {\it 2.1.} replacement \(B_{1} \Rightarrow B_{3}\),
 {\it 2.2.} replacement \(V_{3} \Rightarrow V_{4}\),
 {\it 2.3.} replacement \(A_{1} \Rightarrow A_{3}\).

 Evidently, it is reasonable to evaluate the above-mentioned modification operations
 (e.g., cost, profit) and to consider an optimization model.
 Later, corresponding optimization problems (e.g., knapsack problem,
 multiple choice problem) will be examined.
 An example of the resultant solution
 \(S^{1}\) (modification of ``system kernel'' \(K\))
 is shown in Fig. 4.10.

\begin{center}

\end{center}

 On the the hand,
 building the ``system kernel''
 can be based on
  multicriteria selection and/or expert judgment.
 Let us consider the following  basic structure of ``system kernel'':
 \(B\),  \(U\),  \(R\),  \(V\),  \(O\),
 \(F\),  \(D\),  \(G\).
 For each system component above, it is possible to consider
 a selection procedure to choose the ``best'' system element
 (while taking into account elements of the initial solution
 or additional elements as well).
%
% Thus,
 For example,
  we can obtain the following ``system kernel'':
 \[K^{*} =  B_{1} \star U_{2} \star R_{2} \star V_{3} \star E_{2} \star O_{2} \star F_{2}  \star D_{2} \star
 G_{2}.\]
 Further, the system correction process is based on the following
 operations:

 {\it 1.} addition:
  \(A_{1}\),  \(P_{1}\),  \(L_{1}\);

 {\it 2.} deletion:
  \(E_{2}\);

 {\it 3.} replacement:
 \(B_{1} \Rightarrow
% \longrightarrow
 B_{3}\),
 \(U_{2} \Rightarrow U_{1}\),
 \(O_{2} \Rightarrow O_{1}\).

 A resultant solution \(S^{2}\) is presented in Fig. 4.11.

\begin{center}

\end{center}

% ----------------------------------------------------------------
\section{Metrics and Proximities}

% ----------------------------------------------------------------
%%\subsection{Metrics and proximities}

 Let us consider similarity measure between objects
 (in our case:  sets,
% multisets,
  rankings, trees, graphs)
  \(A_{1}\) and  \(A_{2}\).
 It is often desired that the
  distance measure (function) \( d ( A_{1},A_{2} )\)
 fulfills the following properties of a metric:

 1.  \( d ( A_{1},A_{2} ) \geq 0\)  (nonnegativity),

 2.  \( d ( A_{1},A_{1} ) = 0\) (identity),

 3.  \( d ( A_{1},A_{2} ) = 0  \Leftrightarrow  A \cong B \)
 (uniqueness),

 4.  \( d ( A_{1},A_{2} )  = d ( A_{2},A_{1} )\) (symmetry),

 5.  \( d ( A_{1},A_{2} ) + d ( A_{2},A_{3} ) \geq  d ( A_{1},A_{3} )\)
 (triangle inequality).

 If the function satisfies \(d ( A_{1},A_{2} ) \leq 1\)
 it is said to be a {\it normalized} metric.

%%%%%%%%%%%%%%%%%%%%%%%%%%%%%%%%
 In many applied domains,
 the above-mentioned conditions are too restrictive
 and
 a more {\it weak} set of properties is used.
 As  a result,
 \( d ( A_{1},A_{2} )\) corresponds to
 more {\it weak} situations, for example:
 (i) quasi-metrics,
 (ii) proximities
 (without property 5,
 e.g., proximity for rankings in \cite{lev98}).

 Metrics/proximities play
 the basic role in many important problems over structures,
 for example:
 approximation, modification, aggregation.
 There exist three basic approaches to
%  model
% measure
 similarity/proximity of objects/structures:

% Three basic kinds of metric/proximity between some two
% objects/structures exists:

% (1) distance as function that satisfies three Freshe's axioms
% (e.g., \cite{}, \cite{}, \cite{}, \cite{});

 (1) traditional metrics/distances
 (e.g., \cite{bogart73}, \cite{eghe03},
  \cite{gordon90}, \cite{gower85}, \cite{mirkin70});

% (2) editing distance (a lenght/cost of transformation
% process of an object into another)
% (e.g., \cite{},  \cite{leven66}, \cite{}, \cite{});
% V. Banfa and P.A. Pevzner,

 (2) minimum cost transformation of an object/strcuture into
 another one (edit distance)
 (e.g.,
 \cite{bille05},
 \cite{bunke97}, \cite{hanen95}, \cite{leven66},
  \cite{marzal93},
 \cite{tai79},
 \cite{vidal95},
 \cite{zhang96});
 and

% (3) distance of proximity as common part of initial
% objects/strcutures
% (e.g., \cite{bunke}, \cite{lev03}, \cite{lev06}).

 (3) maximum common substructure or maximum agreement substructure
 (e.g., \cite{akatsu00}, \cite{amir97}, \cite{bunke97}, \cite{bunke98},
 \cite{lev03}, \cite{lev06}, \cite{ray02}, \cite{wallis01}).

%%%%%%%%%%%%%%%%%%%%%%%%%%%%%%%%%%%%%%%%%%%%%%%%%%%%%%%%%%%%%%%%%%%
%  Basic kinds of metrics and proximities for
% the examined structures are presented in Table 2.

% ----------------------------------------------------------------
\subsection{Metric/Proximity for Sets}

%\section{Simplified Cases}

 Let \(A =  \{1,...,i,...,n\}\) be a set of elements.
 Let us consider ... for two subsets
 \(A^{1} \subseteq A\) and  \(A^{2} \subseteq A\).
 The most simple case of metric by elements
  (i.e., distance)
 is the following:
 (while taking into account assumption
 \(  | A^{1} \bigcup A^{2} | \neq 0  \)):
 \[ \rho_{e} (A^{1},A^{2})  = 1 - \frac{ |A^{1} \bigcap A^{2}| }
 { | A^{1} \bigcup A^{2} | }.\]
 Further, let \(w_{i} \in (0,1]\) be a weight of element \(i\in
 A\).
 Then proximity (i.e., metric, distance) by element weights is as follows
 (while taking into account assumption
 \( \sum_{i\in (A^{1}\bigcup A^{2})} w_{i}   \neq 0  \)):
 \[ \rho_{w} (A^{1},A^{2})  = 1 - \frac{\sum_{i\in (A^{1} \bigcap A^{2})} w_{i}}
 {\sum_{i\in (A^{1}\bigcup A^{2}}) w_{i} }.\]
 Now let us consider a simple numerical example (Table 5.1)
 that involves
 initial set
  \(A = \{1,2,3,4,5,6,7,8\}\) and
  two subsets:
 \(A^{1} = \{1,2,4,5\}\) and
 \(A^{2} = \{1,2,3,4,5,6,7\}\).
 Elements weights are the following:
 \(0.5\), \(0.6\), \(0.4\), \(1.0\), \(0.7\), \(0.2\),
 \(0.1\), and \(1.0\).
 The resultant proximities  are:
 \(\rho_{e}  (A^{1},A^{2}) = 1 - \frac{3}{7} = 0.571\) and
 \(\rho_{w}  (A^{1},A^{2}) = 1 - \frac{1.8}{3.5} = 0.496\).

\begin{center}

\end{center}

 In the case of vector weights
 \(\overline{w}_{i} = ( w^{1}_{i},...,w^{\xi}_{i},...,w^{r}_{i}  )
 \),
 a simplified vector distance (or proximity) is:
 \[ \rho_{\overline{w}} (A^{1}, A^{2}) = (
 \rho^{1}_{\overline{w}} (A^{1}, A^{2}),...,
 \rho^{\xi}_{\overline{w}} (A^{1}, A^{2}),...,
 \rho^{r}_{\overline{w}} (A^{1}, A^{2})) =\]
 \[(1 - \frac{\sum_{i\in (A^{1} \bigcap A^{2})} w^{1}_{i}}
 {\sum_{i\in (A^{1}\bigcup A^{2}}) w^{1}_{i} },...,
 1 - \frac{\sum_{i\in (A^{1} \bigcap A^{2})} w^{\xi}_{i}}
 {\sum_{i\in (A^{1}\bigcup A^{2}}) w^{\xi}_{i} },...,
 1 - \frac{\sum_{i\in (A^{1} \bigcap A^{2})} w^{r}_{i}}
 {\sum_{i\in (A^{1}\bigcup A^{2}}) w^{r}_{i} })  .\]
%

% ----------------------------------------------------------------
\subsection{Proximity for Strings/Sequences}

 This problem of proximity analysis
 is important in
 code design,
  genom studies, information processing,
  and mathematical linguistics.
 Generally, three kinds of proximity for strings/ sequences
 are examined:
 (1) common part of initial strings as
  substring, superstring
 (e.g.,
  \cite{apost87}, \cite{apost97},
 \cite{hig00},
  \cite{gusfield99},
 \cite{jiang95}, \cite{jiang95a},
  \cite{tim90}),
 (2) median string
 (e.g., \cite{kohonen85}, \cite{nicolas03}), and
 (3) editing distance
 (a lenght/cost of an transformation/editing)
 (e.g., \cite{apost97}, \cite{hanen95}, \cite{leven66},
 \cite{robles04}, \cite{wagner74}).

 Here basic research issues
% problems
 are targeted to the following:
 (a) complexity analysis
 (e.g., \cite{tim90});
 (b) design of polynomial algorithms
 (e.g., \cite{hanen95}, \cite{jiang95a}; and
 (c) design of approximation algorithms
 (e.g., \cite{jiang95}).

% Three basic kinds of metric/proximity between some two
% objects/structures exists:
%
% (1) distance as function that satisfies three Freshe's axioms
% (e.g., \cite{}, \cite{}, \cite{}, \cite{});
%
% (3) distance of proximity as common part of initial
% objects/strcutures
% (e.g., \cite{bunke},  \cite{}).

% ----------------------------------------------------------------
\subsection{Proximity for Rankings}

 Here several types of metrics/proximities can be used, for example:
% are briefly considered:
%
 {\it 1.} Kendall-Tau distance \cite{kendall62};
% M. Kendall,
%%%%%%%%%%%%%%%%%%%%%%%%%%%%%%%%%%%%%%%%%%%%%%%%%%%
% Here
% Kemeni optimal aggregation is widely used
% (i.e., searching for the Kemeni-Snell median
% through minimum sum of Kendall-Tau distances)
%
% (e.g., \cite{barth89}, \cite{kemeny59}, \cite{kemeny60}).
%%%%%%%%%%%%%%%%%%%%%%%%%%%%%%%%%%%%%%%%%%%%%%%%%%%
%
 {\it 2.} distances for partial rankings
 (e.g., \cite{bansal09}, \cite{fagin06});
 {\it 3.} vector-like proximity (\cite{lev88}, \cite{lev98a}, \cite{lev98}).
% Here multiple choice problem is used to obtain
% a resultant interval ranking.
%
% Usually the proximity measures have been used as scalar functions which
% satisfy to Freshe's axioms for  metrics (pseudo-metrics).
%
%Kendall-Tau distance
% Kendall's proximity measure is used the most widely (Kendall, 1962).
%
 Further, Kendall-Tau distance and vector-like
 proximity are briefly described (the description is based on material from
 \cite{lev98}).

 Let \(\|g_{ij} \|, (i,j\in A)\) be an adjacency matrix for graph \(G\):
    \[g_{ij} = \left\{ \begin{array}{ll}
               1, & \mbox{if $ i \succ j, $}\\
               0, & \mbox{if $ i \sim  j, $}\\
              -1, & \mbox{if $ i \prec j. $}
               \end{array}
               \right. \]
 Kendall-Tau distance (metric) for graphs \(G^{1}\) and \(G^{2}\)
 is the following:
   \[\rho_{K}(G^{1},G^{2}) = \sum_{i<j} \mid g^{1}_{ij} - g^{2}_{ij} \mid ,\]
 where  \(g^{1}_{ij}\), \(g^{2}_{ij}\) are elements of adjacency matrices
 of graphs \(G^{1}\) and \(G^{2}\) accordingly.
% Often this metric has been used for rankings (linear and layered structures).

% Further, we describe a vector-like proximity for rankings (Levin, 1988;
% Belkin and Levin, 1990).
%%%%%%
%    In this case a measurement scale is a simplex in which for a set of
% measured objects a poset maybe constructed.

%------------------------------------------------
% \subsection{Definitions}

% Now let us consider
 Basic definitions for vector-like proximity are the following
  (\cite{lev88}, \cite{lev98a},  \cite{lev98}).
 Let \(\Psi (S)\) be a set of all layered structures on \(A\).

~~

 {\bf Definition 5.1.} Let us call the first order error \( \forall i \in A\),
  and the second order error
  \(\forall (i,j) \in \{A \star A | i \neq j\) \(\forall S,Q \in \Psi (S)\}\)
  as follows:
    \[\delta^{\pi}_{i}(S,Q)=\pi_{i}(S)-\pi_{i}(Q),\]
    \[\delta^{\pi}_{ij}(S,Q)=\pi_{i}(S)-\pi_{j}(S)-(\pi_{i}(Q)-\pi_{j}(Q)),\]
 where \(\pi_{i}(S)=l\) \(\forall i \in A(l)\) in \(S\).
 Thus,
 for an estimate of a discordance between the structures
 \(S,Q \in \Psi (S)\) with respect to \(i\) and \((i,j)\),
  an integer-valued scale with the following ranges is obtained:

 (i) \( -(m-1) \leq r \leq m-1\) for \(\delta^{\pi}_{i}(S,Q)\), and

 (ii) \(-2(m-1) \leq r \leq 2(m-1)\) for \(\delta^{\pi}_{i,j}(S,Q)\).

~~

 {\bf Definition 5.2.} Let
%
%%%%%%%%%%%%%%%%%%%%%%%%%%%%%%%%%%%%%%%%%%%%%%%%%%%%%%%%%%%%%%%
%   \begin{equation}
%       x(S,Q)=(x^{-(m-1)},...,x^{-1},x^{1},...,x^{m-1}),
%   \end{equation}
%
%%%%%%%%%%%%%%%%%%%%%%%%%%%%%%%%%%%%%%%%%%%%%%%%%%%%%%%%%%%%%%%
%
%   \begin{equation}
%       y(S,Q)=(y^{-2(m-1)},...,y^{-1},y^{1},...,y^{2(m-1)}),
%   \end{equation}
%%%%%%%%%%%%%%%%%%%%%%%%%%%%%%%%%%%%%%%%%%%%%%%%%%%%%%%%%%%%%%
%
%
%%%%%%%%%%%%%%%%%%%%%%%%%%%%%%%%%%%%%%%%%%%%%%%%%%%%%%%%%%%%%%%
%
   \[  x(S,Q)=(x^{-(m-1)},...,x^{-1},x^{1},...,x^{m-1}),\]
%
%%%%%%%%%%%%%%%%%%%%%%%%%%%%%%%%%%%%%%%%%%%%%%%%%%%%%%%%%%%%%%%
%
    \[   y(S,Q)=(y^{-2(m-1)},...,y^{-1},y^{1},...,y^{2(m-1)}),\]
%
%%%%%%%%%%%%%%%%%%%%%%%%%%%%%%%%%%%%%%%%%%%%%%%%%%%%%%%%%%%%%%
%
 be vectors of an error (proximity) \(\forall S,Q \in \Psi (S)\)
 with respect to components \(i\) (\(1\)st order), and the pairs
 \((i,j)\) (\(2\)nd order).
 The components of above-mentioned vectors are defined as follows:
  \[x^{r}=|\{i \in A|\delta^{\pi}_{i}(S,Q)=r\}| / n,\]
  \[y^{r}=2|\{(i,j) \in \{A*A|i\neq j\}|\delta^{\pi}_{ij}(S,Q)=r\}| / (n(n-1)).\]
%
% It maybe reasonable to define similar vectors of the higher order also.
% Moreover it is possible to examine the weighted errors of the first and the
% second order with taking into account the dependence on corresponding number
% \(l\) of layer \(A(l)\) for the definition of vector components.

 Now denote a set of arguments for the components of vectors \(x\) and \(y\)
 as follows:
    \(\Omega =\{-k,...,k\}\),
    negative values as \(\Omega ^{-}\), and
    positive ones as \(\Omega ^{+}\).
     In addition, we will use the vectors \(x\) with
 aggregate components of the following type (similarly, for \(y\)):
  \[x^{k_{1},k_{2}}=\sum_{r=k_{1}}^{k_{2}} x^{r},
  x^{\leq -k}= \sum_{r=-(m-1)}^{-k} x^{r},
  x^{\geq k}= \sum_{r=k}^{m+1} x^{r}, k>0,
  x^{|r|}=x^{r}+x^{-r}.\]

 {\bf Definition 5.3.} Let
 \(|x(S,Q)|=\sum_{r\in \Omega }x^{r}\), \(|y(S,Q)|=\sum_{r \in \Omega} y^{r}\)
 be modules of the vectors.

% Afterhere we will consider
 Vector \(x\) will be used as a basic one.

~~

 {\bf Definition 5.4.} We will call vectors truncated ones if

   (1) the part of terminal components is rejected, e.g.,
 \[x(S,Q) = (x^{-k_{1}},x^{-(k_{1}-1)},...,x^{-1},x^{1},...,x^{k_{2}-1},
x^{k_{2}}),\]
 and one or both of following conditions are satisfied:
 \(k_{1}<m-1\), \(k_{2}<m-1\);

    (2) aggregate components are used as follows:
 \[x(S,Q)=(x^{\leq k_{1}},...,x^{k_{a}-1},x^{k_{a},k_{b}},x^{k_{b}+1},...,
x^{\geq k_{2}}),\]
%
%%%%%%%%%%%%%%%%%%%%%%%%%%%%%%%%%%%%%%%%%%%%%%%%
%  \begin{equation}
%     x(S,Q)=(x^{|1|},...,x^{|r|},...,x^{|k|}).
%  \end{equation}
%%%%%%%%%%%%%%%%%%%%%%%%%%%%%%%%%%%%%%%%%%%%%%%%%%%%%%
%
%%%%%%%%%%%%%%%%%%%%%%%%%%%%%%%%%%%%%%%%%%%%%%%%
%
    \[ x(S,Q)=(x^{|1|},...,x^{|r|},...,x^{|k|}).\]
%
%%%%%%%%%%%%%%%%%%%%%%%%%%%%%%%%%%%%%%%%%%%%%%%%%%%%%%
%
 {\bf Definition 5.5.} Let us call vector \(x\) (\(y\)):

 (a) the two-sided one, if \( |\Omega ^{+}|\neq 0\) and \(|\Omega ^{-}|=0\);

 (b) the one-sided one, if \(|\Omega ^{+}|=0\) or \(|\Omega ^{-}|=0\);

 (c) the symmetrical one,
     if \(-r \in \Omega ^{-}\) exists \(\forall r \in \Omega ^{+}\),
     and vice versa;

 (d) the modular one, if it is defined with respect of definition 5.4 (5.3).

 Moreover,
  a pair of linear orders on the components of vectors
 \(x\)  and \(y\) (definition 5.2) is obtained:
 {\it component} \(1\)(\(-1\)) \(\prec\) ... \(\prec \) {\it component} \(k\)(\(-k\)).

%%%%%%%%%%%%%%%%%%%%%%%%%%%%%%%%%%%%%%%%%%%%%%%%%%%%%%%%%%%%%%%%%%%%%
% Clearly, if the components are aggregate ones, the orders will be analogues
% ones. Fig. 4 depicts examples of vector domains, \(\alpha\), \(\beta\),
% and \(\gamma\) denote examples of indifference sets (Fig. 4a).
%%%%%%%%%%%%%%%%%%%%%%%%%%%%%%%%%%%%%%%%%%%%%%%%%%%%%%%%%%%%%%%%%%%%%

~~~

 {\bf Definition 5.6.}
 \(x_{1}(S,Q) \succeq x_{2}(S,Q)\), \(\Omega (x_{1})=\Omega (x_{2})\),
 \(\forall S,Q \in \Psi (S)\), if any decreasing of weak components
 \(x_{1}\) in the comparison with \(x_{2}\) is compensated by corresponding
 increasing of it's 'strong' components (\(r,p \in \Omega ^{+}\) or
 \(-r,-p \in \Omega ^{-}\)):
%
%%%%%%%%%%%%%%%%%%%%%%%%%%%%%%%%%%%%%%%%%%%%%%%%%%%%%%%%%%%%%%%%%%%%%%
%    \begin{equation}
%    \sum_{r\geq u}^{r} x_{1}^{r} - \sum_{r\geq u}^{r} x_{2}^{r} \geq 0, \forall u \in \Omega ^{+} (\forall -u \in \Omega ^{-}, -r\leq -u).
%    \end{equation}
%%%%%%%%%%%%%%%%%%%%%%%%%%%%%%%%%%%%%%%%%%%%%%%%%%%%%%%%%%%%%%%%%%%%%
%
%%%%%%%%%%%%%%%%%%%%%%%%%%%%%%%%%%%%%%%%%%%%%%%%%%%%%%%%%%%%%%%%%%%%%%
%
 \[\sum_{r\geq u}^{r} x_{1}^{r} - \sum_{r\geq u}^{r} x_{2}^{r} \geq 0, \forall u \in \Omega ^{+} (\forall -u \in \Omega ^{-}, -r\leq
 -u).\]

 Now let us consider an illustrative numerical example:

 (a) initial set
 \(A = \{1,2,3,4,5,6,7,8,9\}\),

 (b) ranking
 \(S^{1} =
 \{
 A^{1}_{1} = \{2,4\},
 A^{1}_{2} = \{9\},
 A^{1}_{3} = \{1,3,7\},
 A^{1}_{4} = \{5,6,8\}
 \}\)
 and

 (c) ranking
 \(S^{2} =
 \{
 A^{2}_{1} = \{7,9\},
 A^{2}_{2} = \{1,3\},
 A^{2}_{3} = \{2,5,8\},
 A^{2}_{4} = \{4,6\}
 \}
 \).

 Corresponding adjacency matrices are as follows:
%
%%%%%%%%%%%%%%%%%%%%%%%%%%%%%%%%%%%%%%%%%%%%%%%%%%%%%%%%%%%%
%
 \[|g_{ij}(S^{1})|= \left(
\begin{array}{rrrrrrrrr}
             . & -1 & 0 & -1 & 1 & 1 & 0 & 1 & -1 \\
             1 &  . & 1 &  0 & 1 & 1 & 1 & 1 &  1 \\
             0 & -1 & . & -1 & 1 & 1 & 0 & 1 & -1 \\
             1 &  0 & 1 &  . & 1 & 1 & 1 & 1 &  1 \\
            -1 & -1 &-1 & -1 & . & 0 &-1 & 0 & -1 \\
            -1 & -1 &-1 & -1 & 0 & . &-1 & 0 & -1 \\
             0 & -1 & 0 & -1 & 1 & 1 & . & 1 & -1 \\
            -1 & -1 &-1 & -1 & 0 & 0 &-1 & . & -1 \\
             1 & -1 & 1 & -1 & 1 & 1 & 1 & 1 &  .
               \end{array}  \right)
            \]

 \[|g_{ij}(S^{2})|= \left( \begin{array}{rrrrrrrrr}
             . &  1 & 0 &  1 & 1 & 1 &-1 & 1 & -1 \\
            -1 &  . &-1 &  1 & 0 & 1 &-1 & 0 & -1 \\
             0 &  1 & . &  1 & 1 & 1 &-1 & 1 & -1 \\
            -1 & -1 &-1 &  . &-1 & 0 &-1 &-1 & -1 \\
            -1 &  0 &-1 &  1 & . & 1 &-1 & 0 & -1 \\
            -1 & -1 &-1 &  0 &-1 & . &-1 &-1 & -1 \\
             1 &  1 & 1 &  1 & 1 & 1 & . & 1 &  0 \\
            -1 &  0 &-1 &  1 & 0 & 1 &-1 & . & -1 \\
             1 &  1 & 1 &  1 & 1 & 1 & 0 & 1 &  .
               \end{array}  \right)
            \]
 Now it is very easy to calculate Kendall-Tau distance between
 \(S^{1}\) and   \(S^{2}\):
 \[\rho_{K}(S^{1},S^{2})=31.\]
%
%
% Proposed
 Vector-like proximities allow more prominent description of
 dissimilarity between structures
  \(S^{1}\) and   \(S^{2}\):
 \[\pi(S^{1})=(\pi_{1}(S^{1}),...,\pi_{i}(S^{1},...,\pi_{4}(S^{1})=
               (3,1,3,1,4,4,3,4,2),\]
 \[\pi(S^{2})=(\pi_{1}(S^{2}),...,\pi_{i}(S^{2},...,\pi_{4}(S^{2})=
               (2,3,2,4,3,4,1,3,1),\]
 \[\delta^{\pi}_{i}(S^{1},S^{2})=(1,-2,1,-3,1,0,2,1,1),\]
 \[\delta^{\pi}_{ij}(S^{1},S^{2})= \left( \begin{array}{rrrrrrrrr}
             . &  3 & 0 &  4 & 0 & 1 &-1 & 0 &  0 \\
            -3 &  . &-3 & -1 &-3 &-2 &-4 &-3 & -3 \\
             0 &  3 & . & -4 & 0 & 1 & 1 & 0 &  0 \\
            -4 &  1 & 4 &  . &-4 &-3 &-5 &-4 & -4 \\
             0 &  3 & 0 &  4 & . & 1 &-1 & 0 &  0 \\
            -1 & -2 &-1 &  3 &-1 & . &-1 &-1 & -1 \\
             1 &  4 &-1 &  5 & 1 & 1 & . & 1 &  1 \\
             0 &  3 & 0 &  4 & 0 & 1 &-1 & . &  0 \\
             0 &  3 & 0 &  4 & 0 & 1 &-1 & 0 &  .

               \end{array}  \right)
            \]
\[x(S^{1},S^{2})=(x^{-3},x^{-2},x^{-1},x^{1},x^{2},x^{3})=
(1,1,0,5,1,0),\]
\[y(S^{1},S^{2})=(y^{-6},y^{-5},y^{-4},y^{-3},y^{-2},y^{-1},
y^{1},y^{2},y^{3},y^{4},y^{5},y^{6})=\]
\[(0,1,5,5,1,6, 6,0,1,1,0,0).\]

 Note,
% that here we do not use in \(x\) and \(y\)
%
 the coefficients
 \(\frac{1}{n}\) and \(\frac{2}{n(n-1)}\)
 were not used in \(x\) and \(y\).

%%%%%%%%%%%%%%%%%%%%%%%%%%%%%%%%%%%%%%%%%%%%%%%%%%%%%%%%%%%%%%%%%%%%%%%%%%%%%%
% In addition, we may examine vector-like proximity with aggregate components,
% for example:
%
% \[x(S^{1},S^{2})=(x^{\leq -1}, x^{\geq 1})=(2,6).\]
%
% This vector is shown in Fig. 4a by *.
% Clearly, that the vector demonstrates a change of elements of \(A\), mainly,
% to the top layer.
%
% Thus we face a new problem:
% construct for an applied task the best vector-like proximity
% or a set of the proximities.
%%%%%%%%%%%%%%%%%%%%%%%%%%%%%%%%%%%%%%%%%%%%%%%%%%%%%%%%%%%%%%%%%%%%%%%%%%%%%

%%%%%%%%%%%%%%%%%%%%%%%%%%%%%%%%%%%%%%%%%%%%%%%% Proximity for trees

% ----------------------------------------------------------------
\subsection{Proximity for Trees}

 The following approaches to
 proximity (similarity/ dissimilarity) of trees are considered
 (e.g., \cite{connor11}, \cite{dul03},
 \cite{akatsu00}, \cite{amir97},
 \cite{farach95}, \cite{hoang11},
  \cite{selkow77},
 \cite{tanaka94},
 \cite{valiente01}, \cite{valiente02}):

 (i) metrics (distance)
 (e.g., \cite{connor11}, \cite{dul03}, \cite{tanaka94},
 \cite{valiente02}) including some special kinds of distances:
  alignment distance \cite{jiang95b},
  isolated subtree distance \cite{tanaka88},
  top-down  distance
 (\cite{selkow77}, \cite{yang91}), and
  bottom-up distance
 \cite{valiente01};

 (ii) tree edit distance
%  (or tree-to-tree correction problem)
 (e.g.,
 \cite{chen01},
 \cite{sasha89},
% \cite{selkow77},
 \cite{tai79},
 \cite{torsello01}, \cite{zhang96a},
 \cite{zhang89},
 \cite{zhang92}); and

 (iii) common subtree, median tree or tree agreement, consensus
 (e.g., \cite{akatsu00}, \cite{amir97},
 \cite{farach95}, \cite{finden85},
 \cite{hoang11},
 \cite{jansson05},
 \cite{semple00}, \cite{steel93},
 \cite{wang01}).

 It is reasonable to note, trees are ordered structures.
 Thus, efficient (polynomial) computing algorithms have been
 suggested for building metrics/proximities of some kinds of trees
 (e.g., \cite{ele03},
  \cite{torsello01}, \cite{torsello05}, \cite{valiente01}).

% Some time, the approaches are equivalent ones
% (\cite{}, \cite{}, \cite{}).

% ----------------------------------------------------------------
%\section{Definition}

% Afterhere
 Further, our simplified version of two-component proximity
 for rooted labelled trees
  is considered and used.
 Let \(T'=(A',E')\)
 and \(T''=(A'',E'')\)
 be two rooted labelled trees
 (the root is the same in both trees)
 where
 \(A'\) and  \(A''\) are the vertices,
 \(E'\) and   \(E''\) are the arcs.
 Let us consider {\it dominance parameter}
 \(\forall (a,b) \in (E'\bigcap E'' )\).
 The following three {\it dominance cases} can be examined:
 (i)) \(a \rightarrow b\) (\(a\) dominates \(b\), i.e., \(a \succ b\)),
 (ii)  \(b \rightarrow a\) (\(b\) dominates \(a\), i.e., \(a \prec b\)), and
 (iii) \(a\) and \(b\) are independent.
  Then {\it dominance parameter} is:
%
%  \[d_{(a,b)\in (E'\bigcap E'')} = \]
%
    \[d_{(a,b)\in (E'\bigcap E'')} =
     \left\{ \begin{array}{ll}
               d^{1}, & \mbox{if $  a \rightarrow b, $}\\
               d^{2}, & \mbox{if $  b \rightarrow a, $}\\
              d^{3}, & \mbox{if  $  a,  b ~ are~ independent. $}
               \end{array}
               \right. \]
 As a result, {\it change parameter}
 \(\forall (a,b) \in (E'\bigcap E'' )\)
 can be define as the following:
    \[p_{(a,b)\in (E'\bigcap E'')} =
     \left\{ \begin{array}{ll}
              0, & \mbox{if $  d_{(a,b)\in (E'\bigcap E'')} ~is~ not~ changed, $}\\
              1, & \mbox{if  $ d_{(a,b)\in (E'\bigcap E'')} ~is~ changed. $}
               \end{array}
               \right. \]
%
%   \[p_{ (a,b)\in (E'\bigcap E'') } =
%          \left\{ \begin{array}{ll}
%               0, & \mbox{if $  d_{(a,b)\in (E'\bigcap E'') ~is~ not~ changed, $}\\
%               1, & \mbox{if  $  d_{(a,b)\in (E'\bigcap E'') ~is~ changed. $}
%               \end{array}
%               \right. \]
%
%%%%%%%%%%%%%%%%%%%%%%%%%%%%%%%%%%%%%%%%%%%%%%%%%%
 Finally, the proximity of two trees is:
 \[\overline{\rho} (T',T'') = ( \rho (A',A''),  \rho(E',E'')
 ),\]
 where
 \[ \rho_{e} (A',A'')  = 1 - \frac{ |A' \bigcap A'' | }
 { | A' \bigcup A'' | },\]
%
%
%  \[\delta(E',E'') = ( \sum_{(a,b) \in (E'\bigcap E'' )}  p_{(a,b)}(E',E'') /  |(E'\bigcap E'') | .\]
%
%
 \[ \rho (E',E'')  = \frac{\sum_{(a,b) \in (E'\bigcap E'' )}  p_{ (a,b) \in (E' \bigcap  E'')} }
 {|(E'\bigcap E'') |  }.\]
 Evidently, the following properties are satisfied
 (Freshe axioms 1 and 2):

  (1) \( 0 \leq \rho (E',E'') \leq 1\),
   (2) \( \rho (E,E) = 0 \).

 Note the described approach can be applied for digraphs as well.
 Now let us consider illustrative numerical examples for trees.

~~

 {\bf Example 5.1:}
 Trees \(T'\) and \(T''\) are presented in Fig. 5.2.
 Clearly, the general proximity is: \(\overline{\rho} (T',T'') = ( 0, 1 ) \).

\begin{center}

\end{center}

 {\bf Example 5.2:}
 Trees \(T'\) and \(T''\) are presented in Fig. 5.3.
 Here \(A' = A''\) and \( \rho(A',A'') = 0\).
 Table 5.2 and table 5.3 contain
 corresponding {\it dominance factors}
 for \(T'\) and \(T''\) (changes are depicted via 'ovals').
 Finally, the general proximity is:
  \(\overline{\rho } (T',T'') = ( 0, \frac{6}{28} ) \).

\begin{center}
%
\end{center}

%%%%%%%%%%%%%%%%%%%%%%%%%%%%%%%%%%%%%%%%% EXAMPLE 3

 {\bf Example 5.3:}
 Trees \(T'\) and \(T''\) are presented in Fig. 5.4:
 \(A' = \{ 1,2,3,4,5,6,7,8 \}\),
 \(A'' = \{ 1,2,3,6,8,9,10 \}\),
 and
 \(A' \bigcap A'' = \{ 1,2,3,6,8 \}\).
 The proximity for vertex sets is: \(\rho (A',A'') = \frac{1}{2}  \).

 Table 5.4 and table 5.5 contain
 corresponding {\it dominance factors}
 for \(A' \bigcap A''\) in \(T'\) and  in \(T''\) (changes are depicted via 'ovals').
 Finally, the general proximity is: \(\overline{\rho} (T',T'') = (\frac{1}{2}, \frac{1}{5} ) \).

\begin{center}
%
\end{center}

% ----------------------------------------------------------------
\subsection{Proximity for Morphological Structures }

 Generally,
 morphological structures are composite ones
 and it is reasonable
 to consider the
 corresponding their proximity
 as vector-like proximity.
 General structure
% \(\Lambda\)
 ({\bf \(\Lambda\)})
 consists of the following parts:
 (i) tree-like system model {\bf T},
 (ii) set of leaf nodes {\bf P},
 (iii) sets of DAs for each leaf node {\bf D},
 (iv) DAs rankings (i.e., ordinal priorities) {\bf R}, and
 (v) compatibility estimates between DAs {\bf I}.
 Thus, the vector-like proximity for two morphological structures
 \(\Lambda^{\alpha},\Lambda^{\beta}\)
 can be examined as follows:

 \(\overline{\rho}\) (\(\Lambda^{\alpha},\Lambda^{\beta})=\)
 (\(\rho^{t}\)({\bf T}\(^{\alpha}\),{\bf T}\(^{\beta}\)),
 \(\rho^{t}\)({\bf P}\(^{\alpha}\),{\bf P}\(^{\beta}\)),
 \(\rho^{t}\)({\bf D}\(^{\alpha}\),{\bf D}\(^{\beta}\)),
 \(\rho^{t}\)({\bf R}\(^{\alpha}\),{\bf R}\(^{\beta}\)),
 \(\rho^{t}\)({\bf I}\(^{\alpha}\),{\bf I}\(^{\beta}\))).

 Let us consider a simplified example:
 structure
 (\(\Lambda\))
 is examined as
 a composition of two parts:
 (a) tree  \(T = (A,E)\) (i.e., set of vertices and set of arcs),
 (b) rankings  for each leaf node \(i\)
 ( \( \bigcup_{i} R_{i} \) ).
%
%%%%%%%%%%%%%%%%%%%%%%%%%%%%%%%%%%%%%
  In our case,
  the proximity of two morphological structures is:
 \[\overline{\rho } (\Lambda',\Lambda'') =
  ( \rho (A',A''),  \rho (E',E''),
 \rho_{r} (\Lambda',\Lambda'')
%  \bigcup_{i} \rho (R'_{i},R''_{i})
 ).\]
 Here it is assumed
% that
 sets of design alternatives for leaf vertices are not
 changed and tables of compatibility estimates are not considered.
 An example illustrate the approach.

%~~

%%%%%%%%%%%%%%%%%%%%%%%%%%%%%%%%%%%%%%%%% EXAMPLE 4

 {\bf Example 5.4:}
 Trees \(T'\) and \(T''\) are presented in Fig. 5.5:
 \(A' = \{ 1,2,3,4,5,6,7,8 \}\),
 \(A'' = \{ 1,2,3,4,5,6,7,9 \}\),
 and
 \(A'\bigcup A'' = \{ 1,2,3,4,5,6,7  \}\).
 The proximity for vertex sets is: \(\rho (A',A'') = \frac{2}{9} \).
 Table 5.6 and Table 5.7 contain
 corresponding {\it dominance factors}
 for \(A' \bigcup A''\) in \(T'\) and  in \(T''\) (changes are depicted via 'ovals').
 Finally, the proximity for trees is:
 \(\overline{\rho } (T',T'') = ( \frac{2}{9}, \frac{2}{21} ) \).

\begin{center}

\end{center}

 Here the comparison process for rankings is based on
 rankings for common leaf vertices:
 \( L(T',T'')= \{ 4,5,6,7 \}\).
 Thus, for each vertex above
 it is possible to use a metric/proximity.
 In our case (Fig. 14), the following metric
 (normalized Kendall-Tau distance) values are obtained:
 vertex \(4\): \(\delta_{4} (\Lambda',\Lambda'') = \frac{1}{3}\),
 vertex \(5\): \(\delta_{5} (\Lambda',\Lambda'')= \frac{1}{2}\),
 vertex \(6\): \(\delta_{6} (\Lambda',\Lambda'')= \frac{1}{3}\),  and
 vertex \(7\): \(\delta_{7} (\Lambda',\Lambda'')= \frac{1}{6}\).
 The resultant general proximity for rankings can be computer as
 an average value:
% follows:
%
 \[\rho_{r} (\Lambda',\Lambda'') = \frac{\sum_{i\in L(T',T'')} \delta_{i} (\Lambda',\Lambda'') }{ | L(T',T'')| } = 0.33.\]
%

%\bigcup_{i} \rho (R'_{i},R''_{i})

 Finally, the general proximity is:
 \[\overline{\rho } (\Lambda',\Lambda'') =
  ( \rho (A',A''),  \rho (E',E''), \rho_{r} (\Lambda,\Lambda'')  ) = (0.22, 0.1 , 0.33).\]
 Note, proximity of compatibility estimates can be added into the
 proximity vector above as well.

\begin{center}

\end{center}

%%%%%%%%%%%%%%%%%%%%%%%%%%%%%%%%%%%%%%%%%%%%%%%%%%%%%%%%%%%%%%%%%%%

% ----------------------------------------------------------------
\section{Underlying Problems}

% ----------------------------------------------------------------
\subsection{Multicriteria Ranking}

 Let ~\(A = \{ A_{1},...,A_{j},...,A_{n}\}\)~ be a set of
 alternatives
 and
  ~\(C = \{ C_{1},...,C_{p},...,C_{m}\}\)~ be a set of
   criteria.
 An estimate vector
   ~\(z_{j} = \{ z_{j1},...,z_{jp},...,z_{jm}\}\)
 as a result of an assessment upon criteria above
 is given for each alternative  \(A_{j}\).
 In addition, criteria weights
  \(\{ \mu_{p} \}\)
  are used as well.
 The problem consists in comparison of the alternatives
 and forming for each alternative an ordinal quality estimate
 (priority).
 This problem belongs to a class of ill-structured problems
 by classification of H. Simon \cite{sim58}.
 In the paper, a modification of outranking technique Electre
  \cite{roy96} has been used
 (the modification was suggested in  \cite{levmih88}).

% ----------------------------------------------------------------
\subsection{Knapsack Problem}

 The basic problem is (e.g., \cite{gar79}, \cite{keller04}):
% \cite{mar90}):
%
 \[\max\sum_{i=1}^{m} c_{i} x_{i}\]
 \[s.t.~~\sum_{i=1}^{m} a_{i} x_{i} \leq b,
 ~x_{i} \in \{0,1\}, ~ i=\overline{1,m} \]
%
%
%
% \[\max\sum_{i=1}^{m} c_{i} x_{i}
% ~~~s.t.~~\sum_{i=1}^{m} a_{i} x_{i} \leq b,
% ~x_{i} = 0 \cup 1, ~ i=\overline{1,m} \]
%
 and additional resource constraints
 ~~\(\sum_{i=1}^{m}a_{i,k} x_{i} \leq b_{k}; ~ k=\overline{1,l};\)~
 where \(x_{i}=1\) if item \(i\) is selected,
 \(c_{i}\) is a value (``utility'') for item \(i\), and
 \(a_{i}\) is a weight (or required resource).
 Often nonnegative coefficients are assumed.
 The problem is NP-hard \cite{gar79}
% , \cite{mar90})
 and can be solved by the following methods:
 (i) enumerative methods
 (e.g., Branch-and-Bound, dynamic programming),
  (ii) approximate schemes with a limited
 relative error (e.g., \cite{keller04}),
% , \cite{mar90}),
  and
  (iii) heuristics.

 A basic version of knapsack problem with minimization of the objective
 function is:
 \[\min \sum_{i=1}^{m} c_{i} x_{i}\]
 \[s.t.~~\sum_{i=1}^{m} a_{i} x_{i} \geq b,
 ~x_{i} \in \{0,1\}, ~ i=\overline{1,m}. \]

 For multiple criteria statements
 it is reasonable to search for Pareto-efficient solutions.
% and analogical approaches can be used.
%
 Mainly,
 heuristics, evolutionary algorithms, dynamic programming, and
 local search methods
 are applied to
 multicriteria knapsack problems
 (e.g.,
 \cite{bazgan09},
 \cite{ehrgott05}, \cite{gand00},
 \cite{klam00},
 \cite{lust10},
 \cite{zitz99}).
 A recent survey on multicriteria knapsack problem
 is contained in  \cite{lust10}.

%-------------------------------------

\subsection{Knapsack Problem and Compatibility}

 Now let us consider item dependence  in
 knapsack problem \cite{lev98}.
 Here the following
%  structural requirements
  is considered:
   binary compatibility
%    (Ins)
     of items as a symmetric binary relation
    (i.e., the selected subset has to contain only compatible
    items, \(1\) corresponds to compatibility of items and
    \(0\) correspond to incompatible case).
%
%    (2) dependence of items as an additional constraint of the following
%    kinds:
% \(x_{i_{1}}\geq x_{i_{2}}\) \(\forall\) \(i_{1}\),\(i_{2}\);
% or a dependence digraph (graph)
% \(Q=(I,D)\) consisting of these pairs
% (set \(D\)).
%
 Thus,  the consideration of compatibility for knapsack problem leads
 to searching for a
 clique  (``profit clique'') with the following properties:

 (i) maximum total profit of the selected items:
 ~~ \(\max \sum_{i=1}^{n} c_{i}\),

 (ii) restricted total weight of the selected items:
 ~~ \( \sum_{i=1}^{n} a_{i} \leq b\).

 Table 6.1 and Table 6.2 contain descriptions of numerical examples for the following problems
 (this example is a modification of the example from \cite{lev98}):

\begin{center}

\end{center}

 (a) basic knapsack problem,

 (b) ``profit clique''
 (i.e., knapsack problem and compatibility of the selected items),
 and

% limited clique (maximizing the number of selected compatible items,
% and constraint to total item weight as in KP); and

 (c) maximal clique (maximizing the number of selected compatible items).

 The following notations are used:

 (1) \(M\) is the number of items in a solution
 (i.e., the number of the selected items);

 (2) \(b=13\) is the right-side constraint (additional constraints are
  not applied);

 (3) symbol \(\star \) corresponds to a selected item; and

 (4) characteristics  the solution are:
 \( C = \sum_{i=1}^{m} c_{i} x_{i}, ~ A = \sum_{i=1}^{m} a_{i}
 x_{i}\).

% Here we point out an important KP with item dependence as a
% tree-like digraph \(Q\) \cite{john83}.
% Johnson and Niemi have examined two cases \(Q\):
%   (i) {\em  outtree}  (arcs direct from the tree root), and
%   (ii) {\em in-tree} (arcs direct to the tree root).
% For the first case, fully approximation scheme is proposed.

 Clearly, the considered problems are NP-hard.
 Thus, heuristics are used.
% Evidently,
 Multicriteria versions of
 ``profit clique'' problem can be examined and used as well.

\begin{center}

\end{center}

% ----------------------------------------------------------------
\subsection{Multiple Choice Knapsack Problem}

% ----------------------------------------------------------------
%\subsubsection{Basic multiple choice knapsack problem}

%%%%%%%%%%%%%%%%%%%%%%%%%%%%%%%%%%%%%%%%%%%%%%%%%%%%%%%%%%%%%%%%%
%% The basic multiple choice knapsack problem is:
%
%% \[max ~~u ~=~ \sum_{i=1}^{n} ~ p_{i} ~x_{i}  \]
%
%% \[s.t. ~~~~ \sum_{i=1}^{n} ~ w_{i} ~ x_{i} ~~\leq ~~c,~~~ x_{i} \in \{0,1\} ~~~i=\overline{1,n}.\]
%%%%%%%%%%%%%%%%%%%%%%%%%%%%%%%%%%%%%%%%%%%%%%%%%%%%%%%%%%%%%%%%%

 %
 In the case of a multiple choice problem,
%(e.g., \cite{mar90}),
 the items (e.g., actions) are divided into groups
 and we select elements from each group
 while taking into account a total resource constraint (or constraints)
 (e.g., \cite{gar79}, \cite{hire07}, \cite{keller04},
  \cite{parra02}):
 \[\max\sum_{i=1}^{m} \sum_{j=1}^{q_{i}} c_{ij} x_{ij}\]
  \[s.t.~\sum_{i=1}^{m} \sum_{j=1}^{q_{i}} a_{ij} x_{ij} \leq
 b,
  ~~~\sum_{j=1}^{q_{i}} x_{ij}=1, ~i=\overline{1,m},
 ~~x_{ij} \in \{0,1\}.\]
%
%--------------------------------------------------------------
% In the case of multiple criteria description
%  ~\(\{ c_{i,j} \}\)
% for each element
% ~\( \forall (i,j)\)~
% (i.e., multi-objective multiple choice problem),
% the vector goal function
% ~\((~f^{1},...,f^{r}~)\)~
% is as follows \cite{levs06}:
% \[~~~ (~\max \sum_{i=1}^{m} \sum_{j=1}^{q_{i}} c^{1}_{ij} x_{ij}~,...,
% % ~\max \sum_{i=1}^{m} \sum_{j=1}^{q_{i}} c^{p}_{ij} x_{ij}~,...,
%  ~\max \sum_{i=1}^{m} \sum_{j=1}^{q_{i}} c^{r}_{ij} x_{ij}~) \]
%
%>>>>>>>>>>>>>>>>>>>>>>>>>>>>>>>>>>>>>>>>>>>>>>>>>>>>>>>>>>>>>
%
% In the case of multicriteria description
% each element (i.e., \((i,j)\)) has a vector profit:
%
%
%  ~\( \overline{c_{i,j}}    = (  c^{1}_{i,j}, ..., c^{\xi}_{i,j}, ... , c^{r}_{i,j} )
%  \).
%
% ----------------------------------------------------------------
%\subsubsection{Multicriteria multiple choice knapsack problem}
%
%
 In the case of multicriteria description,
 each element (i.e., \((i,j)\)) has a vector profit:
  ~\( \overline{c_{i,j}}    = (  c^{1}_{i,j}, ..., c^{\xi}_{i,j}, ... , c^{r}_{i,j} )
  \).
  A version of multicriteria multiple choice problem
 was presented in
% by Levin and Safonov
 (\cite{lev09}, \cite{levsaf10a}, \cite{levsaf11}):
%
%%  \cite{levs06}{levsaf10}):
%
%
 \[\max\sum_{i=1}^{m} \sum_{j=1}^{q_{i}} c^{\xi}_{ij} x_{ij}, ~~ \forall \xi =
 \overline{1,r}\]
 \[s.t.~\sum_{i=1}^{m} \sum_{j=1}^{q_{i}} a_{ij} x_{ij} \leq
 b,
 ~~~\sum_{j=1}^{q_{i}} x_{ij}=1, ~~i=\overline{1,m},
 ~~x_{ij} \in \{0, 1\}.\]
 Evidently, in this case it is reasonable to search for
 Pareto-efficient  solutions
 (by the vector objective function above).
%
%>>>>>>>>>>>>>>>>>>>>>>>>>>>>>>>>>>>>>>>>>>>>>>>>>>>>>>>>>>>>>
%
%  Generally, it is necessary
%   here
%  to search for the
%  Pareto-efficient (by the vector objective function above)  solutions.
  Here
%  In this case,
% Levin and Safonov
% \cite{levs06}
% described
% show
 the following solving schemes
 can be used
  (\cite{levsaf10a}, \cite{levsaf11}):
%% \cite{levs06}:
%
 ~(i)
% enumerative algorithm based on
  dynamic programming,
 ~(ii) heuristic based on preliminary multicriteria ranking of elements
 to get their priorities
 and
 step-by-step packing the knapsack (i.e., greedy approach),
 ~(iii) multicriteria ranking of elements to get their ordinal
 priorities and usage of approximate solving scheme (as for knapsack
 problem) based on discrete space of system excellence
 (i.e., quality lattice as in HMMD \cite{lev98}).
% (it is described in the section on HMMD).
%
 In the article, greedy heuristic above is used later.

%%%%%%%%%%%%%%%%%%%%%%%%%%%%%%%%%%%%%%%%%%%%%%%%%%%%%%%%%%%%%%%%%%%%
% ----------------------------------------------------------------
%\subsection{Assignment Problem}

\subsection{Morphological Design}

% The considered morphological design approach
% generalizes multiple choice
% problem (\cite{lev98},
%  \cite{lev05},
% \cite{lev06}).

% Here
% Hierarchical Morphological Multicriteria Design (HMMD) approach,
% suggested by Levin
%  (e.g., \cite{lev98},
%  \cite{lev05},
%  \cite{lev06}),
% is based on the morphological clique problem.
% (e.g., \cite{lev98},
%  \cite{lev05},
%   \cite{lev06}).
%
 A brief description of HMMD is a typical one as follows
  (e.g., \cite{lev98},
  \cite{lev05},
   \cite{lev06},
   \cite{lev09}).
%
%
% Here
 The composite (modular, decomposable) system
 under examination
 consists
 of the components and their interconnections or compatibilities.
 Basic assumptions of HMMD are the following:
 ~(a) a tree-like structure of the system;
 ~(b) a composite estimate for system quality
     that integrates components (subsystems, parts) qualities and
     qualities of interconnections
      (hereinafter referred as 'IC')
     across subsystems;
 ~(c) monotonic criteria for the system and its components;
 and
 ~(d) quality of system components and IC are evaluated
   on the basis of coordinated ordinal scales.
 The designations are:
  ~(1) design alternatives (DAs) for
%  leaf
  nodes of the model;
  ~(2) priorities of DAs (\(r=\overline{1,k}\);
      \(1\) corresponds to the best level);
  ~(3) ordinal compatibility estimates for each pair of DAs
  (\(w=\overline{0,l}\); \(l\) corresponds to the best level).
 The basic phases of HMMD are:
  ~{\it 1.} design of the tree-like system model (a preliminary phase);
  ~{\it 2.} generating DAs for model's leaf nodes;
  ~{\it 3.} hierarchical selection and composing of DAs into composite
    DAs for the corresponding higher level of the system
    hierarchy (morphological clique problem); and
  ~{\it 4.} analysis and improvement of the resultant composite DAs (decisions).
 Let ~\(S\) be a system consisting of ~\(m\) parts (components):
 ~\(P(1),...,P(i),...,P(m)\).
 A set of design alternatives
 is generated for each system part above.
  The problem is:

~~

 {\it Find a composite design alternative}
 ~~ \(S=S(1)\star ...\star S(i)\star ...\star S(m)\)~~
 {\it of}~ DAs ({\it one representative design alternative} \(S(i)\)
 {\it for each system component/part} ~\(P(i)\), \(i=\overline{1,m}\))~
 {\it with non-zero}~ IC ~{\it estimates between design alternatives.}

~~

%%%%%%%%%%%%%%%%%%%%%%%%%%%%%%%%%%%%%%%%%%%%%%%%%%%%%%%%%%%%%%%%%%%%%%%%%%%%
  A discrete space of the system excellence on the basis of the
 following vector is used:
 ~~\(N(S)=(w(S);n(S))\),
 ~where \(w(S)\) is the minimum of pairwise compatibility
 between DAs which correspond to different system components
 (i.e.,
 \(~\forall ~P_{j_{1}}\) and \( P_{j_{2}}\),
 \(1 \leq j_{1} \neq j_{2} \leq m\))
 in \(S\),
 ~\(n(S)=(n_{1},...,n_{r},...n_{k})\),
 ~where ~\(n_{r}\) is the number of DAs of the \(r\)th quality in ~\(S\)
 ~(\(\sum^{k}_{r=1} n_{r} = m\)).
 As a result,
 we search for composite system decisions which are nondominated by
 ~\(N(S)\).
 Here an enumerative solving scheme
 (e.g., dynamic programming)
  is used (usually \(m \leq 6\)).
%   \cite{lev98}.

%    Thus, the following layers of system excellence can be considered:
%  ~(i) ideal point;
%  ~(ii) Pareto-efficient points;
%  ~(iii) a neighborhood of Pareto-efficient DAs
% (e.g., a composite decision of this set can be
% transformed into a Pareto-efficient point on the basis of an
% improvement action(s)).
%
% Clearly, the compatibility component of vector ~\(N(S)\)
% can be considered on the basis of a poset-like scale too
% (as \(n(S)\))
%% as it was suggested
%% by Levin
%  (\cite{levf01}, \cite{lev06}).
%
% In this case, the discrete space of
% system excellence will be an analogical lattice.
% \cite{lev06}.
%

%%%%%%%%%%%%%%%%%%%%%%%%%%%%%%%%%%%%%%%%%%%%%%%%%%%%%%%%%%%%%%%%%%%%%%%%%%%%%%%%
%
 Fig. 6.1 and Fig. 6.2 illustrate the composition problem
 (by a numerical example for a system consisting of three parts
 ~\(S = X \star Y \star Z\)).
 Priorities of DAs are shown in Fig. 6.1 in parentheses and are
 depicted in Fig. 6.2;
  compatibility estimates are pointed out in Fig. 6.2).
 In the example, the resultant composite decision is
 (Fig. 6.1, Fig. 6.2):
 ~\(S_{1}=X_{4}\star Y_{2}\star Z_{2}\), ~\(N(S_{1}) =
 (2;1,1,1)\).
%
% ~\(S_{2}=X_{3}\star Y_{1}\star Z_{3}\), ~\(N(S_{2}) = (3;1,0,2)\).

\begin{center}

\end{center}

%%%%%%%%%%%%%%%%%%%%%%%%%%%%%%%%%%%%%%%%%%%%%%%%%%%%%%%%%%%%%%%%%%%
%\subsection{Additional Problems}
%
% In the future, it may be reasonable to consider
% various versions of
% additional underlaying problems, for example:
%
% (i) covering problems,
%
% (ii) clustering problems,
%
% (iii) alignment problems.
%%%%%%%%%%%%%%%%%%%%%%%%%%%%%%%%%%%%%%%%%%%%%%%%%%%%%%%%%%%%%%%%%%%%%%%%%

% ----------------------------------------------------------------
%\section{Median/Consensus Problems and Extended Median Problem}

\section{Median/Consensus Problems and Aggregation Problems}

% ----------------------------------------------------------------
\subsection{Sets}

 Evidently, sets are the basic structures.
% Fig. 19 depicts a numerical example for two initial sets \(A\) and
% \(B\).
%
%%%%%%%%%%%+++++++++++++++++++++++++++++++++++++++++++++++++++
%
%\subsection{Multi-set Case}
%
 Let us consider the case of \(m\) sets.
 The extended median/consensus for \(m\) sets
 ~\( \{A_{1},...,A_{i},...,A_{m}\}\)
 is the following.
 Analogically,
 ~\(\forall e \in \bigcup_{i=1}^{m} A_{i})\)~ two attributes are
 examined:
 ``profit''/``utility'' ~\(c_{e} \geq 0\),
 required resource  ~\(b_{e} \geq 0\).
 Let
 ~~\(R_{\{A_{i}\}} = \widetilde{S}_{\{A_{i}\}} ~ \subseteq ~ \bigcap_{i=1}^{m} A_{i}\)
 ~~ (or  ~\(R_{\{A_{i}\}} = M_{\{A_{i}\}}\))~~
 be a basic ``consensus'' set
 ~~ and ~~~~
 ~ \(\sum_{e\in R_{AB}} ~ \leq b\), where \(b\) is a total
 resource.
 The problem of ``extended median/consensus'' is:
 \[ \max ~ \sum_{e\in W \subseteq  (( \bigcup_{i=1}^{m} A_{i} ) \backslash R_{\{A_{i}\}})} ~~~~~ c_{e} \]
 \[ s.t. ~ \sum_{e\in W \subseteq  (( \bigcup_{i=1}^{m} A_{i} ) \backslash R_{\{A_{i}\}})} ~~~~~ b_{e} ~ \leq ~ b. \]
 This model corresponds to  basic knapsack problem.

% ----------------------------------------------------------------
%\subsection{Set Aggregation}

%\subsection{Two-set Case}

 Further, aggregation problems for sets will be examined.
 First, let us consider two-set case for sets \(A\), \(B\).
 Here the following proximity/metric is used:
  ~\(\rho(A,B) \geq 0\),   ~\(\rho(A,B) = \rho(B,A)\).
 For example, the following simple metric can be used:
 ~\( \rho(A,B) =  | A \bigcap B |  /  | A \bigcup B | \).

 A median-like subset
 (median-like consensus model by Kendall etc. \cite{kendall62})
 is:
 \[M_{AB} = ~arg ~\min_{\{M\}} ~ ( \rho(M,A) + \rho(M,B) ). \]

 The case of the extended median/consensus for two sets \(A\) , \(B\)
 is similar.
 Here
 two attributes are examined
 \(\forall e \in (A \bigcup B)\):
 ``profit''/``utility'' ~\(c_{e} \geq 0\),
 required resource  \(b_{e} \geq 0\).
 Let ~\(R_{AB} = \widetilde{S}_{AB}\)~  (or  \(R_{AB} = M_{AB}\))
 be a basic ``consensus'' set and
 ~\(\sum_{e\in R_{AB}} ~ \leq b\), where \(b\) is a total
 resource.
 The problem of building the``extended median/consensus'' is:
 \[ \max ~ \sum_{e\in W \subseteq  ((A \bigcup B) \backslash R_{AB})} ~~~~~ c_{e} \]
 \[ s.t. ~ \sum_{e\in W \subseteq  ((A \bigcup B) \backslash R_{AB})} ~~~~~ b_{e} ~ \leq ~ b. \]
 This model corresponds to basic knapsack problem.

%%%%%%%%%%%+++++++++++++++++++++++++++++++++++++++++++++++++++

%\subsection{Multi-set Case}

 Now let us consider the multi-set case.
 The extended median/consensus for \(m\) sets
 ~\( \{A_{1},...,A_{i},...,A_{m}\}\)
 is the following.
 Analogically,
 ~\(\forall e \in \bigcup_{i=1}^{m} A_{i})\)~ two attributes are
 examined:
 ``profit''/``utility'' ~\(c_{e} \geq 0\),
 required resource  ~\(b_{e} \geq 0\).
 Let
 ~~\(R_{\{A_{i}\}} = \widetilde{S}_{\{A_{i}\}} ~ \subseteq ~ \bigcap_{i=1}^{m} A_{i}\)
 ~~ (or  ~\(R_{\{A_{i}\}} = M_{\{A_{i}\}}\))~~
 be a basic ``consensus'' set
 ~~ and ~~~~
 ~ \(\sum_{e\in R_{AB}} b_{e} ~ \leq b\), where \(b\) is a total
 resource.
 The problem of ``extended median/consensus'' is:
 \[ \max ~ \sum_{e\in W \subseteq  (( \bigcup_{i=1}^{m} A_{i} ) \backslash R_{\{A_{i}\}})} ~~~~~ c_{e} \]
 \[ s.t. ~ \sum_{e\in W \subseteq  (( \bigcup_{i=1}^{m} A_{i} ) \backslash R_{\{A_{i}\}})} ~~~~~ b_{e} ~ \leq ~ b. \]
 This model corresponds to basic knapsack problem.

% ----------------------------------------------------------------
%\subsection{Aggregation of multisets}

\subsection{Rankings}

%\Lambda

 Median/consensus of rankings is one of the basic problems in decision
 making
 (e.g., \cite{barth89},
 \cite{cook92},
 \cite{kemeny59}, \cite{kemeny60},
  \cite{barth89}, \cite{kemeny59}, \cite{kemeny60}).
%
%%%%%%%%%%%%%%%%%%%%%%%%%%%%%%%%%%%%%%%%%%%%%%%%%%%%
% Cook-Kress
%
%  Here two types of metric/proximity for rankings are briefly considered:
%
% 1. Kendall-Tau distance \cite{kendall62}.
% M. Kendall,
%
% Here Kemeni optimal aggregation is widely used
% (i.e., searching for the Kemeni-Snell median
% through minimum sum of Kendall-Tau distances)
%
% (e.g., \cite{barth89}, \cite{kemeny59}, \cite{kemeny60}).
%
% 2. Distances for partial rankings
% (e.g., \cite{bansal09}, \cite{fagin06}).
%
% 3. Vector-like proximity (\cite{lev88}, \cite{lev98a}, \cite{lev98}).
% Here multiple choice problem is used to obtain
% a resultant interval ranking.
%
% Here aggregation problems are often based on the following combinatorial
% problems:
%  (i) median/consensus
%  (ii)  multiple choice problem (\cite{lev98}, \cite{})
%  (iii) assignment problem (\cite{}, \cite{dinu06}) Cheborarev
%%%%%%%%%%%%%%%%%%%%%%%%%%%%%%%%%%%%%%%%%%%%%%%%%%%%
% ADDITIONS !!!
%
% Some heuristics for the consensus ranking problem \cite{beck83}.
%
% consensus ranking in security (NB!) \cite{kruger08}.
%
%
 Here three methods are briefly described:
 (i) median consensus method based on assignment problem
 (e.g., \cite{cookseif78}, \cite{cook92}, \cite{cook96},
 \cite{kruger08});
 (ii) heuristic approach
% Beck and Lin, 1983
 (e.g., \cite{beck83}, \cite{kruger08}, \cite{tavana03});
 and
 (iii) method based on multiple choice problem
 (e.g., \cite{lev98a}, \cite{lev98}).

~~

 {\it Method 1.} The median consensus method based on distance
 (usually: Kendall-Tau distance)
 and assignment problem
 has been studied by Cook et al.
% (Cook-Seiford)
 (e.g., \cite{cookseif78}, \cite{cook92}, \cite{cook96}).
 Our version of the approach
 (for layered sets)
  is the following.
 Let \(A = \{1,...,i,...,n\}\) be the initial set of elements
 (alternatives, objects).
 The number of layers equals \(m\) (\(k=\overline{1,m}\)).
 There are \(\mu\) initial rankings of set \(A\):
 \(S^{1}\),..., \(S^{\lambda}\),..., \(S^{\mu}\).
 Thus, \(S^{\lambda} = \bigcup_{k=1}^{m}  A_{k}^{\lambda} \).
 Let \(r_{i}^{\lambda}\) (\(i\in A\)) be the priority of
 \(i\) in \(S^{\lambda}\), i.e.,
 the number of corresponding layer:
 \(r_{i}^{\lambda} = k\) if \( i \in A_{k}^{\lambda}\).

 The resultant ranking (consensus) is
 \(S^{a} = \{A_{1}^{a},...,A_{k}^{a},..., A_{m}^{a} \}\),
 and corresponding consensus priorities are
 \(r_{i}^{a}\) \(\forall i \in A\).
 The following binary variable will be used:
    \[x_{ik} = \left\{ \begin{array}{ll}
               1, & \mbox{if $ r_{i}^{a} = k $ or $ i\in A_{k}^{a} $,}\\
               0,  & \mbox{otherwise.}
               \end{array}
               \right. \]

 The assignment problem for finding the consensus is
 (case of layered set):
 \[\min \sum_{i=1}^{n} \sum_{k=1}^{m}  ( \sum_{\lambda=1}^{\mu} | r_{i}^{\lambda} - k  |  ) x_{ik}\]
 \[s.t.
% ~~\sum_{i=1}^{n}  x_{ik} = 1,
% k=\overline{1,m};
%
 ~~\sum_{k=1}^{m}  x_{ik} = 1, i=\overline{1,n};
 ~x_{ik} \in \{0,1\}.
%  ~ i=\overline{1,m}, ~ j=\overline{1,n}.
   \]
 Generally, polynomial algorithms exist for basic assignment problem
 (e.g., \cite{gar79}, \cite{kuh57}).
 The obtained version of assignment problem is more simple and can be solved by
 an evident greedy algorithm:
 selection of the closest layer \(\forall i \in A\).
%
%%%%%%%%%%%%%%%%%%%%%%%%%%%%%%%%%%%%%%%%%%%%%%%%%%%%%%%%%
 Let us consider a numerical example (Table 7.1):

\begin{center}

\end{center}

 (a) initial set of elements \(A = \{1,2,3,4,5,6,7,8,9\}\),

 (b) three rankings (four layer are examined):

 \(S^{1} = \{A_{1}^{1} = \{ 2,4\},  A_{2}^{1} = \{ 9\}, A_{3}^{1} = \{ 1,3,7\}, A_{4}^{1} = \{ 5,6,8\}
 \}\),

  \(S^{2} = \{A_{1}^{2} = \{ 2,3\},  A_{2}^{2} = \{ 4,9\}, A_{3}^{2} = \{ 1,7\}, A_{4}^{2} = \{ 5,6,8\}
 \}\),

  \(S^{3} = \{A_{1}^{3} = \{ 2,4\},  A_{2}^{3} = \{ 3,9\}, A_{3}^{3} = \{ 1,5\}, A_{4}^{3} = \{
  6,7,8\} \}\).

  The resultant ranking based on assignment problem above is (Table 10):

  \(S^{a} = \{A_{1}^{a} = \{ 2,4\},  A_{2}^{a} = \{ 3,9\}, A_{3}^{a} = \{ 1,7\}, A_{4}^{a} = \{ 5,6,8\}
 \}\).

~~

 {\it Method 2.} A basic heuristic approach
 has been suggested in
% Beck and Lin, 1983
 \cite{beck83}.
 The method is widely used (e.g., \cite{tavana03}).
 Let us consider a simplified version of heuristic
 to find the corresponding solution
 \(\overline{S^{a}}\):
    \[\overline{r_{i}^{a}} = \left\{ \begin{array}{ll}
               \lceil (\sum_{\lambda=1}^{\mu} r_{i}^{\lambda} ) / \mu \rceil , & \mbox{if $
                (\sum_{\lambda=1}^{\mu} r_{i}^{\lambda} ) / \mu  -  \lceil (\sum_{\lambda=1}^{\mu} r_{i}^{\lambda} ) / \mu \rceil < 0.5  $,}\\
               \lceil (\sum_{\lambda=1}^{\mu} r_{i}^{\lambda} ) / \mu \rceil + 1,  & \mbox{otherwise.}
               \end{array}
               \right. \]
 Thus, solution
 \(\overline{S^{a}} \)
 is (Table 10):

  \(\overline{S^{a}} = \{ \overline{A_{1}^{a}} = \{ 2,4\},  \overline{A_{2}^{a}} = \{ 3,9\},
  \overline{A_{3}^{a}} = \{ 1,7\}, \overline{A_{4}^{a}} = \{ 5,6,8\}
 \}\).

~~

 {\it Method 3.}
% multiple choice problem \cite{lev98})
%%%%%%%%%%%%%%%%%%%%%%%%%%%%%%%%%%%%%%%%%%%%%%%%%%%%%%%%%%%%%%%%%
%  From book Levin, 1998
%\section{Aggregation of layered structures}
%
% Now a more complicated aggregation process based
% on multiple choice problem is examined
% (e.g, \cite{lev98}).
%
% n aggregation of layered structures above to obtain .
% Generally,
%  the problem consists in building of a consensus for a set of initial
% structures. We will use
%%%%%%%%%%%%%%%%%%%%%%%%%%%%%%%%%%%%%%%%%%%%%%%%%
%
% [F:  \{G^{\lambda} = (A,E^{\lambda}), \lambda=1,...,\Lambda \}
%   \Longrightarrow
%   G^{a}=(A,E^{a}).\]
%
% In our opinion, it is reasonable to consider the following standard problems:
% \(\{L\} \Rightarrow S\),
% \(\{T\} \Rightarrow P\),
% \(\{S\} \Rightarrow S\),
% \(\{S\} \Rightarrow S_{f}\),
% \(\{S_{f}\} \Rightarrow S_{f}\).
%
% There exist the following three approaches to the synthesis problem above:
%
% (i) axiomatic;
%
% (ii) criterial (e.g., concordance criteria, majority rules); and
%
% (iii) modeling (i.e., usage of physical model in a space) (Kuzjmin, 1982).
%  \cite{kuz82}
%
% Let us examine an
%  modeling
% approach to problem
%
 Now a more complicated aggregation process based
 on multiple choice problem is examined
 (initial rankings \(\{S^{\lambda} | \lambda = \overline{1,\mu} \}\)
 are mapped into a fuzzy aggregated ranking
 \(S^{a}\))
 \(\{S^{\lambda}\} \Rightarrow S^{a}\)  \cite{lev98}:
 \[h(S^{a} \mid s^{a} \in \{S \mid \eta (S,S^{\lambda}) \preceq \eta_{o}
 ~~ \forall \lambda=1,...,\mu \} \longrightarrow max , \]
 where \(h\) is an attribute (quality) of the resultant ('overage') structure
 \(S^{a}\);
 \(\eta_{o}\) is a vector-like proximity.
 The problem is depicted in Fig. 7.1.

\begin{center}
\begin{picture}(45,31)
\put(3,00){\makebox(0,0)[bl]{Fig. 7.1.
 Aggregation
% of structures
 \cite{lev98}}}

% (\(S^{a}\) is denoted by}}

%\put(85,02){\circle*{0.8}} \put(85,02){\circle{1.5}}
% \put(85,02){\circle{2.2}}

%\put(87,00){\makebox(0,0)[bl]{) \cite{lev98}}}

\put(20,12){\vector(0,-1){7.6}} \put(20,12){\circle*{1}}
\put(20,12){\oval(15,15)}
\put(21,8){\makebox(0,8)[bl]{\(\eta_{o}\)}}

\put(26,19){\vector(0,1){7.6}} \put(26,19){\circle*{1}}
\put(26,19){\oval(15,15)}

\put(17,21){\vector(-1,0){7.6}} \put(17,21){\circle*{1}}
\put(17,21){\oval(15,15)}

\put(05,24){\makebox(0,8)[bl]{\(S^{1}\)}}
\put(34,24){\makebox(0,8)[bl]{\(S^{\lambda}\)}}
\put(27,5){\makebox(0,8)[bl]{\(S^{\mu}\)}}

\put(21,17){\circle*{0.8}} \put(21,17){\circle{1.5}}
\put(21,17){\circle{2.2}}

\put(22.5,17){\line(1,0){12.5}}

\put(35.5,16){\makebox(0,8)[bl]{\(S^{a}\)}}

%\put(40,21){\makebox(0,8)[bl]{\(S^{a}\)}}

\end{picture}
\end{center}

% Now let us examine problem \(\{S\} \Rightarrow S_{f}\) in details.
%
 The following notations are used:
 \(a_{il}\) is the number of initial structures,
   in which element \(i \in A_{l}\), \(l=\overline{1,m}\) (layers);
 vector \(\xi_{i} =  (\xi_{i1},...,\xi_{il},...,\xi_{im})\) defines
 frequencies of belonging of element \(i\)
 to layers
 \(\{A_{1},...,A_{l},...,A_{m}\}\),
% (\(1,...,m\)),
 where   \(\xi_{il} = \frac{a_{il}}{\mu}\)
 (it is a membership function of element \(i\) to layer \(l=1,..,m\)).
 Let us denote \(S^{a}_{f}\) as a set of intervals \(\{d_{i}\}\).
 The resultant quality of \(S^{a}_{f}\)
 is  based on the following entropy-like function:
  \[ \sum^{n}_{i=1} H_{i} = \sum^{n}_{i=1} \frac{1}{d^{2}_{i}-d^{1}_{i} + 1}
 \longrightarrow max.\]
 Next, a modular vector as the proximity is used:
 ~~ \( z_{o} = (z^{1},...,z^{k},0,...,0)\).
 Finally, the problem is:
%
%  \[ \sum^{n}_{i=1} H_{i} (S_{f}) \longrightarrow max,
%
% ~~ z(S^{\lambda},S_{f}) \preceq z_{o}, \forall \lambda = 1,...,\mu . \]
%
%
  \[ \sum^{n}_{i=1} H_{i} (S^{a}) \longrightarrow max,
  ~~ z(S^{\lambda},S^{a}) \preceq z_{o}, \forall \lambda = 1,...,\mu . \]
 With respect to zero-valued components \(z^{k+1},...,z^{k^{+}}\),
 it is possible to define a set of admissible variants to intervals
 (\(d_{i\theta} \mid \theta=1,...,q_{i} \)).
 Thus,
 we reduce the model to the following modification of
 multiple-choice problem
  (\cite{gar79}, \cite{keller04}):
 \[ \sum^{n}_{i=1} \sum^{q_{i}}_{\theta=1} H_{i\theta}(d_{i\theta})
 \kappa_{i\theta}
 \longrightarrow max, \]
 \[\sum_{r\geq p} \sum^{n}_{i=1} \sum^{q_{i}}_{\theta=1} b^{r}_{i\theta}
 \kappa_{i\theta}
 \leq \sum_{r\geq p} z^{r}, p=1,...,k,
 ~~ \sum^{q_{i}}_{\theta=1} \kappa_{i\theta} = 1, i=1,...,n, \kappa_{i\theta} \in \{
  0,1\} , \]
 where \(b^{r}_{i\theta}\) is the sum of components \(\xi _{i}\),
 which are differed from \(d^{1}_{i\theta}\) (\(d^{2}_{i\theta}\)) by \(r\).
 A version of the described aggregation scheme has been implemented in
 DSS COMBI
% (Levin and Michailov, 1988).
 (\cite{lev98}, \cite{levmih88}).

%%%%%%%%%%%%%%%%%%%%%%%%%%%%%%%%%%%%%%%%%%%%%%%%%%%%%%%%%%%%%%
% In the case, when we use two-side vector \(z_{o}\), the first constraint
% of the model has to be transformed into two constraints for negative and
% positive components accordingly.
%
% Note that examined modification of multiple-choice problem is interesting
% for cases,
% when resources in knapsack-like problems
% are ordered by their importance and maybe replaced.
%%%%%%%%%%%%%%%%%%%%%%%%%%%%%%%%%%%%%%%%%%%%%%%%%%%%%%%%%%%%%%%%%%%%%%%%%%%%%

% ----------------------------------------------------------------
\subsection{Trees}

 Here initial information consists in
 a set of trees.
 Usually four basic approaches are considered:

 (1) maximum common subtree
 (e.g., \cite{akatsu00},
 \cite{finden85});

 (2) median/agreement tree
 (e.g., \cite{amir97},
 \cite{berry04},
 \cite{farach95},
 \cite{gui06},
 \cite{jansson05},
 \cite{phillips96},
 \cite{steel93});

%consensus phillips

 (3) compatible tree
 (e.g.,
 \cite{berry04},
 \cite{gui06}, \cite{hamel96}); and

 (4) maximum agreement forest
 (e.g., \cite{cha05},
 \cite{hallett07}, \cite{rodr07},
 \cite{whidden10}, \cite{whidden11}).

 Mainly, the problems above correspond to class of NP-hard problems
 (e.g.,
 \cite{hamel96},
 \cite{phillips96}).
 As a result, heuristics, approximation schemes, and enumerative
 methods are used.

 Thus,
 the aggregation problem
 for set of initial trees
 \( \{T\} = \{ T^{1},...,T^{i},...,T^{m} \}\)
  can be considered as follows
 (addition strategy I):

~~

 {\it Stage 1.} Searching for a median-like tree (i.e., ``kernel''):
 \[T^{agg} = ~arg ~\min_{\{T\}} ~ ( \sum^{m}_{i=1} \rho(T,T^{i}) ). \]

 {\it Stage 2.} Generation of a set of additional elements
 (nodes and/or edges).

 {\it Stage 3.} Addition of elements to \(T^{agg}\)
 (knapsack-like problem).

~~

 Evidently,
 in the case of vector-like proximity
 \(\overline{\rho}(T',T'')\),
 \(T^{agg}\)
 has to be searched for as Pareto-efficient solution(s).
 On the other hand,
 it is reasonable to consider some heuristic algorithms for
 building the ``kernel'', for example:
  \[K =  \bigcup_{i=1}^{m-1} (  T^{i} \bigcap T^{i=1} ).\]

\subsection{Morphological Structures}

% Generally, the following basic kinds of sets are examined:
% sets, multisets, lists, and complete lists
%
%  (e.g., \cite{dovier98}, \cite{knuth98}, \cite{syr01}).
%
%
% Further, sets and multisets will be  considered.
%%%%%%%%%%%%%%%%%%%%%%%%%%%%%%%%%%%%%%%%%%%%%%%%%%%%%%

  The basic aggregation problem
 for set of initial structures
 \( \{\Lambda\} = \{ \Lambda{1},...,\Lambda^{i},...,\Lambda^{m} \}\)
  can be considered as follows
 (addition strategy I):

~~

 {\it Stage 1.} Searching for a median-like tree (i.e., ``kernel''):
 \[\Lambda^{agg} = ~arg ~\min_{\{\Lambda\}} ~ ( \sum^{m}_{i=1} \rho(\Lambda,\Lambda^{i}) ). \]

 {\it Stage 2.} Generation of a set of additional elements
 (nodes and/or edges).

 {\it Stage 3.} Addition of elements to \(\Lambda^{agg}\)
 (knapsack-like problem).

~~

 Evidently,
 in the case of vector-like proximity
 \(\rho(\Lambda',\Lambda'')\),
 \(\Lambda^{agg}\)
 has to be searched for as Pareto-efficient solution(s).
 Generally,
 morphological structures (morphological structures with
 compatibility) are very complicated composite structures:
 ~~~~ {\bf \(\Lambda\)}\(= \langle \)
% \(\overline{\rho}\) (\(\Lambda^{\alpha},\Lambda^{\beta})=\)
%
 {\bf T},{\bf P},{\bf D},{\bf R},{\bf I}
 \( \rangle\).
%%%%%%%%%%%%%%%%%%%%%%%%%%%%%%%%%%%%%%%%%%%%%%%%%%%%%%%%%%%%%%%%%%%
% where the following parts are considered
% (\cite{lev98}, \cite{lev06}, \cite{levprob07}):
%
% (i) tree-like system model {\bf T},
%
% (ii) set of leaf nodes as basic system parts/components {\bf P},
%
% (iii) sets of DAs for each leaf node {\bf D},
%
% (iv) DAs rankings (i.e., ordinal priorities) {\bf R}, and
%
% (v) compatibility estimates between DAs {\bf I}.
%%%%%%%%%%%%%%%%%%%%%%%%%%%%%%%%%%%%%%%%%%%%%%%%%%%%%%%%%%%%%%%%%%%
%
%
 Here, it may be reasonable
 to consider the following  heuristic solving scheme:

~~

 {\it Stage 1.} Aggregation of sets of systems parts
 \(\{\){\bf P}\(\}\).

 {\it Stage 2.} Aggregation of sets of DAs
 \(\{\){\bf D}\(\}\).

 {\it Stage 3.} Aggregation of sets of compatibility estimates
 \(\{\){\bf I}\(\}\).

 {\it Stage 4.} Aggregation of tree-like models
 \(\{\){\bf T}\(\}\).
%

% ----------------------------------------------------------------
\section{Illustrative Applied Numerical Examples}

 The list of examined applied examples is presented in Table 8.1.

\begin{center}

\end{center}

%\subsection{Aggregation of trees}

\subsection{ZigBee Communication Protocol}
%
%\subsection{Morphological Structures (Example for ZigBee Protocol)}
%
 Here an applied example for aggregation of ZigBee protocol
 solutions from \cite{levand10} is briefly described.
 A general framework for aggregation of two solutions is depicted in Fig. 8.1.
  A brief description of Zigbee protocol components is the
 following \cite{levand10}:

   {\it 1.} Interference avoidance \(A\):~
     {\it 1.1.} Startup Procedure of Channel Acquisition \(X\).
     {\it 1.2.} Channel Hopping \(J\).

 {\it 2.} Automated/distributed address management \(B\).

 {\it 3.} Group addressing \(I\).

 {\it 4.} Centralized data collection \(C\):~
       {\it 4.1.} Low-overhead data collection by ZigBee Coordinator
       \(G\).
       {\it 4.2} Low-overhead data collection by other devices
       \(H\).
       {\it 4.3.} Many-to-one routing \(Q\).
%       \(Q_{1}\) (Whole network discovers the aggregator in one pass), and
%
%
       {\it 4.4.} 6LoWPAN multicast/broadcast support \(V\).
%       \(P_{1}\)
%
       {\it 4.5.} Source routing \(P\).
%       \(P_{1}\)
%        (Aggregator responds to all senders in an economical manner).

 {\it 5.} Network scalability \(D\).
% \(D_{2}\) (An addressing algorithm that
% relaxes the limits on network size. Networks with hundreds to
% thousands of devices are supported).

 {\it 6.} Message size \(E\).
%  \(E_{2}\) (Large messages, up to the buffer capacity of the sending and receiving
% devices, are supported using Fragmentation and Reassembly).

 {\it 7.} Standardized commissioning \(Z\).
% \(K_{1}\) (Standardized startup
% procedure and attributes support the use of commissioning tools in
% a multi-vendor environment).

 {\it 8.} Robust mesh networking:
%  \(F\):
  6LoWPAN approach \(U\).
%
%      {\it 1.8.1.} Fault tolerant routing algorithms \(R\).
%      \(R_{1}\) (Response to changes in the network and in the RF
%      environment),
%
%        {\it 8.2.} Neighborhood tables \(T\).
%        \(T_{1}\) (Kept by every device).

 {\it 9.} Cluster Library support \(L\).
% \(L_{1}\) (Standardizes application behavior across profiles).

  {\it 10.} Web services support \(W\).
%  \(W_{1}\) (condensed HTTP with tokenized XML data).

 The above-mentioned system components
 have implementation alternatives
 (e.g., \(B_{1}\), \(B_{2}\)).

 The following initial solutions are considered:

 (a) an expert-based forecast  \(\Theta^{1}\)
  (\( \widetilde{S}_{4} \)  from  \cite{levand10})
 (Fig. 8.2),

 (b) forecast based on using knapsack problem
 \(\Theta^{2}\)
 (\( \widehat{\Phi} \)   from  \cite{levand10})
 (Fig. 8.3).

\begin{center}
%
\end{center}

 The corresponding substructure of protocol \(\widetilde{\Theta}\)
 and superstructure of protocol \(\overline{\Theta}\)
 are shown in  Fig. 8.4 and in Fig. 8.5.

\begin{center}
%%
\end{center}

%%%%%%%%%%%%%%%%%%%%%%%%%%%%%%%%%%%%%%%%%%%%%%%%%%%%%%%%%%%%%%%

 Further, let us consider aggregation processes.
%
%%%%%%%%%%%%%%%%%%
%
 A list of addition operations (for strategy I)
 is presented in Table 8.2
 (designation of operation from
 \cite{levand10}
 is depicted in parentheses).
 Here and hereafter the following attributes (criteria) for an assessment of the
 addition operations (deletion operations) are used \cite{levand10}:
 (1)  cost
% (i.e., required resource)
 \(\Upsilon_{1}\);
 (2) required time for implementation \(\Upsilon_{2}\);
 (3) performance \(\Upsilon_{3}\);
 (4) decreasing a cost of  maintenance \(\Upsilon_{4}\);
 (5) scalability  \(\Upsilon_{5}\);
 (6) reliability \(\Upsilon_{6}\);
 (7) mobility \(\Upsilon_{7}\); and
 (8) usability value \(\Upsilon_{8}\).
 An ordinal scale [1,5] is used for each criterion:
 \(1\) corresponds to ``strong negative effect'',
  \(2\) corresponds to ``negative effect'',
   \(3\) corresponds to ``no changes'',
 \(4\) corresponds to ``positive effect'', and
 \(5\) corresponds to ``strong positive effect''.
 Priorities \(\{r_{i}\}\) are obtained via multicriteria ranking
 (Electre-like method \cite{levand10}).

\begin{center}
%
\end{center}

  Thus, the addition problem (simplified knapsack problem) is:
 \[\max \sum_{i=1}^{5} c_{i} x_{i} ~~~
 \]
  \[
  s.t. \sum_{i=1}^{5}  a_{i} x_{i} \leq b, x_{i} \in \{0,1\}.\]
 Cost estimates are (by criterion \(\Upsilon_{1}\)) used as
 \(\{a_{i}\}\),
  priorities \(\{r_{i}\}\) are used as (transform to)
  \(\{c_{i}\}\),
  and
%  constraint
   \(b = 8.00\).

 A resultant solution for strategy I is depicted in Fig. 8.6
 (\(x_{1} = 1\), \(x_{2} = 1\), \(x_{3} = 0\), \(x_{4} = 0\), \(x_{5} = 1\)).
 Here compatibility estimates between design alternatives for
 system components are not considered.

\begin{center}
%
\end{center}

  A list of deletion operations (for strategy II)
 is presented in Table 8.3 (designation of operation from
 \cite{levand10}
 is depicted in parentheses).

\begin{center}
%%
\end{center}

  Thus, the deletion problem (knapsack problem with minimization
 of the objective function) is:
 \[\min \sum_{i=1}^{6} c_{i} x_{i}\]
  \[s.t. \sum_{i=1}^{6}  a_{i} x_{i} \geq b, x_{i} \in \{0,1\}.\]
 Cost estimates are (by criterion \(\Upsilon_{1}\)) used as
 \(\{a_{i}\}\),
  priorities \(\{r_{i}\}\) are used as (transform to)
  \(\{c_{i}\}\),
  and \(b=8.00\).

\begin{center}
%%
\end{center}

 A resultant solution based on strategy II is depicted in Fig. 8.7
 (\(x_{1} = 0\), \(x_{2} = 1\), \(x_{3} = 1\), \(x_{4} = 1\),
 \(x_{5} = 0\), \(x_{6} = 0\), \(x_{7} = 1\)).
 Here compatibility estimates between design alternatives for
 system components are not considered.

%%%%%%%%%%%%%%%%%%%%%%%%%%%%%%%%%%%%%%%%%%%% EXAMPLE - TELEMETRY SYSTEM

\subsection{On-Board Telemetry System}

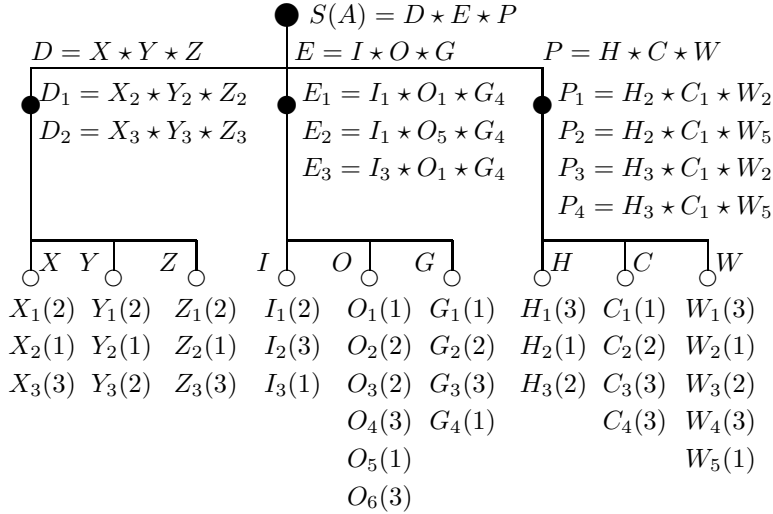
 Here a numerical example for on-board telemetry
 system is considered from \cite{levkhod07}.
 The initial morphological structure for on-board equipment is
 the following  (Fig. 8.8):

 {\bf 1.} On-board equipment \(S=D \star E \star P\).

 {\it 1.1.} Power supply \(D = X \star Y \star Z \):
 {\it 1.1.1.} stabilizer \(X\):
 \(X_{1}\) (standard),
 \(X_{2}\) (transistorized),
 \(X_{3}\) (high-stability);
 {\it 1.1.2.} main source \(Y\):
 \(Y_{1}\) (Li-ion),
 \(Y_{2}\) (Cd-Mn),
 \(Y_{3}\) (Li);
 {\it 1.1.3.} emergency cell \(Z\):
 \(Z_{1}\) (Li-ion),
 \(Z_{2}\) (Cd-Mn),
 \(Z_{3}\) (Li).

 {\it 1.2.} Sensor element \(E = I \star O \star G \):
 {\it 1.2.1.} acceleration sensors  \(I\):
 \(I_{1}\) (ADXL),
 \(I_{2}\) (ADIS),
 \(I_{3}\) (MMA);
 {\it 1.2.2.} position sensors \(O\):
 \(O_{1}\) (SS12),
 \(O_{2}\) (SS16),
 \(O_{3}\) (SS19),
 \(O_{4}\) (SS49),
 \(O_{5}\) (SS59),
 \(O_{6}\) (SS94);
 {\it 1.2.3.} global positioning system (GPS) \(G\):
 \(G_{1}\) (EB),
 \(G_{2}\) (GT),
 \(G_{3}\) (LS),
 \(G_{4}\) (ZX).

 {\it 1.3.} Data processing system  \(P = H \star C \star W \):
  {\it 1.3.1.} data storage unit \(H\):
 \(H_{1}\) (SRAM),
 \(H_{2}\) (DRAM),
 \(H_{3}\) (FRAM);
 {\it 1.3.2.} processing unit (CPU) \(C\):
 \(C_{1}\) (AVR),
 \(C_{2}\) (ARM),
 \(C_{3}\) (ADSP),
 \(C_{4}\) (BM);
 {\it 1.3.3.} data write unit \(W\):
 \(W_{1}\) (built-in ADC),
 \(W_{2}\) (external ADC I2C),
 \(W_{3}\) (external ADC SPI),
 \(W_{4}\) (external ADC 2W),
 \(W_{5}\) (external ADC UART(1)).

\begin{center}

\end{center}

 In \cite{levkhod07},
 24 resultant solutions have been obtained
 via HMMD
 (\(A_{1},...,A_{24}\)).
 In the example, 6 initial solutions are examined
 (in parentheses designation from  \cite{levkhod07} in pointed out):

 \(S_{1} (A_{1})  = D_{1} \star E_{1} \star P_{1}  =
  ( X_{2} \star Y_{2} \star Z_{2} ) \star
  ( I_{1} \star Q_{1} \star G_{4} ) \star
  ( H_{2} \star C_{1} \star W_{2} )\),

  \(S_{2} (A_{24})  = D_{2} \star E_{3} \star P_{4}  =
  ( X_{3} \star Y_{3} \star Z_{3} ) \star
  ( I_{3} \star Q_{1} \star G_{4} ) \star
  ( H_{3} \star C_{1} \star W_{5} )\),

   \(S_{3} (A_{11})  = D_{1} \star E_{3} \star P_{3}  =
  ( X_{2} \star Y_{2} \star Z_{2} ) \star
  ( I_{3} \star Q_{1} \star G_{4} ) \star
  ( H_{3} \star C_{1} \star W_{2} )\),

   \(S_{4} (A_{19})  = D_{2} \star E_{2} \star P_{3}  =
  ( X_{3} \star Y_{3} \star Z_{3} ) \star
  ( I_{1} \star Q_{5} \star G_{4} ) \star
  ( H_{3} \star C_{1} \star W_{2} )\),

   \(S_{5} (A_{23})  = D_{2} \star E_{3} \star P_{3}  =
  ( X_{3} \star Y_{3} \star Z_{3} ) \star
  ( I_{3} \star Q_{1} \star G_{4} ) \star
  ( H_{3} \star C_{1} \star W_{2} )\),

   \(S_{6} (A_{12})  = D_{1} \star E_{3} \star P_{4}  =
  ( X_{2} \star Y_{2} \star Z_{2} ) \star
  ( I_{3} \star Q_{1} \star G_{4} ) \star
  ( H_{3} \star C_{1} \star W_{5} )\).

 In Fig. 8.9 and Fig. 8.10,
 supersolution and subsolution are depicted.

\begin{center}

\end{center}

 The obtained subsolution contains only two elements.
 Thus,
 the design of ``system kernel''
 is based on the special method with \(\alpha = 0.6\) (Fig. 8.11).

\begin{center}
%
\end{center}

%++++++++++++++++++++++++++++++++++++++++++++++++++++++++++++

 Further, the addition strategy as simplified multiple choice problem is
 used:
 \[\max \sum_{i=1}^{3} \sum_{j=1}^{2} c_{ij} x_{ij}\]
  \[s.t. \sum_{i=1}^{3} \sum_{j=1}^{2}  a_{ij} x_{ij} \leq b,
  \sum_{j=1}^{2}  x_{ij} = 1 \forall i=\overline{1,3};
  x_{ij} \in \{0,1\}.\]
  A list of addition operations
 is presented in Table 8.4 \cite{levkhod07}.
  Here cost estimates are based on expert judgment,
  priorities \(\{r_{i}\}\) were computed via
  Electre-like method in \cite{levkhod07}
 and transformed to
  \(\{c_{i}\}\),
%  and
%  constraint
   \(b = 9.00\).

 A resultant solution is depicted in Fig. 8.12
 (\(x_{11} = 1\), \(x_{12} = 0\), \(x_{21} = 1\), \(x_{22} = 0\),
 \(x_{31} = 1\), \(x_{32} = 0\)).
 Note, compatibility estimates between design alternatives for system components
  \(X\), \(Y\), \(Z\) are not considered.

\begin{center}
%
\end{center}

%%%%%%%%%%%%%%%%%%%%%%%%%%%%%%%%%%%%%%%%%%%% CONTINUATION OF EXAMPLE FOR NOTEBOOK

 \subsection{Continuation of Example for Notebook}

 Let us examine the final part of the example for notebook.
 Here the combined  aggregation strategy (strategy III) is considered for two cases:

 (1)  ``system kernel'' as an extension of
 substructure:

  \(K' = B_{1} \star R_{1} \star  V_{3} \star J_{1} \star E_{1} \star O_{1} \star D_{1}
 \star A_{1} \star G_{1} \star L_{1} \star Q_{2} \);

 (2) ``system kernel''
 \(K^{*} \)
 based on multicriteria selection
 of the ``best'' design alternatives for each system component.

%
% \(K^{*} = B_{1} \star U_{2} \star R_{2}  \star  V_{3} \star E_{2} \star O_{2}
% \star F_{2} \star D_{2} \star G_{2} \).

%
%
 Let us consider case 1.
 Here the aggregation strategy
 as modification of ``system kernel'' \(K'\)
  can be applied.
 A set of candidate modification operations are the  following:

 {\it 1.} addition operations:
 {\it 1.1.} addition for \(U\):
 \(U_{1}\) or   \(U_{2}\) or  \(U_{3}\),
  {\it 1.2.} addition for \(F\):
 \(F_{1}\) or  \(F_{2}\),
 {\it 1.3.} addition for \(P\):
 \(P_{2}\) or   \(P_{3}\) or  \(P_{4}\);

 {\it 2.} correction  operations:
 {\it 2.1.} replacement \(B_{1} \Rightarrow B_{2}\),
 {\it 2.2.} replacement \(V_{3} \Rightarrow V_{4}\),
 {\it 2.3.} replacement \(A_{1} \Rightarrow A_{3}\).

 Table 8.5 contains the list of modification operations above,
 their estimates (ordinal expert judgment) and corresponding binary variables.

\begin{center}
%
\end{center}

 The following simplified multiple choice problem is
 used  (\(c_{ij} = 3-r_{ij}\), \(b = 11.00\)):
 \[\max \sum_{i=1}^{6} \sum_{j=1}^{q_{i}} c_{ij} x_{ij}\]
  \[s.t. \sum_{i=1}^{6} \sum_{j=1}^{q_{i}}  a_{ij} x_{ij} \leq b,
  \sum_{j=1}^{q_{i}}  x_{ij} = 1 \forall i=\overline{1,6};
  x_{ij} \in \{0,1\}.\]
 A resultant computer solution \(S^{1c}\) is depicted in Fig. 8.13
 (\(x_{11} = 1\), \(x_{12} = 0\), \(x_{13} = 0\),
 \(x_{21} = 0\), \(x_{22} = 1\),
 \(x_{31} = 0\), \(x_{32} = 1\), \(x_{33} = 0\),
 \(x_{41} = 0\), \(x_{42} = 1\),
 \(x_{51} = 0\), \(x_{52} = 1\),
 \(x_{61} = 1\), \(x_{62} = 0\)).
 Here a greedy algorithm was used.
 Note, compatibility estimates between design alternatives for system components
%  \(X\), \(Y\), \(Z\)
  are not considered.
%
%
% An example of the resultant solution
% \(S^{1}\) (modification of ``system kernel'' \(K\)) is shown in Fig. 92.

\begin{center}

\end{center}

 Now let us consider case 2.
 Here building of  ``system kernel''
 is based on
  multicriteria selection and/or expert judgment.
   The basic structure of ``system kernel'' is:
 \(B\),  \(U\),  \(R\),  \(V\),  \(O\),
 \(F\),  \(D\),  \(G\).
 For each system component above, it is possible to consider
 a selection procedure to choose the ``best'' system element
 (while taking into account elements of the initial solution
 or additional elements as well).
%
% Thus,

 Table 8.6 contains design alternatives for the selected components
 of ``system kernel'' structure above including ordinal estimates
 (expert judgment, the smallest estimates correspond to the best situation)
 and the resultant priorities.
 The following criteria were used:
 cost (\(\Upsilon_{1}\)),
 usefulness (\(\Upsilon_{2}\)),
 experience (\(\Upsilon_{3}\)),
 prospective features (\(\Upsilon_{4}\)).
 As a result, the following  ``system kernel'' \(K^{*} \)  is obtained:
 \(K^{*} =  B_{2} \star U_{2} \star R_{3} \star V_{3} \star E_{1} \star O_{1} \star F_{2}  \star D_{1} \star
 G_{1}\).

 Further, the system correction process is based on the following
 operations:

 {\it 1.} addition:
   {\it 1.1.} \(A_{1}\),
   {\it 1.2.}  \(P_{1}\),
   {\it 1.3.} \(L_{1}\);

 {\it 2.} deletion: {\it 2.1.}
  \(E_{1}\);

 {\it 3.} replacement:
 {\it 3.1.} \(B_{2} \Rightarrow
% \longrightarrow
 B_{3}\),
 {\it 3.2.} \(U_{2} \Rightarrow U_{1}\),
 {\it 3.3.} \(O_{1} \Rightarrow O_{3}\).

  Table 8.7 contains the list of modification operations above,
 their estimates (ordinal expert judgment) and corresponding binary variables.

 The following multiple choice problem is used
  (\(c_{ij} = 3-r_{ij}\), \(b = 9.00\)):
 \[\max \sum_{i=1}^{7} \sum_{j=1}^{2} c_{ij} x_{ij}\]
  \[s.t. \sum_{i=1}^{7} \sum_{j=1}^{2}  a_{ij} x_{ij} \leq b,
  \sum_{j=1}^{2}  x_{ij} = 1 \forall i=\overline{1,7};
  x_{ij} \in \{0,1\}.\]
 A resultant computed solution \(S^{2c}\) is depicted in Fig.
 8.14
 (\(x_{11} = 1\), \(x_{12} = 0\),
 \(x_{21} = 0\), \(x_{22} = 1\),
 \(x_{31} = 1\), \(x_{32} = 0\),
 \(x_{41} = 1\), \(x_{42} = 0\),
 \(x_{51} = 1\), \(x_{52} = 0\),
 \(x_{61} = 0\), \(x_{62} = 1\),
 \(x_{71} = 1\), \(x_{72} = 0\)).
 Here a greedy algorithm was used.
 Note, compatibility estimates between design alternatives for system components
%  \(X\), \(Y\), \(Z\)
  are not considered.

\begin{center}
%
\end{center}

%%%%%%%%%%%%%%%%%%%%%%%%%%%%%%%%%%%%%%%%%%%%%%%%%%%%%%%%%%%%%%%%%%%%%
%%%%%%%%%%%%%%%%%%%%%%%%%%%%%%%%%%%%%%%%%% EXAMPLE: SECURITY SYSTEM

\subsection{Integrated Security System}

 Here a numerical example for integrated security
 system is considered from \cite{levleus09}.
 The initial morphological structure is the following  (Fig. 8.15):

 {\bf 0.} Integrated
% safety/
 security system ~\(S = A \star B \star O \).

 {\bf 1.} Closed-circuit television (CCTV)
 \(A = J \star D \):

 1.1. Cameras  J:~
 conventional \(J_{1}(2)\),
 ``Tirret''
% camera(s)
 \(J_{2}(2)\),
 ``varifocal'' \(J_{3}(3)\), and
 ``auto-house'' \(J_{4}(1)\).

 1.2. Lighting
  \(D\):~
 natural \(D_{1}(3)\),
 natural and guard \(D_{2}(2)\), and
 natural, guard, and alarm  \(D_{3}(1)\).

 {\bf 2.} Access control
   \(B = G \star U \star V \):

  {\it 2.1.} Access to territory
    G:~
  card  \(G_{1}(1)\),
  radio-pendant
   \(G_{2}(3)\), and
  biometry  \(G_{3}(2)\).

 {\it 2.2.} Access to building
 U:~
 card  \(U_{1}(1)\),
 radio-pendant \(U_{2}(3)\), and
 biometry \(U_{3}(2)\).

 {\it 2.3.} Access to premises
  V:~
 card  \(V_{1}(1)\),
 code \(V_{2}(2)\),
 biometry \(V_{3}(2)\),
 \(V_{4} = V_{1} \& V_{2} (2)\),
 \(V_{5} = V_{1} \& V_{3} (3)\), and
 \(V_{6} = V_{1} \& V_{2} \& V_{3} (3)\).

 {\bf 3.} Burglar alarm
 \(O = X \star Y \star Z \):

  {\it 3.1.} Border-line 1 based on some principles
  \(X\):~
    single physical principle
  \(X_{1}(1)\),
 two physical principles
    \(X_{2}(2)\), and
 three physical principles
     \(X_{3}(3)\).

 {\it 3.2.} Border-line 2 based on
 some principles
 \(Y\):~
 single physical principle \(Y_{1}(1)\),
 two physical principles \(Y_{2}(3)\), and
 three physical principles
 \(Y_{3}(2)\).

{\it 3.3.} Border-line 3 based on some principles
 \(Z\):~
  single physical principle \(Z_{1}(1)\),
  two physical principles \(Z_{2}(3)\), and
  three physical principles \(Z_{3}(2)\).

 In \cite{levleus09},
 two resultant solutions have been obtained
 (via HMMD):

 \(S_{1} = A_{1} \star B_{1} \star O_{1}  =
  ( J_{2} \star D_{1} ) \star
  ( G_{1} \star U_{1} \star V_{1} ) \star
  ( X_{3} \star Y_{3} \star Z_{3} )\),

 \(S_{2} = A_{1} \star B_{1} \star O_{1}  =
  ( J_{2} \star D_{1} ) \star
  ( G_{1} \star U_{1} \star V_{1} ) \star
  ( X_{1} \star Y_{1} \star Z_{1} )\).

\begin{center}

\end{center}

 In Fig. 8.16 and Fig. 8.17,
 supersolution and
 subsolution are depicted.

\begin{center}
%
\end{center}

 Here, the obtained subsolution can be used as ``system kernel'',
 i.e.,
 \(K = \widetilde{\Theta}\).

%++++++++++++++++++++++++++++++++++++++++++++++++++++++++++++

 Further, the addition strategy as simplified multiple choice problem is
 used:
 \[\max \sum_{i=1}^{3} \sum_{j=1}^{2} c_{ij} x_{ij}\]
  \[s.t. \sum_{i=1}^{3} \sum_{j=1}^{2}  a_{ij} x_{ij} \leq b,
  \sum_{j=1}^{2}  x_{ij} = 1 \forall i=\overline{1,3};
  x_{ij} \in \{0,1\}.\]
  A list of addition operations
 is presented in Table 8.8 \cite{levleus09}.
  Here cost estimates are based on expert judgment,
  priorities \(\{r_{i}\}\) were computed via
  Electre-like method in \cite{levleus09}
 and transformed to
  \(\{c_{i}\}\),
%  and
%  constraint
   \(b = 7.00\).

 A resultant solution is depicted in Fig. 8.18
 (\(x_{11} = 1\), \(x_{12} = 0\), \(x_{21} = 1\), \(x_{22} = 0\),
 \(x_{31} = 1\), \(x_{32} = 0\)).
 Note, compatibility estimates between design alternatives for system components
  \(X\), \(Y\), \(Z\) are not considered.

\begin{center}
%
\end{center}

%%%%%%%%%%%%%%%%%%%%%%%%%%%%%%%%%%% Example: Aggregation of Common Course

\subsection{Common Educational Course}

 Here the following initial sets are considered (Table 8.9):

  (i) initial set of educational modules for a basic course on
 combinatorial optimization
 \(A = \{1,...,i,...,n\} = \{1,2,3,4,5,6,7,8,9,10,11,12,13,14,15,16 \}\),

 (ii) initial set of educational modules for a course on
 combinatorial optimization for students in
 ``communication systems''
 \(A^{1} = \{1,2,3,4,5,7,8,9,10,11,12,13,14 \} \subseteq A\), and

  (iii) initial set of educational modules for a course on
 combinatorial optimization for students in
 ``information systems''
 \(A^{2} = \{1,2,3,4,5,6,7,9,10,12,15 \} \subseteq A\).

 The aggregation problem is:

  Design a common course for students in
 ``communication systems'' and
 ``information systems'' \(A^{0} \subseteq A\)
  while taking into account
  weights of educational modules
  (\(\{w_{i}\}\), \(\{w^{1}_{i}\}\), \(\{w^{2}_{i}\}\) ).

%%%%%%%%%%%%%%%%%%%%%%%%%%%%%%%%%%%%%%%%%%%%%%%%%%%%%%%

 Here two solving strategies are considered:

 {\it 1.} Addition strategy (strategy I).

 {\it 2.} Median-based strategy.

 Let us examine the first case.
 The following ``system kernel'' is considered:
 \(K =  A^{1}  \bigcap  A^{2}  = \{1,2,3,4,5,7,9,10,12\}\).
 The set of elements for addition is:
 \[ B = \{ i\in A \backslash ( A^{1}  \bigcap  A^{2}  )  | ( w^{1}_{i} \geq 0.73 ) \cup ( w^{2}_{i} \geq 0.73 ) \} =
  \{6,8,11,14,15 \}. \]
 Here, three design versions
 (alternatives)
 are considered \(i \in B\):
 \(V^{i}_{1}\) (None),
 \(V^{i}_{2}\) (compressed version), and
 \(V^{i}_{3}\) (normal version)
 (\(j=\overline{1,3}\)).
 Table 8.10 contains the description of the addition elements and corresponding versions.

\begin{center}

\end{center}

  Thus, the addition problem (multiple choice problem) is:
 \[\max \sum_{\kappa=1}^{5} \sum_{j=1}^{3}  c_{\kappa j} x_{\kappa j}\]
  \[s.t. \sum_{\kappa=1}^{5} \sum_{j=1}^{3}   a_{\kappa j} x_{\kappa} \leq b,
 \sum_{j=1}^{3}  x_{\kappa j} \leq 1 \forall \kappa,
  x_{\kappa j} \in \{0,1\}.\]
 Here
 \(c_{\kappa 3} = \max ( w^{1}_{\kappa},w^{2}_{\kappa }  )\),
  estimate of \(a_{\kappa j}\) is based on expert judgment (Table 8.10),
  and \(b=3.50\).

 A resultant solution based on strategy I is depicted in Fig. 8.19
 (\(x_{11} = 0\), \(x_{12} = 0\), \(x_{13} = 1\),
 \(x_{21} = 0\), \(x_{22} = 0\), \(x_{23} = 1\),
  \(x_{31} = 1\), \(x_{32} = 0\), \(x_{33} = 0\),
   \(x_{41} = 0\), \(x_{42} = 1\), \(x_{43} = 0\),
  \(x_{51} = 0\), \(x_{52} = 0\), \(x_{53} = 1\)).
 Evidently,
 normal version \(V^{i}_{3}\) is used
 for elements of ``system kernel'' \(K\).
 The solving process was based on a greedy algorithm.
 Here compatibility estimates between design alternatives for
 system components were not considered.

\begin{center}

\end{center}

 In the second case,
 Pareto-efficient median-solutions
 \(\{A^{0}\}\)
 are searched for
 through
 the following two-component vector criterion
 \( \overline{\rho} = ( \rho^{1} (A^{0},A^{1}), \rho^{2} (A^{0},A^{2}) )  \)
 (by weights \( \{w^{1}_{i}\} \)
 and \( \{w^{2}_{i}\} \)):
 \[ \min \rho^{1} (A^{0},A^{1}) =
 1 - \frac{\sum_{i\in (A^{0} \bigcap A^{1})} w^{1}_{i}}
 {\sum_{i\in (A^{0}\bigcup A^{1}}) w^{1}_{i} },
% \]
%
%  \[
  \min \rho^{2} (A^{0},A^{2}) =
 1 - \frac{\sum_{i\in (A^{0} \bigcap A^{2})} w^{2}_{i}}
 {\sum_{i\in (A^{0}\bigcup A^{2}}) w^{2}_{i} }
  .\]
 Generally, the problem of searching for the median belongs to
 class of NP-hard problems.
 Let us consider an approximation heuristic.
 Initial elements for the median are the following
 (i.e., deletion of element \(16\)):
 \[\{ 1,2,3,4,5,6,7,8,9,10,11,12,13,14,15 \}.\]
 Thus, cardinality of search space is:
 \( | \{A^{0}\}  | = 2^{15}\).

 First, let us assume that elements \(\{1,2,5\}\)
 will be included into
 each median-solution
% (expert judgment)
 (to decrease problem dimension)
 and, as a result,
 \( | \{A^{0}\}  | = 2^{12}\).

 Second, elements from set
 \(\{ 3,4,9,10,11,12 \}\)
 will be included into each median solution as well
% (expert judgment)
 and,
 as a result,
 \( | \{A^{0}\}  | = 2^{6} = 64\).

 Finally, only six submedian-subsolutions are selected for examination:

%
% \(A^{01} = \{ 6,7,8,13,14,15 \}\),
%
 \(A^{01} = \{ 7,8,13,14,15 \}\),
 \(A^{02} = \{ 6,8,13,14,15 \}\),
 \(A^{03} = \{ 6,7,13,14,15 \}\),

 \(A^{04} = \{ 6,7,8,14,15 \}\),
 \(A^{05} = \{ 6,7,8,13,15 \}\),
 \(A^{06} = \{ 6,7,8,13,14\}\).

 Table 92 contains the subsolutions
 (\( \kappa = \overline{1,6}\)),
  corresponding values of vector
 \( \overline{\rho}(A^{0 \kappa}) =
 ( \rho^{1 \kappa} (A^{0 \kappa},A^{1}), \rho^{2 \kappa} (A^{0 \kappa},A^{2}) )
 \),
 and
 information on inclusion into the layer of Pareto-efficient
 solutions.
 Here parts of \(A^{1}\) and \(A^{2}\) are considered, which
 correspond to \( \{ 6,7,8,13,14,15 \}\):
 \( \{ 7,8,13,14 \}\) (for \(A^{1}\))
 and
 \( \{ 6,7,13,15 \}\) (for \(A^{2}\)).

 Thus, the following preference relation is obtained over
 the six subsolutions above:
 \(A^{01} \succeq A^{06} \) and
  \(A^{04} \succeq A^{05} \succeq A^{03} \succeq A^{02} \).

  Finally, subsolutions \(A^{01}\) and \(A^{04}\) are Pareto-efficient ones
  (Table 8.11, Fig. 8.20).

\begin{center}

\end{center}

 The resultant (extended)
 Pareto-efficient solutions
 \(\widehat{A^{01}}\)  and  \(\widehat{A^{04}}\)
 are presented in Fig. 8.21 and in Fig. 8.22
 (normal versions of course components are used).

\begin{center}
%
\end{center}

%%%%%%%%%%%%%%%%%%%%%%%%%%%%%%%%%%%%%%%%%%%%%%%%%%%%%%%%%%%%%%%
%%%%%%%%%%%%%%%%%%%%%%%%%%%%%%%%%%% EXAMPLE:  STUDENTS ART PLAN

\subsection{Plan of Students Art Activity}

 The considered numerical example is a small part
 of the example of students plan from  \cite{lev98}.
 The initial morphological structure is
 the following  (Fig. 8.23):

 {\bf 1.} Plan of students art activity \(S = I \star J \star U\):

 {\it 1.1} dance \(I\):
 \(I_{1}\) (none),
 \(I_{2}\) (ball dance),
 \(I_{3}\) (ensemble);

 {\it 1.2} music \(J\):
 \(J_{1}\) (none),
 \(J_{2}\) (classic),
 \(J_{3}\) (jazz),
 \(J_{4}\) (singing);

 {\it 1.3} theatre \(U\):
 \(U_{1}\) (none),
 \(U_{2}\) (actor),
 \(U_{3}\) (producer),
 \(U_{4}\) (technical worker),
 \(U_{5}\) (author).

  The following criteria were used for assessment of the design
 alternatives:
  cost \(C_{1}\),
  opportunity to meet new friends \(C_{2}\),
  opportunity to meet boy friend or girl friend \(C_{3}\),
  accordance to personal inclinations  \(C_{4}\),
  usefulness for future career  \(C_{5}\),
  usefulness for health \(C_{6}\), and
  usefulness for future life \(C_{7}\).
 The resultant priorities of design alternatives
 (via Electre-like technique)
 are shown in parentheses (Fig. 60).
 Thus, three resultant solutions are the following
 (via HMMD):

 \(S_{1}=I_{2}\star J_{3} \star U_{2} \),
 \(S_{2}=I_{3}\star J_{3} \star U_{2} \), and
 \(S_{3}=I_{3}\star J_{2} \star U_{5} \).

 In Fig. 8.24 and Fig. 8.25,
 supersolution and subsolution are depicted.
 Note, the subsolution contains only one element.
 Thus,
 the design of ``system kernel''
 is based on the special method with \(\alpha = 0.6\) (Fig. 8.26).
 Finally, the obtained ``system kernel'' \(K\)
 can be considered as the resultant solution.

\begin{center}
%
\end{center}

\subsection{Combinatorial Investment}

 The considered numerical example was presented in \cite{lev98}.
 The initial morphological structure is
 the following  (Fig. 8.27):

\begin{center}
%%
\end{center}

 {\bf 1.} composite portfolio \(S = A \star B \star L\):

 {\it 1.1} short-time investment \(A\):
 \(A_{1}\) (state bonds),
 \(A_{2}\) (bank deposit),
 \(A_{3}\) (speculation on the stock exchange),
 \(A_{4}\) (oil shares);

 {\it 1.2} middle-time investment \(B\):
 \(B_{1}\) (state bonds),
 \(B_{2}\) (bank deposit),
 \(B_{3}\) (immovables),
 \(B_{4}\) (jewelry),
 \(B_{5}\) (shares in biotechnology);

 {\it 1.3} long-time investment \(L\):
 \(L_{1}\) (state bonds),
 \(L_{2}\) (bank deposit),
 \(L_{3}\) (antique),
 \(L_{4}\) (shares of airspace companies).

  The following criteria were used for assessment of the design
 alternatives:
  possible profit \(\Upsilon_{1}\),
  risk \(\Upsilon_{2}\),
  prestige \(\Upsilon_{3}\),
  possibility for continuation \(\Upsilon_{4}\),
  possibility to establish a new company \(\Upsilon_{5}\),
  obtaining a new experience \(\Upsilon_{6}\),
  possibility to organize a new market \(\Upsilon_{7}\),
  possibility to obtain ``name''  \(\Upsilon_{8}\), and
  connection with previous activity \(\Upsilon_{9}\).

 The resultant priorities of design alternatives
 (via Electre-like technique)
 are shown in parentheses (Fig. 8.27).
 Thus, four resultant solutions are the following
 (via HMMD):

 \(S_{1}=A_{4}\star B_{3} \star L_{1} \),
 \(S_{2}=A_{2}\star B_{5} \star L_{1} \),
 \(S_{3}=A_{2}\star B_{3} \star L_{4} \), and
 \(S_{4}=A_{2}\star B_{5} \star L_{4} \).

 In Fig. 8.28 and Fig. 8.29,
 supersolution and
 subsolution are depicted.
 Note, the subsolution does not contain  elements
 (empty).
 Thus,
 the design of ``system kernel''
 is based on the special method with \(\alpha = 0.6\)
 and selection of the best elements for system parts:
 \(B_{5}\) for components \(B\) and
 \(L_{4}\) for components \(L\) (Fig. 8.30).
 Finally, the obtained ``system kernel'' \(K\)
 can be considered as the resultant solution.

\begin{center}
%
\end{center}

% ----------------------------------------------------------------
\subsection{Configuration of Applied Web-based Information System}
 Here aggregation of solution for applied Web-based information system
 is considered \cite{lev11inf}.
 The tree-like model of the system is depicted in Fig. 8.31.

\begin{center}
%
\end{center}

 DAs for system components are the following \cite{lev11inf}:

 (1) server for DBs \(J\):
    PC (\(J_{1}\)),
    Supermicro (\(J_{2}\)), and
    Sun (\(J_{3}\));

 (2) server for applications \(E\):
    on server of DBs (\(E_{1}\)),
    Sun (\(E_{2}\)),
    Supermicro (\(E_{3}\)), and
    PC (\(E_{4}\));

 (3) Web-server \(W\):
    Apache HTTP-server (\(W_{1}\)),
    Microsoft IIS (\(W_{2}\)),
    Bea Weblogic (\(W_{3}\)),
    Web Sphere (\(W_{4}\)), and
    Weblogic cluster (\(W_{5}\));

 (4) DBMS \(D\):
    Oracle (\(D_{1}\)),
    Microsoft SQL (\(D_{2}\)), and
    designed SQL (\(D_{3}\)); and

 (5) operation system \(O\):
    Windows 2000 server (\(O_{1}\)),
    Windows 2003 (\(O_{2}\)),
    Solaris (\(O_{3}\)),
    FreeBSD (\(O_{4}\)), and
    RHEL AS (\(O_{5}\)).

 In \cite{lev11inf},
 the design problem was solved for three basic applied situations:
 (a) communication provider (Fig. 8.32),
 (b) corporate application (Fig. 8.33), and
 (c) academic application (Fig. 8.34).

\begin{center}

\end{center}

  Now let us consider an aggregation problem:

 {\it To obtained an aggregated solution while taking into account all
 three application situations.}

 Here two simplified heuristic solving strategies are examined:

 (i) aggregation of composite solutions
 for three considered cases;
% \cite{lev11inf} (Fig. 70):
%
% (a) communication provider,
%
% (b) corporative application, and
%
% (c) academic application.

 (ii) aggregation of information at the first stage of the solving
 process (aggregation of rankings for DAs of system components
  \(J\), \(E\), \(W\), \(D\), \(O\))
  and usage of HMMD to solve the design problem.

 Let us describe the first strategy.
 Here the solving process is based on
 median/kernel-based strategy  (Fig. 8.35).
 The resultant solution is
 (an approximate median/kernel):

  \(S =
% A^{3}_{1}\star B^{3}_{1}=
 (J_{2}\star E_{2})\star (W_{1}\star D_{3}\star O_{2})\).

\begin{center}
%
\end{center}

 The second solving strategy
 is based on preliminary aggregation of rankings
 (Fig. 8.36, Fig. 8.37, Fig. 8.38, Fig. 8.39, Fig. 8.40)
 and a new design.

\begin{center}
%%
\end{center}

 After the usage of HMMD for the aggregated rankings of DAs
 (for system components \(J\), \(E\), \(W\), \(D\), \(O\)),
  the following four resultant composite DAs are obtained
  (Fig. 8.41, compatibility estimates are contained in \cite{lev11inf}):

 (a) \(S^{agg}_{1}=
 A_{1}\star B_{1}=
 (J_{2}\star E_{2})\star (W_{1}\star D_{2}\star O_{5})\),

 (b) \(S^{agg}_{2}=
 A_{1}\star B_{2}=
 (J_{2}\star E_{2})\star (W_{1}\star D_{3}\star O_{5})\),

 (c) \(S^{agg}_{3}=
 A_{2}\star B_{1}=
 (J_{3}\star E_{2})\star (W_{1}\star D_{2}\star O_{5})\), and

 (d) \(S^{agg}_{4}=
 A_{2}\star B_{2}=
 (J_{3}\star E_{2})\star (W_{1}\star D_{3}\star O_{5})\).

\begin{center}

\end{center}

%%%%%%%%%%%%%%%%%%%%%%%%%%%%%%%%%%%%%%%%%%%%%%%%%%%%%%%%%%%%%%%%%%%%
%%%%%%%%%%%%%%%%%%%%%%%%%%%%%%%%%%%%%%%%%%%%%%%%%%%%%%%%%%%%%%%%%%%%
% ----------------------------------------------------------------
\subsection{Modular Educational Course on Design}

 Here aggregation of three educational courses
 (morphological structures) is examined:
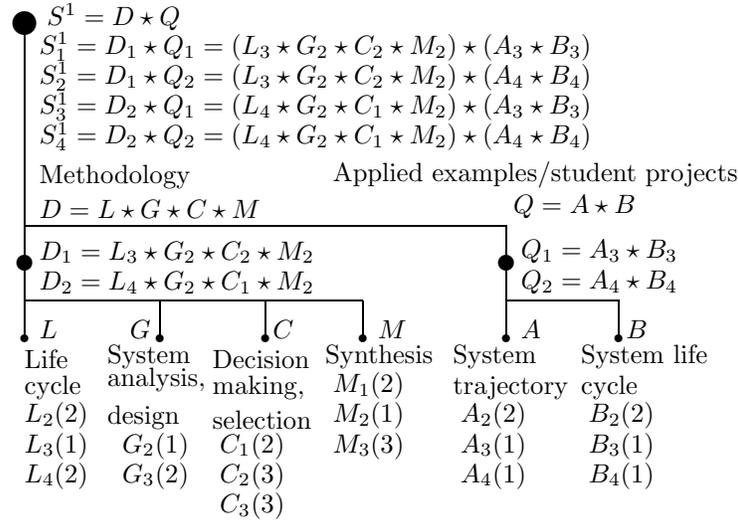
  (1) course on systems engineering (structure \(\Lambda^{1}\))
 (\cite{lev98}, \cite{lev00}) (Fig. 8.42);
 (2) course on information engineering (structure \(\Lambda^{2}\))
 (\cite{lev96}, \cite{lev98}) (Fig. 8.43); and
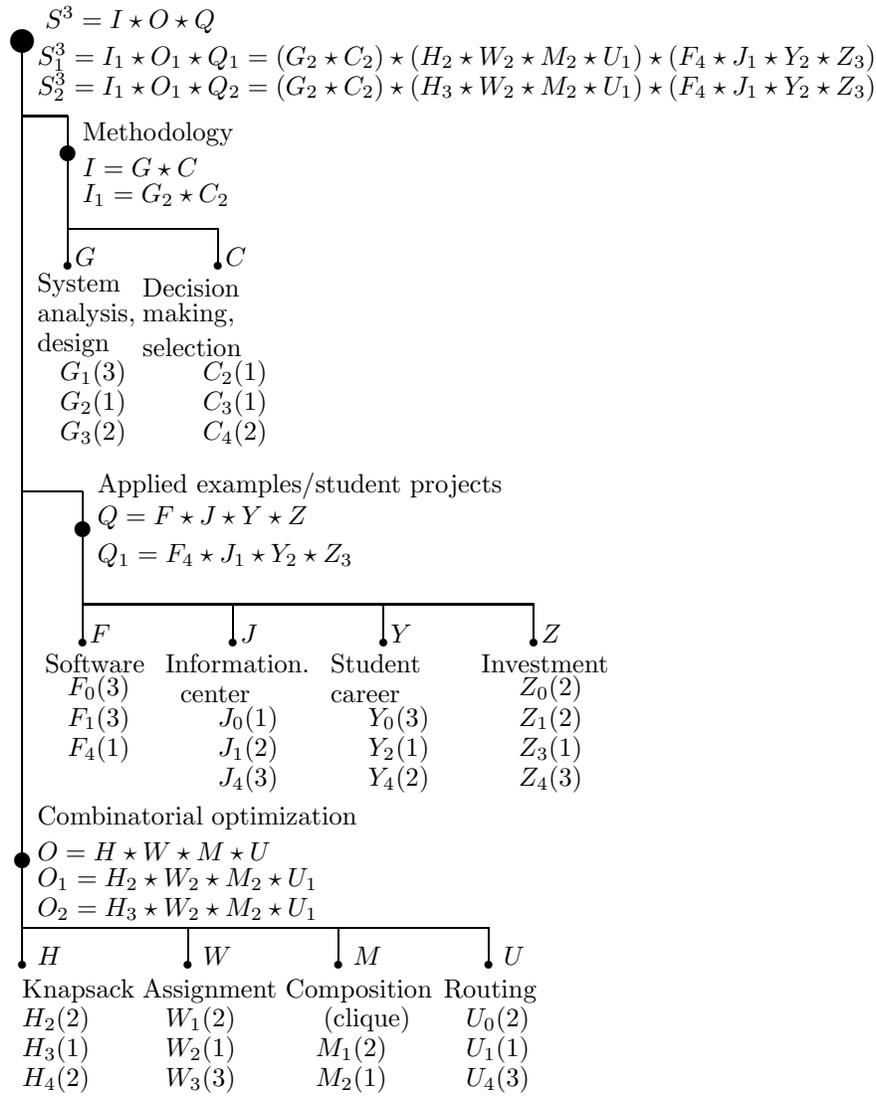
 (3) course on hierarchical design (structure \(\Lambda^{3}\))
 (\cite{lev98}, \cite{lev06}, \cite{lev11ed}) (Fig. 8.44).
 The following general types DAs (i.e., of the corresponding educational module)
 are examined for each leaf node of the
 presented hierarchical models (Fig. 8.42, Fig. 8.43, and Fig. 8.44):
  ``absence'' of the educational module \(X_{0}\),
 compressed information on the educational module \(X_{1}\),
 teaching at a medium level \(X_{2}\),
  serious teaching \(X_{3}\), and
 serious teaching with a special student research work/project
 \(X_{4}\).
 Compatibility estimates for
 three examined morphological structures above
% (\(\Lambda^{1}\),  \(\Lambda^{2}\), \(\Lambda^{3}\))
 are presented
 in Tables (expert judgment):
 (i) structure \(\Lambda^{1}\): Table 8.12, Table 8.13;
 (ii) structure \(\Lambda^{2}\): Table 8.14, Table 8.15, Table 8.16; and
 (iii) structure \(\Lambda^{3}\): Table 8.17, Table 8.18, Table 8.19.

 After usage of HMMD,
 the following composite DAs are obtained:

%~~

 {\bf I.} For structure \(\Lambda^{1}\) (Fig. 8.42):

 \(D_{1}=L_{3}\star G_{2}\star C_{2} \star M_{2} \),
 \(N(D_{1}) = (2;3,0,1)\),

 \(D_{2}=L_{4}\star G_{2} \star C_{1} \star M_{2}\),
 \(N(D_{2}) = (1;2,2,0)\);

  \(Q_{1}=A_{3}\star B_{2}\),
 \(N(Q_{1}) = (3;2,0,0)\),
 \(Q_{2}=A_{4}\star B_{4} \),
 \(N(Q_{2}) = (3;2,0,0)\);

%%%%%%%%%%%%%%%%%%%%%%%%%%%%%%%%%%%%%%%%%%%%%%

\(S^{1}_{1}= D_{1}\star Q_{1}
 = ( L_{3}\star G_{2}\star C_{2} \star M_{2}) \star ( A_{3}\star
 B_{3})\),

 \(S^{1}_{2}= D_{1}\star Q_{2}
 = ( L_{3}\star G_{2}\star C_{2} \star M_{2}) \star ( A_{4}\star
 B_{4})\),

 \(S^{1}_{3}= D_{2}\star Q_{1}
 = ( L_{4}\star G_{2}\star C_{1} \star M_{2}) \star ( A_{3}\star
 B_{3})\),

 \(S^{1}_{4}= D_{2}\star Q_{2}
 = ( L_{4}\star G_{2}\star C_{1} \star M_{2}) \star ( A_{4}\star
 B_{4})\).

%~~

%%%%%%%%%%%%%%%%%%%%%%%%%%%%%%%%%%%

 {\bf II.} For structure  \(\Lambda^{2}\) (Fig. 8.43):

 \(I_{1}=L_{3}\star G_{2}\star C_{2} \star E_{2} \),
 \(N(I_{1}) = (3;2,1,0)\);

 \(O_{1}=H_{3}\star W_{2} \star M_{2}\),
 \(N(O_{1}) = (2;3,0,0)\),
 \(O_{2}=H_{2}\star W_{2} \star M_{2}\),
 \(N(O_{2}) = (3;2,1,0)\);

 \(Q_{1}=V_{3} \star J_{3} \star Y_{2} \star R_{3} \),
 \(N(Q_{1}) = (3;4,0,0)\);

%%%%%%%%%%%%%%%%%%%%%%%%%%%%%%%%%%%%%%%%%%%%%%

 \(S^{2}_{1}= I_{1} \star   O_{1} \star   Q_{1}
 = ( G_{2}\star C_{2}\star E_{2})
  \star
  ( H_{3} \star W_{2} \star M_{2}  )
 \star
 ( V_{3} \star J_{3} \star Y_{2} \star R_{3}  )\),

  \(S^{2}_{2}= I_{1} \star   O_{2} \star   Q_{1}
 = ( G_{2}\star C_{2}\star E_{2})
  \star
  ( H_{2} \star W_{2} \star M_{2}  )
 \star
 ( V_{3} \star J_{3} \star Y_{2} \star R_{3}  )\).

%%%%%%%%%%%%%%%%%%%%%%%%%%%%%%%%%%%

%~~

 {\bf III.} For structure \(\Lambda^{3}\) (Fig. 8.44):

 \(I_{1}= G_{2}\star C_{2} \),
 \(N(I_{1}) = (3;2,0,0)\);

 \(O_{1}=H_{2}\star W_{2} \star M_{2} \star U_{1}  \),
 \(N(O_{1}) = (3;3,1,0)\),

 \(O_{2}=H_{3}\star W_{2} \star M_{2} \star U_{1}  \),
 \(N(O_{2}) = (2;4,0,0)\);

 \(Q_{1}=F_{4} \star J_{1} \star Y_{2} \star Z_{3} \),
 \(N(Q_{1}) = (3;3,1,0)\);

%%%%%%%%%%%%%%%%%%%%%%%%%%%%%%%%%%%%%%%%%%%%%%

 \(S^{3}_{1}= I_{1} \star   O_{1} \star   Q_{1}
 = ( G_{2}\star C_{2})
  \star
  ( H_{2} \star W_{2} \star M_{2} \star U_{1}  )
 \star
 ( F_{4} \star J_{1} \star Y_{2} \star Z_{3}  )\),

  \(S^{3}_{2}= I_{1} \star   O_{2} \star   Q_{1}
 = ( G_{2}\star C_{2})
  \star
  ( H_{3} \star W_{2} \star M_{2} \star U_{1}  )
 \star
 ( F_{4} \star J_{1} \star Y_{2} \star Z_{3}  )\).

%~~~

 Evidently, it is possible to aggregate
 the obtained composite solutions
 (Fig. 8.45).
%
%%%%%%%%%%%%%%%%%%%%%%%%%%%%%%%%%%%%%%%%%%%%%%%%%%%%%%%%%%
%
 On the other hand, let us consider   the following
% macro-heuristic
 extended aggregation strategy IV.
 General structure
% \(\Lambda\)
 ({\bf \(\Lambda\)})
 consists of the following parts:
 (i) tree-like system model {\bf T},
 (ii) set of leaf nodes {\bf P},
 (iii) sets of DAs for each leaf node {\bf D},
 (iv) DAs rankings (i.e., ordinal priorities) {\bf R}, and
 (v) compatibility estimates between DAs {\bf I}.
 Thus, a vector proximity for two structures
 \(\Lambda^{\alpha},\Lambda^{\beta}\)
 can be examined as follows:

 \(\overline{\rho}\) (\(\Lambda^{\alpha},\Lambda^{\beta})=\)
 (\(\rho^{t}\)({\bf T}\(^{\alpha}\),{\bf T}\(^{\beta}\)),
 \(\rho^{t}\)({\bf P}\(^{\alpha}\),{\bf P}\(^{\beta}\)),
 \(\rho^{t}\)({\bf D}\(^{\alpha}\),{\bf D}\(^{\beta}\)),
 \(\rho^{t}\)({\bf R}\(^{\alpha}\),{\bf R}\(^{\beta}\)),
 \(\rho^{t}\)({\bf I}\(^{\alpha}\),{\bf I}\(^{\beta}\))).

 As a result,  \(\Lambda^{agg}\) has to be searched for as
  Pareto-efficient solution(s) by the vectors
  ~~\(\overline{\rho}\)(\(\Lambda^{agg},\Lambda^{i})~~ \forall
  i\in \{i\}\)~~
  where index \(i\) corresponds to an initial solution.
 This problem is very complicated.
% Further,
%
 Let us consider a
 simplified solving framework:

%~~

 {\bf Phase 1.} Aggregation of basic initial data for
  initial structures:

 {\it 1.1.} aggregation of morphological structures including the
 following:

 (1.1.1.) aggregation of sets of leaf nodes,

  (1.1.2.) aggregation of sets of DAs for each leaf node,

  (1.1.3.) aggregation of DAs rankings, and

  (1.1.4.) aggregation of compatibility estimates for DAs sets;

   {\it 1.2.} aggregation tree-like structures.

 {\bf Phase 2.} New hierarchical design.

%~~

 Thus, the following stages are considered
 for examined three structures
 (\(\Lambda^{1}\),  \(\Lambda^{2}\), \(\Lambda^{3}\)):

%~~~

 {\it Stage 1.} Aggregation of leaf node sets
 for the initial structures
% \(\Lambda^{1}\),  \(\Lambda^{2}\), and \(\Lambda^{3}\)
 (Fig. 8.46);

 {\it Stage 2.} Aggregation of morphological structure:

 {\it 2.1.} aggregation of
 sets of DAs for each leaf node
 (by each leaf node
 (while taking into account addition of DAs
 which correspond to ``absence'',
 i.e., aggregation of rankings with extension
 of DAs sets)
 (Fig. 8.47, Fig. 8.48, Fig. 8.49, Fig. 8.50, Fig. 8.51,
 Fig. 8.52, Fig. 8.53, Fig. 8.54, Fig. 8.55, Fig. 8.56,
 Fig. 8.57, Fig. 8.58, Fig. 8.59, Fig. 8.60, Fig. 8.61,
 Fig. 8.62);

 {\it 2.2.} aggregation of interconnection (compatibility
 estimates) for DAs sets
 (Tables 8.20, 8.21, 8.22, 8.23;
 selection of a maximal value or expert judgment);

 {\it Stage 3.} Building of an aggregation tree-like structure
 (Fig. 8.63);
 and

 {\it Stage 4.} Hierarchical design of an integrated course
 (as an aggregated solution) (Fig. 8.63).

%~~

 Here the aggregated superstructure
  \(\overline{\Lambda}\) (Fig. 8.63)
  has been obtained via expert judgment.

\begin{center}

\end{center}
%
%

%%%%%%%%%%%%%%%%>>>>>>>>>>>>>>>>>>>>>>>>>>>>>>>>>>>>>>>>>>>>>>>>>>>>>>>>>>>>>>>>>

 After usage of HMMD,
 the following composite DAs are obtained for the resultant
 aggregated structure (\(\overline{\Lambda}\)) (Fig. 8.63):

 \(X_{1}=V_{3}\star F_{4}\star J_{3} \star R_{3}
 \star Y_{2}\star Z_{3}\star A_{3} \star B_{3}\),
 \(N(X_{1}) = (3;8,0,0)\),

 \(O_{1}=H_{3}\star W_{2} \star M_{2} \star U_{1}\),
 \(N(O_{1}) = (2;4,0,0)\);

 \(I_{1}=L_{3}\star G_{2} \star C_{2} \star E_{2}\),
 \(N(O_{1}) = (2;4,0,0)\),

 \(I_{2}=L_{3}\star G_{2} \star C_{3} \star E_{2}\),
 \(N(O_{1}) = (2;4,0,0)\),

 \(I_{3}=L_{4}\star G_{3} \star C_{3} \star E_{2}\),
 \(N(O_{1}) = (3;2,2,0)\);

%%%%%%%%%%%%%%%%%%%%%%%%%%%%%%%%%%%%%%%%%%%%%%

\(\overline{S}_{1}= I_{1} \star O_{1} \star X_{1}
 = ( L_{3}\star G_{2} \star C_{2} \star E_{2} )
  \star
  ( H_{3}\star W_{2} \star M_{2} \star U_{1} )
 \star
  ( V_{3}\star F_{4}\star J_{3} \star R_{3}
 \star Y_{2}\star Z_{3}\star A_{3} \star B_{3} )\),

 \(\overline{S}_{2}= I_{2} \star O_{1} \star X_{1}
 = ( L_{3}\star G_{2} \star C_{3} \star E_{2} )
  \star
  ( H_{3}\star W_{2} \star M_{2} \star U_{1} )
 \star
  ( V_{3}\star F_{4}\star J_{3} \star R_{3}
 \star Y_{2}\star Z_{3}\star A_{3} \star B_{3} )\),

 \(\overline{S}_{2}= I_{3} \star O_{1} \star X_{1}
 = ( L_{3}\star G_{3} \star C_{3} \star E_{2} )
  \star
  ( H_{3}\star W_{2} \star M_{2} \star U_{1} )
 \star
  ( V_{3}\star F_{4}\star J_{3} \star R_{3}
 \star Y_{2}\star Z_{3}\star A_{3} \star B_{3} )\).

%%%%%%%%%%%%%%%%%%%%%%%%%%%%%%%%%%%%%%%%%%%%%%%%%%%%%%%%%%%%%%%%%%
%%%%%%%%%%%%%%%%%%%%%%%%%%%%%%%%%%%%%%%%%%%%%%%%%%%%%%%%%%%%%%%%%%
% ----------------------------------------------------------------
\subsection{Configuration of Car in Electronic Shopping}
%
%%%%%%%%%%%%%%%%%%%%%%%%%%%%%%%%%%%%%%%%%%%%%%%%%%%
%\subsection{Synthesis of Composite Product}

 Recently, many products have a complex configuration and
 a buyer
 can often generate a product configuration that
 is more useful for him/her.
 An approach to electronic shopping based on
 product configuration
 (morphological approach)
 has been suggested in
 \cite{lev08a}.
 Here let us consider a multi-choice scheme
% framework
 for selection of
 structured product/system in electronic shopping
 with a resultant aggregation
 (Fig. 8.64).

\begin{center}

\end{center}

 Now the following example is described (at a conceptual level).
 An initial morphological structure
 \(\Lambda\)
 of a car is the following  (Fig. 8.65)
 (in real application, this structure can be considered
 as a result of processing the selected products/solutions):

 {\bf 0.} Car
 ~\(S = A \star B \star C \).

 {\bf 1.} Main part
 \(A = E \star D \):

 {\it 1.1.} Engine  E:~
 diesel \(E_{1}\),
 gasoline \(E_{2}\),
 electric \(E_{3}\),
 hydrogenous \(E_{4}\), and
 hybrid synergy drive HSD \(E_{5}\);

 {\it 1.2.} Body \(D\):~
 sedan \(D_{1}\),
 universal \(D_{2}\),
 jeep \(D_{3}\),
 pickup \(D_{4}\), and
 sport \(D_{5}\).

 {\bf 2.} Mechanical part
   \(B = X \star Y \star P \star Z \):

  {\it 2.1.} gear box
%   korobka peredach
    X:~
  automate  \(X_{1}\),
  manual \(X_{2}\);

 {\it 2.2.} suspension
% podveska
 Y:~
 pneumatic  \(Y_{1}\),
 hydraulic  \(Y_{2}\), and
 pneumohydraulic \(Y_{3}\);

 {\it 2.3.} drive
%  Privod
 P:~
 front-wheel drive  \(P_{1}\),
 rear-drive \(P_{2}\),
 all-wheel-drive \(P_{3}\).

 {\bf 3.} Safety part
 \(C = O \star G  \):

  {\it 3.1.}
  \(O\):~
  ``absence'' \(O_{0}\),
 electronic \(O_{1}\);

 {\it 3.2.} Safety subsystem
 \(G\):
  ``absence'' \(G_{0}\),
  passive \(G_{1}\),
  active \(G_{2}\).

%%%%%%%%%%%%%%%%%%%%%%%%%%%%%%%%%%%%%%%%%%%%%%%%%%%%%%%%%%

\begin{center}

\end{center}

 The following initial solutions (obtained by users) are considered:

 \(S^{1}_{1}=E_{1}\star D_{1}\star X_{1}\star Y_{1}\star P_{1}
 \star O_{1}\star G_{1} \),
 \(S^{1}_{2}=E_{5}\star D_{1}\star X_{1}\star Y_{1}\star P_{1}
 \star O_{1}\star G_{2} \),

  \(S^{2}_{1}=E_{2}\star D_{1}\star X_{2}\star Y_{1}\star P_{1}
 \star O_{0}\star G_{1} \),
  \(S^{3}_{1}=E_{2}\star D_{3}\star X_{1}\star Y_{2}\star P_{3}
 \star O_{1}\star G_{0} \), and

  \(S^{3}_{2}=E_{2}\star D_{5}\star X_{1}\star Y_{3}\star P_{1}
 \star O_{1}\star G_{1} \).

 An aggregated solution can be considered as simplified aggregation of
 five sets above:

  \(S^{agg}=E_{2}\star D_{1}\star X_{1}\star Y_{1}\star P_{1}
 \star O_{1}\star G_{1} \).

% ----------------------------------------------------------------
%\section{Applied Examples}
%
% Here three applied numerical examples are presented:
%
% (1) design of several scenario-based solutions and their
% aggregation;
%
% (2) aggregation of several solutions which are based
% on fuzzy-set estimation;
% and
%
% (3) aggregation of results which were obtained by multi-source
% information search.
%
% ----------------------------------------------------------------
%\subsection{Scenarios Based Solution Aggregation}
%
% ----------------------------------------------------------------
%\subsection{Fuzzy Set Based Solution Aggregation}
%
% ----------------------------------------------------------------
%\subsection{Aggregation in Multi-Source Information Search}

% Prospective applied examples
%
% (i) communication protocol *** ,
%
% (ii) telemetry system *** ,
%
% (iii) security system * ,
%
% (iv) combinatorial investment * ,
%
% (v) plan of students art activity * ,
%
% (vi) plan of geology exploration,
%
% (vii) wireless sensor,
%
% (viii) multiple source information retrieval,
%
% (ix) structure of web-based information system,
%
% (x) marketing strategy, and
%
% (xi) educational course on design.
%%%%%%%%%%%%%%%%%%%%%%%%%%%%%%%%%%%%%%%%%%%%%%%%%%%%%%%%%%%%%%%%%%%%%%%%

% ----------------------------------------------------------------
% \section{Discussion and Conclusion}

  \section{Conclusion}

 In the article,
 a systemic view point to aggregation
 of modular solutions is firstly presented.
 The aggregation strategies
%  for composite (modular) solutions
 have been suggested and considered:
 (i) extension  strategy (addition of elements to a ``system kernel''),
 (ii) compression strategy (deletion of elements of a superstructure),
 (iii) combined extension/compression strategy
 (addition, deletion, and replacement of elements
 in ``system kernel''),
 and
 (iv) strategy of a new design (while taking into account new
 elements).
 The examined strategies have not been previously discussed in the literature.
 Our
% The suggested
  material corresponds to  the first step in
 the investigated
% considered
% examined
  domain (aggregation of modular solutions).
 The considered aggregation approaches are useful for
 engineering domains, management, computer science and information
 technology.

 It is necessary to point out,
 close aggregation problems
%  for some structures
 have been intensive  studied and used
 in several domains:
 (a) decision making (aggregation of rankings to obtain a consensus,
 aggregation of preferences)
 (e.g.,
% \cite{},
  \cite{cook92}, \cite{cook96},
%  \cite{},
 \cite{yager88});
 (b) integration of organizational structures in organization science
 (e.g.,
 \cite{baligh06}, \cite{crow94}, \cite{dan01}, \cite{dessler80},
 \cite{gho02});
 and
 (c) integration of information
 (database schema integration,
 integration of knowledge base structures,
 integration of catalogs,
 merging and integration of ontologies)
 (e.g.,
  \cite{agrawal01},
 \cite{batini86}, \cite{choi06}, \cite{lin96},
  \cite{noy05}, \cite{pinto01}, \cite{wache01}).
%
%%%%%%%%%%%%%%%%%%%%%%%%%%%%%%%%%%%%%%%%%%%%%%%%%%%%%%%%%%%%
% (c) integration of structured information, for example:
%
% (i) database schema integration
%(e.g., \cite{batini86}, \cite{}, \cite{});
%
% (ii) integration of knowledge base structures
% (e.g., \cite{}, \cite{lin96},  \cite{});
%
% (ii) document summarization
%(\cite{}, \cite{}, \cite{}); and
% (iii) merging and integration of ontologies
% (e.g., \cite{choi06},
% \cite{noy03},
% \cite{noy05}, \cite{pinto01}, \cite{wache01}).

%
 In the article, many illustrative applied examples are presented.
 It is necessary to point out fundamentals of
 the described approaches correspond to  practice
 (e.g., engineering, management).
 Thus, the suggested composite solving schemes and their local
 parts
 have to be based on a combination of mathematical combinatorial models
 and expert judgment of domain experts.
 Our approach for modeling the complex modular solutions/systems
 is based on the following four layers:
 (i) system hierarchy as tree-like structure,
 (ii) system components (leaf nodes of the tree-like mode above),
 (iii) sets of design alternatives for each system components,
 (iv) compatibility between design alternatives of different
 system components.
 Thus,
 models, problems, and algorithmic schemes are examined
 for the above-mentioned structures:
 proximities/metrics,
 substructures (including median/consensus, etc.),
 aggregation strategies/frameworks.
%
%%%%%%%%%%%%%%%%%%%%%%%%%%%%%%%%%%%%%%%%%%%%%%%%%%%%%%
%
 The considered solving strategies are based on combinatorial problems
 (multicriteria ranking/selection,
 knapsack problem, multiple choice problem,
 morphological combinatorial synthesis,
 building median/consensus for structures).

 The suggested aggregation problems may be useful in the following
 situations:
 (a) design processes based on
 uncertainty
 (e.g., fuzzy sets, stochastic models)
 can lead to generating a set (as a grid) of design solutions
 (here the problem under uncertainty can be approximated
 by a set of deterministic problems and, further, their solutions
 can be aggregated);
 (b)
%  Analogically,
 scenario-based methods can lead to a set of design solutions which can be aggregated.
%
% An aggregation of the obtained solutions can be based on
% approaches which were suggested in the article.
%
%
%
%%%%%%%%%%%%%%%%%%%%%%%%%%%%%%%%%%%%%%%%%%%%%%%%%%%%%%%
% It is necessary to point out
 Note,
 complicated methods and models
  are required
 for dynamical modeling of complex
 systems,
  for example:
 (a) various kinds of  Petri nets,
 (b) dynamical graphs.
 Our material does not involve this type of studies.

 In the future, the following research directions can be considered:

 1. usage of the suggested aggregation approaches
  for various kinds of system configuration solutions
  (e.g., set of selected elements, assignment/allocation solutions,
  network-like solutions \cite{lev09});

 2. usage of other types of metrics/proximities for structures;

 3. additional theoretical and applied studies of
 formal approaches to aggregation problems
 based on aggregation of system tree-like models
 and/or general system structures
 (i.e., multicriteria searching for
 median-like tree(s) and/or median-like structure(s)
 as Pareto-efficient structured solution(s));

 4. examination of more complicated design problems
 for several resultant aggregated solutions
 (e.g., as ``product line'');

 5. examination of various models
 (for example, building of median/consensus),
 including multicriteria problem statements and usage
 of special quality domains/ spaces,
 e.g., lattice-like quality domain/space
 (\cite{lev98}, \cite{lev01}, \cite{lev06});

 6. examination and usage of other structures for
   system modeling, e.g., pyramids
    (\cite{brun03},
%    \cite{chris75},
    \cite{hax03});

 7. taking into account uncertainty;

 8. usage of artificial intelligence methods
 in problem solving processes;

 9. investigation of new applications;

 10. usage of the suggested aggregation problems as
 auxiliary
 underlaying subproblems in
 information integration/fusion (for example,
 for approximation of alignment-like problems in bioinformatics,
 image processing; multi-source fusion of structured information);

 11. special examination of aggregation problems for aggregation of
 solutions in combinatorial optimization problems
 (e.g., routing, scheduling,
 traveling salesman problem,
  assignment/allocation, graph coloring,
 covering);

 12. usage of the suggested aggregation approaches in electronic
 shopping
 (and recommender systems)
 as the following two-stage framework:
 (i) an user is selecting
 a preliminary set of modular products (by his/her choice)
 from product catalogs,
 (ii) a computer system
 (i.e., special design support service)
  is constructing an aggregated product
 (selection of the best product from the preliminary product set is a simplified case);

 13. it may be very interesting to apply the suggested aggregation
 approaches to drug design (generally, to combinatorial
 chemistry);  and

 14. usage of the suggested aggregation approaches in education
  (engineering, management, computer science and
 information technology).

% ----------------------------------------------------------------
%\bibliographystyle{amsplain}


\thebibliography{300}

 \bibitem {agrawal01} R. Agarwal, R. Srikant,
 On integrating catalogs.
 In: {\it Proc. 10th Int. World Wide Web Conf. WWW 2001},
 Hong Kong, 603-612, May 2001.

 \bibitem {akatsu00} T. Akatsu, M.M. Halldorsson,
 On the approximation of largest common subtrees
 and largest common point sets.
 {\it Theoretical Computer Science}, 233(1-2), 33-50, 2000.

 \bibitem {amir97} A. Amir, D. Keselman,
 Maximum agreement subtree in a set of evolutionary
 trees: metrics and efficient algorithms.
 {\it SIAM J. Comput.}, 26(6), 1656-1669, 1997.

  \bibitem {amir09} A. Amir, G.M. Landau, J.C. Na,
   H. Park, K. Park, J.S. Sim,
  Consensus optimizing both distance sum and radius.
  In:
  J. Karlgren, J. Tarhio, H. Hyyro (Eds.),
 {\it Int. Conf. on String Processing and Information Retrieval SPIRE 2009},
  LNCS 5721, Springer,
  pp. 234-242, 2009.

 \bibitem {amir11} A. Amir, G.M. Landau, J.C. Na,
   H. Park, K. Park, J.S. Sim,
  Efficient algorithms for consensus string problems
  minimizing both distance sum and radius.
  {\it Theoretical Computer Science}, 412(39), 5239-5246, 2011.

 \bibitem {apost97} A. Apostolico,
  String editing and longest common subsequences.
 In:
 G. Rozenberg, A. Salomaa (Eds.)
 {\it Handbook of Formal Languages}, vol. 2,
 Springer,
 361-398, 1997.

 \bibitem {apost87} A. Apostolico, C. Guerra,
 The longest common subsequence problem revised.
 {\it Algorithmica}, 2(1-4), 315-336, 1987.

 \bibitem {arnold11} M. Arnold, E. Ohlebusch,
 Linear time algorithms for generalizations
 of the longest common substring problem.
 {\it Algorithmica}, 60(4), 806-818, 2011.

 \bibitem {ayr69}  R.U. Ayres,
  {\it Technological Forecasting and Long-Time Planning},
  New York: McGraw-Hill, 1969.

 \bibitem {baligh06} H.H. Baligh,
 {\it Organization Structures: Theory and Design, Analysis and
 Prescription}.
 Springer, 2006.

 \bibitem {bansal09} M.S. Bansal, D. Fernandez-Baca,
 Computing distances between partial rankings.
 {\it Information Processing Letters},
 109(4), 238-241, 2009.

 \bibitem {barth89} J.-P. Barthelemy, A. Guenoche, O. Hudey,
 Median linear orders:
 Heuristics and a branch and bound algorithm.
 {\it EJOR}, 42(3), 313-325, 1989.

 \bibitem {barton11} B.A. Barton,
 Searching a bitstream in linear time for the longest substring
 of any given density.
  {\it Algorithmica}, 61(3), 555-579, 2011.

 \bibitem {batini86} C. Batini, M. Lenzerini, S.B. Navathe,
 A comparative study of methodologies for data base schema
 integration.
 {\it ACM Computing Survey}, 18(4), 323-364, 1986.

 \bibitem {bazgan09} C. Bazgan, H. Hugot,
 Solving efficiently the 0-1 multiobjective knapsack problem.
 {\it Computers and Operations Research},
 36(1), 260-279, 2009.

 \bibitem {beck83} M.P. Beck, B.W. Lin,
 Some heuristics for the consensus ranking problem.
 {\it Computers and Operations Research},
 10(1), 183, 1-7, 1983.

 \bibitem {bellev90} A.R. Belkin, M.Sh. Levin,
 {\it Decision Making: Combinatorial Models of Information
 Approximation}.
 Nauka Publ. House, Moscow, 1990 (in Russian).

 \bibitem {bergroth00} L. Bergroth, H. Hakonen, T. Raita,
 A survey of longest common subsequence algorithms.
 In:
 {\it Proc. of the Seventh Int. Symp. on String Processing and
 Information Retrieval SPIRE'00},
 39-48, 2000.

  \bibitem {berry04} V. Berry, F. Nicolas,
 Maximum agreement and compatible supertrees.
 In: Proc. of the 15th Annual Symp. on
 Combinatorial Pattern Matching CPM-2004,
 LNCS 3109, Springer, Berlin, pp. 205-219, 2004.

 \bibitem {bille05} P. Bille,
 A survey on edit distance and related problems.
 {\it Theoretical Computer Science},
 337(1-3), 217-239, 2005.

  \bibitem {blum09} C. Blum, M.J. Blesa, M.L. Ibanez,
  Beam search for the longest common subsequence problem.
 {\it Comp. and Oper. Res.},
 36(12), 3178-3186, 2009.

 \bibitem {bogart73} K.P. Bogart,
 Preference structures I:
 distance between transitive preference relations.
 {\it J. of Math. Soc.}, 3, 49-67, 1973.

 \bibitem {brin99} S. Brin, L. Page, R. Motwani, T. Winograd,
  {\it The PageRank Citation Ranking: Bringing Order to the Web.}
  Technical report 1999-0120, Computer Science Dept.,
  Stanford Univ., Stanford, CA, 1999.

 % \bibitem {bister90} M. Bister, J. Cornelis, A. Rosenfeld,
% A critical view of pyramid segmentation algorithms.
% {\it Pattern Recognition Letters}, 11(9), 605-617, 1990.

% \bibitem {brun00} L. Brun, W.G. Kropatsch,
% Irregular pyramids with combinatorial maps.
% {\it Proc.  Int. Conf. SSPR/SPR}, LNCS  1451, 256-265, 2000.

 \bibitem {brun03} L. Brun, W.G. Kropatsch,
 Construction of combinatorial pyramids.
 In:
 W. Hancock, M. Vento (Eds.),
 {\it Proc. IAPR Workshop GbPRP 2003},
 LNCS 2726, Springer, 1-12, 2003.

 \bibitem {bunke97} H. Bunke,
 On a relation between graph edit distance and maximum common
 subgraph.
 {\it Pattern Recognition Letters}, 18(8), 689-694, 1997.

 \bibitem {bunke98} H. Bunke, K. Shearer,
 A graph distance metric based on the maximal common subgraph.
 {\it Pattern Recognition Letters},
 19(3-4), 255-259, 1998.

 \bibitem {burkard09} R. Burkard, M. Dell'Amico, S. Martello,
 {\it Assignment Problems}.
 SIAM, Providence, 2009.

 \bibitem {cela98} E. Cela,
  {\it The Quadratic Assignment Problem: Theory and Algorithms}.
  Kluwer Academic Publishers, Dordrecht, 1998.

 \bibitem {char11} T.P. Chartier, E. Krentzer, A.N. Langville,
  K.E. Pedings,
  Sensitivity and stability of ranking vectors.
  {\it SIAM J. on Scientific Computing}, 33(3), 1077-1102, 2011.

  \bibitem {cha05} F. Chataigner,
   Approximating the maximum agreement forest on k trees.
 {\it  Information Processing Letters},
 93(5), 239-244, 2005.

 \bibitem {chen01} W. Chen,
 New algorithm for ordered tree-to-tree correction problem.
 {\it J. of Algorithms}, 40(2), 135-158, 2001.

  \bibitem {choi06} N. Choi, I.-Y. Song, H. Han,
 A survey on ontology mapping.
 {\it SIGMOD Record}, 35(3), 34-41, 2006.

%  \bibitem {chris75} N. Christofides,
% {\it Graph Theory - An Algorithmic Approach}.
% Academic Press, New York, 1975.

  \bibitem {connor11} R. Connor, F. Simeoni, M. Iakovos, R. Moss,
  A bounded distance metric for comparing tree structure.
 {\it Information Systems},
 36(4), 748-764, 2011.
% metric based on Kolmogorov complexity / entropy

  \bibitem {cookseif78}  W.D. Cook, L.M. Seiford,
  Priority ranking and consensus formation.
  {\it Management Science},
   24(16) 1721-1732, 1978.

%   \bibitem {cookseif84}  W.D. Cook, L.M. Seiford,
%  An ordinal ranking model for the highway corridor selection problem.
%  {\it Computers, Environment and Urban Systems},
%   9(4) 271-276, 1984.

 \bibitem {cook92} W.P. Cook, M. Kress,
 {\it Ordinal Information and Preference Structures:
 Decision Models and Applications.}
  Prentice-Hall, Englewood Cliffs, 1992.

  \bibitem {cook96} W.D. Cook, L.M. Seiford, M.Kress,
  A General Framework for Distance-Based Consensus in
  Ordinal Ranking Models.
  {\it European Journal of Operational Research},
   96(2) 392-397, 1996.

 \bibitem{cross00} N. Cross,
 {\it Engineering Design Methods}. 3rd ed., Wiley, 2000.

 \bibitem {crow94} K. Crowston,
 {\it A Taxonomy of Organizational Dependencies and Coordination
 Mechanisms}. Working Paper, No. 174,
 Center of Coordination Science, MIT, Cambridge, Aug. 1994.

  \bibitem {dan01} A. Dan, D.M. Dias, R. Kearney, T.C. Lau,
 T.N. Nguyen, F.N. Parr, M.W. Sachs, H.H. Shaikh,
 Business-to-business integration with
 tpaML and a business-to-business protocol framework.
 {\it IBM Systems Journal},
 40(1), 68-90, 2001.

 \bibitem {hig00} C. de la Higuera, F. Casacuberta,
 Topology of strings: Median string is NP-complete.
 {\it Theoretical Computer Science}, 230(1-2), 39-48, 2000.

  \bibitem {dessler80} G. Dessler,
 {\it Organizational Theory. Integrating Structure and
 Behaviour}. NJ.: Prentice Hall, 1980.

 \bibitem {dinu06} L.P. Dinu, F. Manea,
% Liviu P. Dinu, Florin Manea
 An efficient approach for the rank aggregation problem.
 {\it Theoretical Computer Science}, 359(1-3), 455-461, 2006.
%   NB!

 \bibitem {dovier98} A. Dovier, E.G. Policriti, G. Rossi,
 A uniform axiomatic view of lists, multisets,
 and sets, and the relevant unification algorithms.
 {\it Fundamenta Informaticae}, 36(2/3), 201-234, 1998.

  \bibitem {dul03} S. Dulucq, H. Touzet,
 Analysis of tree edit distance algorithms.
 In: R. Baeza-Yates, E. Chavez, M. Crochemore (Eds.),
 {\it Proc. 14th Annual Symp. on Combinatorial Pattern Matching CPM 2003},
 LNCS 2676, Springer,
 83-95, 2003.
% Mexico, June 25-27, 2003
% NB !!! Strategies for tree editing

  \bibitem {duong09} T.H. Duong, N.T. Nguyen, G.S. Jo,
 A hybrid method for integrating multiple ontologies.
 {\it Cybernetics and  Systems}, 40(2), 123-145, 2009.

 \bibitem {east08} T. Easton,  A. Singireddy,
 A large neighborhood search heuristic for the longest
 common subsequence problem.
 {\it J. of Heuristics}, 14(3), 271-283,
 2008.

 \bibitem {eghe03} L. Eghe, C. Michel,
  String similarity measures for ordered sets
  of documents in information retrieval.
  {\it Information Processing and Management},
  38(6), 823-848, 2003.

  \bibitem {ehrgott05} M. Ehrgott,
 {\it Multicriteria optimization}. Springer, Berlin, 2005.

 \bibitem {ele03} O. Elemento, O. Gascuet,
 An exact and polynomial distance-based
 algorithm to reconstruct
 single copy tandem duplication trees.
 In: R. Baeza-Yates, E. Chavez, M. Crochemore (Eds.),
 {\it Proc. 14th Annual Symp. on Combinatorial Pattern Matching CPM 2003},
 LNCS 2676, Springer,
 96-108, 2003.
% Mexico, June 25-27, 2003

 \bibitem {elzinga11} C. Elzinga, H. Wang, Z. Lin, Y. Kumar,
 Concordance and consensus.
 {\it Information Sciences}, 181(12), 2529-2549, 2011.

 \bibitem {fagin06} R. Fagin, R. Kumar, M. Mahdian,, D. Sivakumar,
 E. Vee,
 Comparing partial rankings.
 {\it SIAM J. Discrete Math.},
 20(3), 628-648, 2006.

 \bibitem {farach95} M. Farach, T. Przytycka, M. Thorup,
 On the agreement of many trees.
 {\it Inf. Process. Lett.}, 55(6), 297-301, 1995.

  \bibitem {ferrer10} M. Ferrer, E. Valveny, F. Serratosa,
 K. Riesen, H. Bunke,
 Generalized median graph computation for means of graph embedding
 in vector space.
 {\it Pattern Recognition}, 43(4), 1642-1655, 2010.

 \bibitem {finden85} C.R. Finden, A.D. Gordon,
 Obtaining common pruned trees.
 {\it J. Classif.}, 2(1) 255-276, 1985.

 \bibitem {gallant80} J. Gallant, D. Maier, J.A. Storer,
 On finding minimal length superstrings.
 {\it J. Comput. Syst. Sci.},
 20(1), 50-58, 1980.

  \bibitem {gand00} X. Gandibleux, A. Freville,
 Tabu search based procedure for solving the 0-1
 multiobjective knapsack problem:
 the two case.
 {\it J. of Heuristics}, 6(3), 361-383, 2000.

 \bibitem {gar79}  M.R. Garey, D.S. Johnson,
   {\it Computers and Intractability.
   The Guide to the Theory of NP-Completeness},
   San Francisco:
   W.H. Freeman and Company,  1979.

  \bibitem {gho02} S. Ghoshal, L. Gratton,
 Integrating the enterprise.
 {\it MIT Sloan Management Review}, 44(1), 31-38, 2002.

 \bibitem {gomes08} F.C. Gomes, C.N. de Mendes,
 P.M. Pardalos, C.V.R. Viana,
 A parallel multistart algorithm for the closest string probl;em.
 {\it Comp. and Oper. Res.},
 35(11), 3636-3643, 2008.

  \bibitem {gordon90} A.D. Gordon,
 Constructing  dissimilarity measures.
 {\it J. Classif.}, 7(2), 257-269, 1990.

 \bibitem {gower85} J.C. Gower,
 Measures of similarity, dissimilarity, and distance.
 In: S. Kotz, N.L. Johnson (Eds.),
 {\it Encyclopedia of Statistical Science}, vol. 5,
 Wiley, New York, 397-405, 1985.

  \bibitem {gramm08} J. Gramm,
 Closest string problem. In: M.-Y. Kao, (Ed.),
 {\it Encyclopedia of Algorithms}, Springer, pp. 156-158, 2008.

  \bibitem {gramm03} J. Gramm, R. Niedermaier, P. Rossmanith,
 Fixed-parameter algorithms for closest string and related problems.
 {\it Algorithmica},
  37(1), 25-42, 2003.

 \bibitem {gui06} S. Guillernot, F. Nocolas,
 Solving the maximum agreement subtree and the maximum compatible
 tree problems on many bounded degree trees.
 In: M. Lewenstein, G. Valiente (Eds.),
 {\it Proc. of the Int. Conf. on Combinatorial Pattern Matching CPM 2006},
 LNCS 4009, Springer, 165-176, 2006.

   \bibitem {gusfield99} D. Gusfield,
 {\it Algorithms on Strings, Trees and Sequences: Computer Science and
 Computational Biology.}
 Cambridge University Press, Cambridge, 1999.

 \bibitem {hallett07} M.T. Hallett, C. McCartin,,
 A faster FPT algorithm for the maximum agreement forest problem.
 {\it Theory of Computing Systems}, 41(3), 539-550, 2007.

  \bibitem {hamel96} A.M. Hamel, M.A. Steel,
 Finding a maximum compatible tree is NP-hard for sequences and
 trees.
 {\it Applied Mathematical Letters},
 9(2), 55-59, 1996.

  \bibitem {hanen95} S. Hannenhalli, P. Pevzner,
 Transforming men into mice (polynomial algorithm for genomic distance problem).
 {\it 36th Ann. Symp. on Foundations of Computer Science FOCS
 1995},
 Milwaukee, Wisconsin, IEEE Computer Society, 581-592, 1995.

% \bibitem {hanen95} S. Hannenhalli, P. Pevzner,
% {\it Transforming Men into Mice}.
% Technical Report CSE-95-012, Dept. of Computer Science and Engineering,
% Pennsylvania State Univ., 1995.

  \bibitem {harel87}  D. Harel,
 STATECHARTS: A visual formalism for complex systems,
  {\it Science of Computer Programming},
   8(3), 231-274, 1987.

 \bibitem {hax03} Y. Haxhimusa, R. Gland, W.G. Kropatsch,
 Constructing stochastic pyramids by MIDES -
 maximal independent directed edge set.
  In:
 W. Hancock, M. Vento (Eds.),
 {\it Proc. IAPR Workshop GbPRP 2003},
 LNCS 2726, Springer, 24-34, 2003.

  \bibitem {her97} F. Herrera, E. Herrera-Viedma, J.L. Verdegay,
 A rational consensus model in group decision making
 using linguistic assessments.
 {\it Fuzzy Sets and Systems}, 88(1), 31-49, 1997.

  \bibitem {her02} E. Herrera-Viedma, F. Herrera, F. Chiclana,
   A consensus model for multiperson decision making with different
   preference structures.
    {\it IEEE Trans. SMC, Part A}, 32(3), 394-402, 2002.

 \bibitem {hire07} C. Hiremath, R.R. Hill,
 New greedy heuristics for the multiple-choice
 multidimensional knapsack problem.
 {\it Int. J. of Operational Research},
 2(4), 495-512, 2007.

  \bibitem {hirsch77} D.S. Hirschberg,
 Algorithms for the longest common subsequence problem.
 {\it J. of the ACM},
 24(4), 664-675, 1977.

  \bibitem {hoang11} V.T. Hoang, W.-K. Sung,
  Improved algorithms for maximum agreement and compatible supertrees.
  {\it Algorithmica} 59(2), 195-214, 2011.

  \bibitem {jansson05} J. Jansson, J.H.-K. Ng,
 K. Sadakane, W.-K. Sung,
 Rooted maximum agreement supertrees.
 {\it Algorithmica}, 43(4), 293-307, 2005.

 \bibitem {jech02} T. Jech,
 {\it Set Theory}, Springer, Berlin, 2002.

  \bibitem {jiang95} T. Jiang, M. Li,
 On the approximation of shortest common supersequences and
 longest common subsequences.
 {\it SIAM J. on Computing}, 24(5), 1122-1139, 1995.

  \bibitem {jiang95a} T. Jiang, V.G. Timkovsky,
 Shortest consistent superstring computable in polynomial time.
 {\it Theor. Comput. Sci.}, 143(1), 113-122, 1995.

 \bibitem {jiang95b} T. Jiang, L. Wang, K. Zhang,
 Alignment of trees - an alternative to tree edit.
 {\it Theoretical Computer Science}, 143(1), 137-148, 1995.

 \bibitem {jiang01} X. Jiang, A. Munger, H. Bunke,
 On median graphs: properties, algorithms, and applications.
 {\it IEEE Trans. PAMI}, 23(10), 1144-1151, 2001.

 \bibitem {jiang04} X. Jiang, H. Bunke, J. Csirik,
 Median strings: A review.
 In: M. Last, A.A. Kandel, H. Bunke, (Eds.),
 {\it Data Mining in Time Series Databases}.
 World Scientific, 173-192, 2004.

 \bibitem {john83} D.S. Johnson, K.A. Niemi,
 On knapsack, partitions, and a new dynamic programming techniques
 for trees.
 {\it Mathematics of Operations Research}, 8(1), 1-14, 1983.

  \bibitem{jon81} J.C. Jones,
 {\it Design Methods}, J. Wiley \& Sons, New York, 1981.

  \bibitem {kaplan05} H. Kaplan, N. Sharir,
 The greedy algorithm for shortest superstrings.
 {\it Inf. Process. Lett.},
 93(1), 13-17, 2005.

  \bibitem {keller04} H. Kellerer, U. Pferschy, D. Pisinger,
 {\it Knapsack Problems}. Berlin, Springer, 2004.

 \bibitem {kelsey10} T. Kelsey, L. Kotthoff,
 The exact closest string problem as a constraint satisfaction
 problem. Electronic preprint,
 CoRR, abs./1005.0089, 2010.

 \bibitem {kemeny59} J. Kemeny,
 Mathematics without Numbers.
 {\it Daedelus}, vol. 88, 577-591, 1959.

% \bibitem {kemeny60} J. Kemeny, L. Snell,
% {\it Mathematical Models in the Social Sciences}.
% Ginn., Boston, 1960.

  \bibitem {kemeny60}
%  \bibitem {kem72}
  J.G. Kemeny, and J.L. Snell,
 {\it Mathematical Models in the Social Sciences},
 Reprint, Cambridge, Mass., The MIT Press, 1972.

 \bibitem {kendall62} M. Kendall,
 {\it Rank Correlation Methods}, 3rd ed., Hafner, New York, 1962.

 \bibitem {kick78} W.J.M. Kickert,
 {\it Fuzzy Theories on Decision Making}.
  Nijhoff, London, 1978.

 \bibitem {klam00} Klamroth, M. Wiecek,
 Dynamic programming approaches
 to the multiple criteria knapsack problem.
 {\it Naval Research Logistics},
 47(1), 57-76, 2000.

 \bibitem {knuth98} D.E. Knuth,
 {\it The Art of Computer Programming},
 Vol. 2: Seminumerical Algorithms. (3rd ed.).
 Addison Wesley,
% . pp. 694,
  1998.
% ISBN 0201896842.
% Knuth also lists other names that were proposed for multisets,
% such as list, bunch, bag, heap, sample, weighted set, collection, and suite.

 \bibitem {kohonen85} T. Kohonen,
 Median strings.
 {\it Pattern Recognition Letters},
 3(5), 309-313, 1985.

 \bibitem {kruger08} H.A. Kruger, W.D. Kearney,
 Consensus ranking - An ICT security awareness case study.
 {\it Computers \& Security},
 27(7-8), 254-259, 2008.
% Print Description of two methods:
% (i) assignment problem, (ii) heuristics

 \bibitem {kuh57} H.W. Kuhn,
 The Hungarian method for the assignment problems.
 {\it Naval Res. Log.}, 2(1-2), 83-97, 1955
 (reprinted in
 {\it Naval Res. Log.}, 52(1), 7-21, 2005).

%%%%%%%%%%%%%%%%%%%%%%%%%%%%%%%%%%%%%%%%%%%%%%%%%%%%%%%% NB ??
%  \bibitem {kuz82} V.B. Kuzjmin,
% {\it Building of Group Decisions in Space of Deterministic
% and Fuzzy Relations.}
% Moscow, Nauka, 1982 (in Russian).

 \bibitem {leven66} V.I. Levenshtein,
 Binary codes capable of correcting deletions, insertions and reversals.
 {\it Cybernetics and Control Theory},
 10(8), 707-710, 1966.

 \bibitem {lev88} M.Sh. Levin,
 Vector-like proximity estimates for structures.
 In:
 {\it All-Russian Conf. ``Problems and Methods of Decision Making
 in Organizational Management Systems''},
 Moscow, Inst. for System Analysis, 116-117, 1988 (in Russian).

 \bibitem {lev96} M.Sh. Levin,
  Hierarchical decision making for education in information engineering.
  In: {\it Proc. of 7th Ann. Eur. Conf. of EAEEIE},
  Oulu, pp. 301-307, 1996.

  \bibitem {lev98a} M.Sh. Levin,
  Towards comparison of decomposable systems.
  In: {\it Data Science, Classification, and Related Methods},
%  Springer-Verlag,
  Springer, Tokyo, pp. 154-161, 1998.

  \bibitem{lev98} M.Sh. Levin,
 {\it Combinatorial Engineering of Decomposable Systems}.
 Kluwer, 1998.

 \bibitem {lev00} M.Sh. Levin,
  Towards systems engineering education.
  In: {\it Proc. of 15th Eur. Meeting on cybern.. and Syst. Res. EMCSR'2011},
 vol. 1,
  Vienna, pp. 257-262, 2000.

  \bibitem {lev01} M.Sh. Levin,
  System synthesis with morphological clique problem:
  fusion of subsystem evaluation decisions,
  {\it Inform. Fusion}, 2(3), 225-237, 2001.

% \bibitem {leved02} M.Sh. Levin,
%  "Combinatorial evolution of composite systems",
%  {\it Proc. of 16th
%  Eur. Meeting on Cybern. and Syst. Res.}, vol. 1, pp. 275-280, 2002.

% \bibitem {lev02} Levin M.Sh.,
%  Towards combinatorial planning of human-computer systems,
%  {\it Applied Intelligence}, 2002, vol. 16, no. 3, pp. 235-247.

  \bibitem {lev03} M.Sh. Levin,
  Common part of preference relations.
  {\it Foundations of Computing \& Dec. Sciences},
  28(4), 223-246, 2003.

 \bibitem {lev05} M.Sh. Levin,
 Modular system synthesis: example for composite packaged software,
 {\it IEEE Trans. on SMC - Part C}, 35(4), 544-553, 2005.

 \bibitem {lev06} M.Sh. Levin,
 {\em Composite Systems Decisions}, New York, Springer, 2006.

% \bibitem {lev06e} M.Sh. Levin,
% ``Course 'System design: structural approach,''
% {\it ASME Int. Design  Engineering Technical Conferences and
% Computers and Information in Engineering Conference
% (IDETC/CIE2006)}, Paper no. DETC2006-99547, Vol. 4a,
% {\it 18th Int. Conf. Design Methodology and Theory DTM2006},
%  USA, pp. 475-484, 2006.

% \bibitem {lev06e} Levin M.Sh.,
% Course 'System design: structural approach,
% {\it ASME Int. Design  Engineering Technical Conferences and
% Computers and Information in Engineering Conference
% (IDETC/CIE2006)},
% {\it Proc. of 18th Int. Conf. Design Methodology and Theory DTM2006},
% Vol. 4a, Paper no. DETC2006-99547,  USA, 2006, pp. 475-484.

 \bibitem{lev07clust} M.Sh. Levin,
 Towards hierarchical clustering,
 In: V. Diekert, M. Volkov, A. Voronkov, (Eds.),
 {\it CSR 2007}, LNCS 4649, Springer, 205-215, 2007.

%% \bibitem {levclust07} M.Sh. Levin,
%% Towards hierarchical clustering,
% in: V. Diekert, M. Volkov, A. Voronkov (Eds.),
%%  {\it CSR-2007},
%% LNCS 4649, Springer, 2007, pp. 205-215.

 \bibitem {levprob07} M.Sh. Levin,
 Combinatorial technological systems problems
 (examples for communication system),
  {\it Proc. of Int. Conf. on Systems Engineering and Modeling ICSEM-2007},
  Israel, pp. 24-32, 2007.

% \bibitem{lev07haifa} M.Sh. Levin,
%  Combinatorial technological systems problems
% (examples for communication system).
% {\it Intl. Conf. on Systems Engineering and Modeling ICSEM-2007},
% 24-32, March 20-23, Israel, 2007.

%%%%%%%%%%%%%%% CSEDU -REF CORRECT NB!!!
% \bibitem {lev07} M.Sh. Levin,
% ''Combinatorial technological systems models
% (Examples for communication system)'',
% {\it 2nd Int. Conf. ICSEM'07},
%  Haifa, 2007, pp. 24-32.
%%%%%%%%%%%%%%%%%%%%%%%%%%%%%%%%%%%%%%%

 \bibitem  {lev08a} M.Sh. Levin,
 Morphological approach to electronic shopping,
 {\it 2008 IEEE Region 8 Int. Conf. ``SIBIRCON-2008''},
  Novosibirsk, pp. 280-285, 2008.

% \bibitem {lev08turku} M.Sh. Levin,
% ``Towards four-layer framework of combinatorial problems,''
% {\it 32rd Annual Int. IEEE Conf. COMPSAC-2008}, Turku,
% 2008, pp. 873-878.



 \bibitem {lev09} M.Sh. Levin,
 Combinatorial optimization in system configuration design,
 {\it Automation and Remote Control}, 70(3), 519-561, 2009.

 \bibitem  {lev10a} M.Sh. Levin,
 Towards communication network development
 (structural systems issues, combinatorial models),
 {\it Proc. of IEEE Region 8 Int. Conf.  Sibircon 2010},
 vol. 1, pp. 204-208, 2010.

%\bibitem {levsib10} M.Sh. Levin,
% Towards communication network development (structural systems
% issues, combinatorial problems).
% {\it IEEE Region 8 Int. Conf. ``Sibircon 2010''}, vol. 1, 204-208, 2010.

   \bibitem{lev11arch} M.Sh. Levin,
  Restructuring in combinatorial optimization. Electronic
  preprint. 11 pp., Febr. 8, 2011.
  http://arxiv.org/abs/1102.1745 [cs.DS]

 \bibitem {lev11ed}  M.Sh. Levin,
  Course on system design (structural approach).
  Electronic preprint, 22 pp.,
 Febr. 19, 2011.
 http://arxiv.org/abs/1103.3845 [cs.DS]

 \bibitem {lev11inf}  M.Sh. Levin,
 Towards configuration of applied Web-based information system.
  Electronic preprint, 13 pp.,
 Aug. 31, 2011.
 http://arxiv.org/abs/3108.3865 [cs.SE]

%%%%%%%%%%%%%%%%%%%%%%%%%%%%%%%%%%%%%%%%%%%%%%%%%%%%%%%%%%%%%%%%%%%%%%%%%%
% \bibitem {lev11ADES} M.Sh. Levin,
%  Four-layer framework for combinatorial optimization problems domain.
%   {\it Advances in Engineering Software}, 42(12), 1089-1098, 2011.
%
% Ref
%%%%%%%%%%%%%%%%%%%%%%%%%%%%%%%%%%%%%%%%%%%%%%%%%%%%%%%%%%%%%%%%%%%%%%%%%%

  \bibitem {levmih88}  M.Sh. Levin, A.A. Mikhailov,
 {\it Fragments of Objects Set Stratification Technology},
  Preprint, Moscow, Inst. for Systems
  Studies (RAS), 60 p., 1988 (in Russian)

% \bibitem {levsaf06}  Levin M.Sh.,  Safonov A.V.,
% Design and redesign of configuration for facility in communication
% network,
% {\it Information Technologies and Computer Systems}, 2006,
%  no. 4, pp. 63-73 (in Russian).

% \bibitem {levfim07} M.Sh. Levin, A.V. Fimin,
% ``Hierarchical design of fire alarm wireless sensor element,''
%  {\it Proc. of 7th Int. Conf. CAD/CAM/PDM-2007},
%  Inst. of Control Problems, Russian Acad. of Sci., Moscow, Russia,
% pp. 33-35, Oct. 2007.

 \bibitem{levkhod07} M.Sh. Levin, I.A. Khodakovskii,
 Structural composition of the telemetry system.
 {\it Automation and Remote Control}, 68(9), 1654-1661, 2007.

%  \bibitem{levrya07} M.Sh. Levin, V.A. Ryabov,
% ``Morphological tree model for communication protocol
% (example),''
% {\it Proc. of 7th Int. Conf. CAD/CAM/PDM-2007,}
%  Inst. of Control Problems, Russian Acad. of Sci., Moscow, Russia,
% pp. 35-39, Oct. 2007.

  \bibitem{levleus09} M.Sh. Levin, A.V. Leus,
  Configuration of integrated security system.
  {\it 7th IEEE Int. Conf. on Industrial Informatics INDIN 2009},
 Cardiff, UK, pp. 101-105, 2009.

% \bibitem {levsaf10} Levin M.Sh., Safonov A.V.,
%  Towards improvement of regional communication network,
% Information Processes, 2010,
% vol. 10, no. 3, pp. 212-223 (in Russian).

% \bibitem {levfim10} Levin M.Sh., Fimin A.V.,
%  Configuration of alarm wireless sensor element,
%  {\it
%  Proc. of
%  Int. Conf. on Ultra Modern Telecommunication
%  'ICUMT 2010'},
%  Moscow, 2010, art. no. 1569336084.
%  pp. 00-00,

  \bibitem {levsaf10a} M.Sh. Levin, A.V. Safonov,
  Towards modular redesign of networked system,
 {\it Proc. of
 Int. Conf. on Ultra Modern Telecommunication 'ICUMT 2010'},
 Moscow,
% art. no. 1569336089.
  pp. 109-114, 2010.

 \bibitem {levand10} M.Sh. Levin, A. Andrushevich, R. Kistler, A. Klapproth,
 Combinatorial evolution of ZigBee protocol.
 {\it IEEE Region 8 Int. Conf. Sibircon 2010},
  vol. 1, 314-319, 2010.

 \bibitem {levsaf11} M.Sh. Levin, A.V. Safonov,
 Improvement of regional telecommunications networks.
 {\it J. of Communications Technology and Electronics},
  56(6), 770-778, 2011.

 \bibitem {levand11} M.Sh. Levin, A. Andrushevich, A. Klapproth,
  Improvement of building automation system.
  K.G. Mehrotra et al. (Eds.),
  {\it Proc. of 24th Int. Conf.  IEA/AIE 2011}, LNCS 6704, Part II,
  Springer,
%  Heidelberg,
  459-468, 2011.

 \bibitem {li02} M. Li, B. Ma, L. Wang,
 On the closest string and substring problem.
 {\it J. of the ACM},
 49(2), 157-171, 2002.

 \bibitem {lin96} J. Lin,
 Integration of weighted knowledge bases.
 {\it Artificial Intelligence},
 83(2), 363-378, 1996.

 \bibitem {lust10} T. Lust, J. Teghem,
 The multiobjective multidimensional knapsack problem:
 a survey and a new approach.
 Techn.  Rep., arXiv: 1007.4063v1, 2010.
 http: //arxiv.org/abs/1007.4063.v1

 \bibitem {ma08} B. Ma, X. Sun,
 More efficient algorithms for closest string and substring
 problems.
 In: M. Vingron, L. Wong (Eds.),
 {\it Int. Conf. on Research in Computational Molecular Biology
 RECOMB 2008},
% LNBI
 LNCS 4955, Springer, pp. 396-409, 2008.

 \bibitem {maier78} D. Maier,
 The complexity of some problems on subsequences and
 supersequences.
 {\it J. of the ACM}, 25(2), 322-336, 1978.

 \bibitem {marzal93} A. Marzal, E. Vidal,
 Computation of normalized edit distance and applications.
 {\it IEEE Trans. PAMI}, 15(9), 926-932, 1993.

 \bibitem {mirkin70} B.G. Mirkin, L.B. Chernyi,
 On measurement of distance between partition of a finite set of
 units.
 {\it Automation and Remote Control},
 31, 786-792, 1970.

 \bibitem {mou12} S.R. Mousavi, F. Tabataba,
 An improved algorithm for the longest common subsequence problem.
 {\it Comp. and Oper. Res.},
 39(3), 512-520, 2012.

 \bibitem {nicolas03} F. Nicolas, E. Rivals,
 Complexities of centre and median string problems.
 In: R. Baeza-Yates, E. Chavez, M. Crochemore (Eds.),
 {\it Proc. 14th Annual Symp. on Combinatorial Pattern Matching CPM 2003},
 LNCS 2676, Springer,
 315-327, 2003.
% Mexico, June 25-27, 2003

%  \bibitem {noy00} N. Noy, M. Musen,
% PROMPT: Algorithm and tool for automated ontology merging and
% alignment.
% In:
% {\it Proc. AAAI2000}, 450-455, AAAI Press, 2000.

% \bibitem {noy03} N. Noy, M. Musen,
% The PROMPT suit:
% Interactive tools for ontology merging and mapping.
% {\it Int. J. of Human-Computer Interaction},
% 00(00), 450-455, 2003.

% \bibitem {noy04} N. Noy, M. Musen,
% Ontology versioning in an ontology management framework.
% {\it IEEE Intelligent Systems}, 19(4), 6-13, 2004.

 \bibitem {noy05} N. Noy, A.H. Doan, A.Y. Halevy,
 Semantic integration.
 {\it AI Magazine},
 26(1), 7-10, 2005.

  \bibitem {parra02} R. Parra-Hernandez, N. Dimopoulos,
  A new heuristic for solving the multichoice
  multidimensional knapsack problem.
  {\it IEEE Trans. SMC - Part A},
 35(5), 708-717, 2002.

 \bibitem {phillips96} C. Phillips, T.J. Warnow,
 The asymmetric tree - A new model for building consensus tree.
 {\it Discrete Applied Mathematics}, 71(1-3), 311-335, 1996.
%  NP-hard problem for 3 and more trees. Polynomial - for 2 trees.

 \bibitem {pinto01} H.S. Pinto, J.P. Martins,
 A methodology for  ontology integration. In:
 {\it Proc. of 1st Int. Conf. Knowledge Capture K-CAP'01},
 Victoria, Canada, Sweden, 113-138, 2001.

 \bibitem {rasa12} S. Rasoul Mousavi, F. Tabataba,
 An improved algorithm for the longest common subsequence problem.
 {\it EJOR}, 39(3), 512-520.

  \bibitem {ray02} J.W. Raymond, E.J. Gardiner, P. Willett,
 RASCAL: calculation of graph similarity using maximal common edge
 subgraph.
 {\it The Computer J.}, 45(6), 631-644, 2002.

  \bibitem {robles04} A. Robles-Kelly, E.R. Hancock,
 string edit distance, random walks and graph matching.
 {\it Int. J. of Pattern Recognition and Artificial
 Intelligence}, 18(3), 315-327, 2004.

 \bibitem {rodr07} E.M. Rodrigues, M.F. Sagot, Y. Wakabayashi,
 The maximum agreement forest problem:
 Approximation algorithms and computational experiments.
 {\it Theoretical Computer Science},
 374(1-3), 91-110, 2007.

  \bibitem {roy96} B. Roy,
  {\it Multicriteria Methodology for Decision Aiding}.
  Kluwer Academic Publishers, Dordrecht, 1996.

 \bibitem {sasha89} D. Sasha, K. Zhang,
 Simple fast algorithms for
 the editing distance between trees and related problems.
 {\it SIAM J. on Computing},
 18(6), 1245-1262, 1989.

 \bibitem {selkow77} S.M. Selkow,
 The tree-to-tree editing problem.
 {\it Information Processing Letters},
 6(6), 184-186, 1977.

  \bibitem {semple00} C. Semple, M. Steel,
 A supertree method for rooted trees.
 {\it Discrete Appl. Math.}, 105(1-3), 147-158, 2000.

 \bibitem {sim99} J.S. Sim, K. Park,
 The consensus string problem for a metric is NP-complete.
 In: R. Raman, J. Simpson (Eds.),
 {\it Proc. of the 10th Australian Workshop on Combinatorial
 Algorithms}, Perth, WA, Australia, pp. 107-113, 1999.

 \bibitem {sim58} H.A. Simon, A. Newell,
 Heuristic problem solving: the next advance in operations
 research.
  {\it Operations Research}, 6(1), 1-10, 1958.

  \bibitem {solnon07}
 % \bibitem {jolion07}
  C. Solnon, J.-M. Jolion,
 Generalized vs set median string for histogram based distances:
 Algorithms and classification results in image domain.
 In: F. Escolano, M. Vento (Eds.),
 {\it Proc. of the 6th Workshop on Graph Based Representation
 in Pattern Recognition GbPrP 2007}, LNCS 4538, Springer,
% IARP-TC-15
 404-414, 2007.

%  \bibitem {sousa06} J.P. Sousa, V. Poladian,
%  D. Garlan, B. Schmerl, W. Shaw,
% Task-based adaptation for ubiquitous computing,
% {\it IEEE Trans. on SMC - Part C}, 36(3), 328-340, 2006.

  \bibitem {steel93} M. Steel, T. Warnow,
 Kaikoura tree theorems:
 Computing the maximum agreement subtree.
 {\it Inf. Process. Lett.}, 48(2), 77-82, 1993.

 \bibitem {syr01} A. Syropoulos,
% Syropoulos, Apostolos,
  Mathematics of Multisets, in
   C. S. Calude et al. (eds.),
  {\it Multiset Processing: Mathematical, Computer
 Science, and Molecular Computing Points of View},
  LNCS 2235, Springer, pp.347-358, 2001.
%  Retrieved from
% "http://en.wikipedia.org/wiki/Multiset"

 \bibitem {tai79} K.-C. Tai,
 The tree-to-tree correction problem.
 {\it J. of the ACM}, 26(3), 422-433, 1979.

 \bibitem {tanaka94} E. Tanaka,
 A metric between unrooted and unordered trees and
 its bottom-up computing method.
 {\it IEEE Trans. PAMI}, 16(12), 1233-1238, 1994.

 \bibitem {tanaka88} E. Tanaka, K. Tanaka,
 The tree-to-tree editing problem.
 {\it Int. J. Pattern Recogn. and Art. Intell.},
 2(2), 221-240, 1988.

  \bibitem {tar88} J. Tarhio, E. Ukkonen,
 A greedy approximation algorithm for constructing shortest common
 superstrings.
 {\it Theor. Comput. Sci.},
 57(1), 131-145, 1988.

 \bibitem {tavana03} M. Tavana,
 CROSS: a multicriteria group-decision-making model for evaluating
 and prioritizing advanced-technology projects at NASA.
 {\it Interfaces}, 33(3), 40-56, 2003.

  \bibitem {tim90} V.G. Timkovsky,
 Complexity of common subsequence and supersequence problems.
% Kibernetika 5, 1-13, 1989 in Russian
 {\it Cybernetics and Systems Analysis}, 25(5), 565-580, 1989.

 \bibitem {torsello01} A. Torsello, E.R. Hancock,
 Efficiently computing weighted tree edit-distance using
 relaxation labeling.
  In: M. Figueiredo, J. Zerubia, A.K. Jain (Eds.),
  {\it Energy Minimization Methods in Computer Vision
  and Pattern Recognition}. LNCS 2134, Springer,
  438-453, 2001.

  \bibitem {torsello05} A. Torsello, D. Hidovic, M. Pelillo,
  Polynomial-time metrics for attributed trees.
  {\it IEEE Trans. PAMI},
  27(7), 1087-1099, 2005.

  \bibitem {valiente01} G. Valiente,
% Gabriel Valiente
  An efficient bottom-up distance between trees.
  {\it Proc. of the 8th Int. Symp. String Processing Information Retrieval
  SPIRE 2001},
 IEEE Computer Science Press,
 Laguna de San Rafael,
  Chile
  212-219, 2001.

 \bibitem {valiente02} G. Valiente,
 {\it Algorithms on Trees and Graphs}.
  Springer, Berlin, 2002.

  \bibitem {vidal95} E. Vidal, A. Marzal, P. Aibar,
  Fast computation of normalized edit distances.
  {\it IEEE Trans. PAMI},
  17(9), 899-902, 1995.

 \bibitem {wache01} H. Wache, T. Vogele, U. Visser, H.
 Stickenschmidt, G. Schuster, H. Neumann, S. Hubner,
 Ontology-based integration of information - A survey of existing
 methods,
 In: {\it Proc. of Int. Workshop Ontologies and Information Sharing
 IJCAI-01},
 Seattle, WA, 108-117, 2001.

 \bibitem {wagner74} R.A. Wagner, M.J. Fisher,
 The string-to-string correction problem.
 {\it J. of the ACM}, 21(1), 168-173, 1974.

  \bibitem {wallis01} W.D. Wallis, P. Shoubridge, M. Kraetz,
 D. Ray,
 Graph distances using graph union.
 {\it Pattern Recognition Letters},
 22(6-7), 701-704, 2001.

 \bibitem {wang01} J.T.-L. Wang, K. Zhang,
 Finding similar consensus between trees: An algorithm and
 distance hierarchy.
 {\it Pattern Recognition}, 34(1), 127-137, 2001.

 \bibitem {wang08} J. Wang, J. Chen, J.M. Huang,
 An improved lower bound on approximation algorithms for the
 closest substring problem.
 {\it Information Processing Letters},
 107(1), 24-28, 2008.

 \bibitem {whidden10} C. Whidden, R.G. Beiko, N. Zeh,
 Fast FPT algorithms for computing rooted agreement forest:
 Theory and experiments.
 In:
 {\it Proc. of the 9th Int. Symp. on Experimental Algorithms
 SEA 2010},
 LNCS 6049, Springer, 141-153, 2010.

 \bibitem {whidden11} C. Whidden, R.G. Beiko, N. Zeh,
 Fixed parameter and approximation
 algorithms for maximum agreement forests.
 Electronic preprint, 12 Aug. 2011,
 arXiv: 1108.2664v1 [q-bio.PE]

 \bibitem {yager88} R.R. Yager,
 On ordered weighted averaging aggregation operators
 in multicriteria decision making.
 {\it IEEE Trans. SMC}, 18(1), 183-190, 1988.

 \bibitem {yang91} W. Yang, Identifying syntactic differences
 between two programs.
 {\it Software - Practice and Experience},
 21(7), 739-755, 1991.

 \bibitem {zhang96a} K. Zhang,
  A constrained edit-distance between unordered labeled trees.
  {\it Algorithmica}, 15(3), 205-222, 1996.

 \bibitem {zhang89} K. Zhang, D. Sasha,
 Simple fast algorithms for the editing distance between trees and
 related problems.
 {\it SIAM J. Comput.}, 18(6), 1245-1262, 1989.

 \bibitem {zhang92} K. Zhang, R. Statman, D. Sasha,
 On the  editing distance between unordered labeled trees.
 {\it Inform. Process. Letters}, 42(3), 133-139, 1992.

  \bibitem {zhang96} K. Zhang, J.T.L. Wang, D. Shasha,
 On the editing distance between undirected acyclic graphs.
 {\it Int. J. Foundations of Computer Science},
 7(1), 43-57, 1996.

% \bibitem {doum00} M. Doumpos, C. Zopounidis, P.M. Pardalos,
% Multicriteria sorting methodology: application to financial decision problems.
%  {\it Parallel Algorithms Appl.}, 15(1-2), 113–-129, 2000.

 \bibitem {zitz99} E. Zitzler, L. Thiele,
 Multiobjective evolutionary algorithms:
 a comparative case study and the strength Pareto approach.
 {\it IEEE Trans. on Evolutionary Computation}, 3(4),
 257-271, 1999.

% \bibitem {zop00} C. Zopounidis, M. Doumpos,
% PREFDIS: a multicriteria decision support system for sorting
% decision problems.
%  {\it Computers and Oper. Res.}, 127(7-8), 779–-797, 2000.

 \bibitem {zap02} C. Zopounidis, M. Doumpos,
  Multicriteria classification and sorting methods:
  a literature review,
  {\it EJOR}, 138(2), 229–-246, 2002.

  \bibitem {zop09} C. Zopounidis, M. Doumpos,
 Multicriteria sorting methods.
 In:
  C.A. Floudas, P.M. Pardalos (Eds.),
  Encyclopedia of Optimization. 2nd ed., Springer, pp. 2379-2396, 2009.

   \bibitem {zwi69} F. Zwicky,
  {\it Discovery Invention, Research Through the Morphological Approach.}
  New York: McMillan, 1969.
\end{document}